\documentclass[iop,apj,12pt,twocolappendix]{emulateapj}

\usepackage{acronym}
\usepackage{booktabs}
\usepackage{dcolumn}
\usepackage{graphicx}
\usepackage{amsmath,amsfonts,amssymb}
\usepackage{mathrsfs}
\usepackage[utf8]{inputenc}
\usepackage[normalem]{ulem}
\usepackage{multirow}

\usepackage[breaklinks,colorlinks,citecolor=blue]{hyperref}

\usepackage{dcolumn}
\newcommand{\myhead}[1]{\multicolumn{1}{c}{#1}}
\newcolumntype{d}[1]{D{.}{.}{#1}}

\usepackage{natbib}
\usepackage{tensind}
\tensordelimiter{?}
\tensorformat{lrc}

\newcommand{\ie}{\textit{i.e.}}
\newcommand{\eg}{\textit{e.g.}}

\usepackage{color}

\newcommand{\eff}{\mathrm{eff}}

\begin{document}

\title{Binary Neutron Star Mergers: \\
Mass Ejection, Electromagnetic Counterparts, and Nucleosynthesis}
\author{David Radice\altaffilmark{1,2},
Albino Perego\altaffilmark{3,4},
Kenta Hotokezaka\altaffilmark{2},
Steven A. Fromm\altaffilmark{5},\\
Sebastiano Bernuzzi\altaffilmark{6,3},
and Luke F. Roberts\altaffilmark{5}}
\altaffiltext{1}{Institute for Advanced Study, 1 Einstein Drive,
Princeton, NJ 08540, USA}
\altaffiltext{2}{Department of Astrophysical Sciences, Princeton University,
4 Ivy Lane, Princeton, NJ 08544, USA}
\altaffiltext{3}{Istituto Nazionale di Fisica Nucleare, Sezione Milano
Bicocca, gruppo collegato di Parma, I-43124 Parma, Italy}
\altaffiltext{4}{Dipartimento di Fisica, Universit\`{a} degli Studi di
Milano Bicocca, Piazza della Scienza 3, 20126 Milano, Italy}
\altaffiltext{5}{NSCL/FRIB and Department of Physics \& Astronomy,
  Michigan State University, 640 S Shaw Lane East Lansing, MI 48824,
  USA}
\altaffiltext{6}{Theoretisch-Physikalisches
  Institut, Friedrich-Schiller-Universit{\"a}t Jena, 07743, Jena,
  Germany}

\begin{abstract}
We present a systematic numerical relativity study of the mass ejection
and the associated electromagnetic transients and nucleosynthesis from
binary neutron star (NS) mergers. We find that a few $10^{-3}\, M_\odot$
of material are ejected dynamically during the mergers. The amount and
the properties of these outflow depend on binary parameters and on the
NS equation of state (EOS). A small fraction of these ejecta, typically
${\sim}10^{-6}\, M_\odot$, is accelerated by shocks formed shortly after
merger to velocities larger than $0.6\, {\rm c}$ and produces bright
radio flares on timescales of weeks, months, or years after merger.
Their observation could constrain the strength with which the NSs bounce
after merger and, consequently, the EOS of matter at extreme densities.
The dynamical ejecta robustly produce second and third $r$-process peak
nuclei with relative isotopic abundances close to solar. The production
of light $r$-process elements is instead sensitive to the binary mass
ratio and the neutrino radiation treatment. Accretion disks of up to
${\sim}0.2\, M_\odot$ are formed after merger, depending on the lifetime
of the remnant. In most cases, neutrino- and viscously-driven winds from
these disks dominate the overall outflow. Finally, we generate synthetic
kilonova light curves and find that kilonovae depend on the merger
outcome and could be used to constrain the NS EOS.

\end{abstract}


\section{Introduction}
\label{sec:introduction}
Merging \acp{NS} are loud \ac{GW} sources and power bright \ac{EM}
transients, as recently demonstrated by GW170817
\citep{theligoscientific:2017qsa, abbott:2018wiz, gbm:2017lvd,
monitor:2017mdv, arcavi:2017a, coulter:2017wya, drout:2017ijr,
evans:2017mmy, hallinan:2017woc, kasliwal:2017ngb, nicholl:2017ahq,
smartt:2017fuw, soares-santos:2017lru, tanvir:2017pws, troja:2017nqp,
mooley:2017enz, ruan:2017bha, lyman:2018qjg}. AT2017gfo, the \ac{EM}
counterpart to GW170817, had a UV/optical/infrared component with
blackbody-like spectrum that has been interpreted as the result of the
radioactive decay of about $0.05\ M_\odot$ of material ejected during
and shortly after the merger \citep{chornock:2017sdf,
cowperthwaite:2017dyu, drout:2017ijr, nicholl:2017ahq, tanaka:2017qxj,
tanvir:2017pws, utsumi:2017cti, perego:2017wtu, villar:2017wcc,
waxman:2017sqv, metzger:2018uni, kawaguchi:2018ptg}. Non thermal
emissions from the radio to the $\gamma$-ray bands was also observed.
The origin of this \ac{EM} component are less clear, but recent
observations point to the presence of synchrotron radiation generated by
a relativistic jetted outflow interacting with the interstellar medium
(ISM; \citealt{monitor:2017mdv, kasliwal:2017ngb, margutti:2017cjl,
mooley:2017enz, lazzati:2017zsj, kim:2017skw,  margutti:2018xqd,
nakar:2018uwt, hotokezaka:2018gmo, resmi:2018wuc, alexander:2018dcl,
barkov:2018viy, mooley:2018qfh, ghirlanda:2018uyx}).

These groundbreaking observations are being used to constrain general
relativity (GR; \citealt{lombriser:2015sxa, mcmanus:2016kxu,
monitor:2017mdv, baker:2017hug, visinelli:2017bny, pardo:2018ipy}),
\acused{GR}cosmological parameters \citep{abbott:2017xzu,
hotokezaka:2018dfi}, the \ac{EOS} of \acp{NS} \citep{margalit:2017dij,
shibata:2017xdx, rezzolla:2017aly, ruiz:2017due, bauswein:2017vtn,
annala:2017llu, radice:2017lry, fattoyev:2017jql, paschalidis:2017qmb,
drago:2018nzf, most:2018hfd, de:2018uhw, tews:2018iwm, malik:2018zcf,
abbott:2018exr, tsang:2018kqj, hinderer:2018pei, wei:2018dyy,
nandi:2018ami}, and the origin of short $\gamma$-ray bursts (SGRBs;
\citealt{monitor:2017mdv, mooley:2017enz, lazzati:2017zsj,
finstad:2018wid, beniamini:2018uwo}), among others.  \acused{SGRB}

\ac{NS} mergers also generate neutron-rich outflows that are thought to
be a likely site for the production of $r$-process nuclei
\citep{lattimer:1974a, symbalisty:1982a, meyer:1989a, eichler:1989ve,
freiburghaus:1999a, goriely:2011vg, korobkin:2012uy, wanajo:2014wha,
just:2014fka, thielemann:2017acv}. This is supported by the
UV/optical/infrared follow up observations to GW170817
\citep[\eg,][]{kasen:2017sxr, rosswog:2017sdn, hotokezaka:2018aui}. The
observed signals are thought to have been produced by the decay of
radioactive isotopes produced during the $r$-process. However,
uncertainty persists on the exact rates and nucleosynthetic yields from
mergers. These needs to be addressed by future observations and
theoretical studies.

Numerical simulations are the cornerstone to the modeling of
multimessenger signatures and nucleosynthetic yields from \ac{NS}
mergers. Binary NS mergers are essentially multi-physics problems,
involving strong-field gravity, strongly interacting matter,
relativistic \mbox{(magneto-)hydrodynamics}, and neutrino transport. Due to the
complexity of the problem, most previous binary \ac{NS} merger
simulations either employed an approximate treatment for the
gravitational field of the \acp{NS} \citep{ruffert:1995fs,
rosswog:1998hy, rosswog:2001fh, rosswog:2003rv, rosswog:2003tn,
oechslin:2006uk, rosswog:2012wb, korobkin:2012uy, bauswein:2013yna}, or
included general relativistic (GR) \acused{GR} effects, but compromised
on the treatment of \ac{NS} matter \citep{shibata:1999wm,
shibata:2003ga, shibata:2005ss, baiotti:2008ra, kiuchi:2010ze,
rezzolla:2010fd, rezzolla:2011da, hotokezaka:2011dh, hotokezaka:2012ze,
palenzuela:2013hu, hotokezaka:2013iia, ruiz:2016rai, dietrich:2016hky,
dietrich:2016lyp}.  In the last few years, however, full-GR simulations
with a microphysical treatment of \ac{NS} matter and with different
levels of approximation for neutrino-radiation effects have also become
available \citep{sekiguchi:2011zd, sekiguchi:2015dma,
palenzuela:2015dqa, radice:2016dwd, lehner:2016lxy, sekiguchi:2016bjd,
foucart:2016rxm, bovard:2017mvn}.

Merger simulations have clarified the mechanisms driving the mass
ejection and highlighted the importance of neutrino effects.
Complementary studies on the long-term evolution of merger remnants also
revealed the importance of the magnetically-, neutrino-, and
viscously-driven secular outflows \citep{dessart:2008zd, metzger:2008av,
metzger:2008jt, lee:2009a, fernandez:2013tya, siegel:2014ita,
just:2014fka, metzger:2014ila, perego:2014fma, martin:2015hxa,
wu:2016pnw, siegel:2017nub, lippuner:2017bfm, fujibayashi:2017xsz,
fujibayashi:2017puw, siegel:2017jug, metzger:2018uni,
fernandez:2018kax}. These might in some cases dominate the
nucleosynthetic yields and produce the bulk of the thermal radiation
from \ac{NS} mergers. However, quantitative questions remain concerning
the dependency of the multimessenger emissions and of the $r$-process
production on the binary parameters and the \ac{EOS}. Indeed, most
previous studies considered only several binary configurations, with the
exception of \citet{hotokezaka:2013iia} and \citet{dietrich:2016hky,
dietrich:2016lyp}, who however used an ideal-gas prescription to
approximate thermal effects in the \ac{NS} matter, and
\citet{bauswein:2013yna} who used temperature dependent microphysical
\ac{EOS}, but adopted an approximate treatment of gravity and neglected
weak reactions.

In this work, we present a systematic study of the mass ejection,
nucleosynthetic yields, and \ac{EM} counterparts of \ac{NS} merger based
on 59 high-resolution numerical relativity simulations. We employ four
microphysical temperature-dependent nuclear \acp{EOS} and include the
impact of neutrino losses. We consider total binary masses between
$2.4\, M_\odot$ and $3.4\, M_\odot$ and mass ratios between $0.85$ and
$1$. Our datasets also includes 13 simulations with an effective
treatment of neutrino reabsorption and 6 simulations with viscosity.  We
quantify the dependency of dynamical ejecta mass and intrinsic
properties on the binary parameters and \ac{NS} \ac{EOS}. We compute
nucleosynthetic yields for the dynamical ejecta and show that second and
third $r$-process peaks are robustly produced with relative isotopic
abundances close to solar, while the production of light $r$-process
elements is sensitive to the binary mass ratio and the treatment of
neutrino radiation. We estimate kilonova light curves and show that
their properties depend on the lifetime of the merger remnant. We also
compute the expected radio signal from the interaction between the
ejecta and the ISM. This signal could be used to probe the strength of
the shocks generated after merger and, thus, indirectly the \ac{EOS} of
matter at extreme densities and temperatures.  Our simulations also
reveal a new outflow mechanism operating in unequal mass binaries and
enabled by viscosity. These ``viscous-dynamical'' ejecta are launched
due to the thermalization of mass exchange streams between the secondary
and the primary \ac{NS} shortly before merger and is discussed in more
detail in a companion paper \citep{radice:2018ghv}.

The rest of this paper is organized as follows. Section
\ref{sec:methods} summarises the numerical methods and the initial data
employed for the simulations. Section~\ref{sec:ejecta} discusses
dynamical and secular mass ejection. Section~\ref{sec:nucleosynthesis}
reports on our $r$-process nucleosynthesis calculations and yields.
Section \ref{sec:em} is dedicated to the discussion of the \ac{EM}
counterparts from binary \ac{NS} mergers, focusing on the kilonova
signal and on the radio remnant powered by the interaction of the ejecta
with the ISM.

Section~\ref{sec:conclusions} is dedicated to discussion and
conclusions. Finally, Appendices \ref{sec:resolution} and
\ref{sec:taudiff} present a discussion of finite resolution effects and
implementation details of the viscosity treatment employed by the
simulations.

\section{Methods}
\label{sec:methods}
\subsection{Initial Data}
We construct initial data in quasi-circular orbit using the
\texttt{Lorene} pseudo-spectral code \citep{lorenecode}. The initial
separation between the \acp{NS} is set to $40\, {\rm km}$, corresponding
to ${\sim}2{-}3$ orbits before merger. The \acp{EOS} used for the
initial data are constructed from the minimum temperature slice of the
\ac{EOS} table used for the evolution assuming neutrino-less
beta-equilibrium. To create the initial data tables we also subtract
from the pressure the contribution of photon radiation, which dominates
at the lowest densities due to the assumption of constant temperature.
When reading the initial data in the evolution code we set the electron
fraction according to the beta equilibrium condition, and we reset the
specific internal energy according to the minimum temperature slice in
the \ac{EOS} table used for the evolution. Small errors in the initial
data and in the mapping from the zero to the finite temperature
\acp{EOS} induce small oscillations in the \acp{NS} prior to merger. We
quantify these in terms of the relative change of the central density of
the stars which we find typically to be $\lesssim 2{-}3\%$.

\subsection{General-Relativistic Hydrodynamics}
We evolve the initial data using the \texttt{WhiskyTHC} code
\citep{radice:2012cu, radice:2013hxh, radice:2013xpa, radice:2015nva}.

\texttt{WhiskyTHC} separately evolves the proton and neutron number
densities
\begin{equation}\label{eq:number}
  \nabla_\mu ( n_{p,n} u^\mu ) = R_{p,n}\,,
\end{equation}
where $n_p = Y_e n$ is the proton number density, $n_n$ is the neutron
number density, $n = n_p + n_n$ is the baryon number density, $u^\mu$
the fluid four-velocity, and $Y_e$ is the electron fraction of the
material. $R_p = - R_n$ is the net lepton number deposition rate due to
the absorption and emission of neutrinos and anti-neutrinos (see Section
\ref{sec:leakage}).

We model \ac{NS} matter as a perfect fluid with stress energy tensor
\begin{equation}
  T_{\mu\nu} = (e + p) u_\mu u_\nu + p g_{\mu\nu}\,,
\end{equation}
where $e$ is the energy density and $p$ the pressure. We solve the
Euler equations for the balance of energy and momentum
\begin{equation}\label{eq:euler}
  \nabla_\nu T^{\mu\nu} = Q u^\mu\,,
\end{equation}
where $Q$ is the net energy deposition rate due to the absorption
and emission of neutrinos (see Section \ref{sec:leakage}).

\texttt{WhiskyTHC} discretizes Eqs.~(\ref{eq:number}) and
(\ref{eq:euler}) using  high-resolution shock-capturing (HRSC) schemes.
For the simulations presented in this work we use a central
Kurganov-Tadmor type scheme \citep{kurganov:2000a} employing the HLLE
flux formula \citep{hlle:88} and non-oscillatory reconstruction of the
primitive variables with the MP5 scheme of \citet{suresh:1997a}. For
numerical reasons we embed the \acp{NS} in a low density medium, with
$\rho_0 = m_b\, n \simeq 6 \times 10^{4}\, {\rm g}\, {\rm cm}^{-3}$,
where $m_b$ is the atomic mass unit. We use the best available numerical
schemes for the treatment of low density regions and for the advection
of the fluid composition. We employ the positivity-preserving limiter
from \citet{radice:2013xpa} to ensure rest-mass conservation even in the
presence of the artificial density floor, and we use the consistent
multi-fluid advection method of \citet{plewa:1999a} to ensure separate
local conservation of the proton and neutron number densities.
Furthermore, we extract the outflows properties when the density is
still several orders of magnitude higher than that of the artificial
atmosphere.

The spacetime is evolved using the Z4c formulation of Einstein's
equations \citep{bernuzzi:2009ex, hilditch:2012fp} as implemented in the
\texttt{CTGamma} code \citep{pollney:2009yz, reisswig:2013sqa}, which is
part of the \texttt{Einstein Toolkit} \citep{loffler:2011ay}.
\texttt{CTGamma} implements fourth-order finite-differencing of the
equations and we use fifth-order Kreiss-Oliger dissipation to ensure the
nonlinear stability of the evolution \citep{kreiss:1973book}. The
coupling between the hydrodynamics and the spacetime evolution is
handled using the method of lines (MoL). We adopt the optimal
strongly-stability preserving third-order Runge-Kutta scheme
\citep{gottlieb:2008a} as time integrator. The timestep is set according
to the speed-of-light \ac{CFL} condition with \ac{CFL} factor $0.15$.
We remark that numerical stability only requires the \ac{CFL} to
be less than $0.25$. However, the smaller value of $0.15$ is necessary
to guarantee the positivity of the density when using the
positivity-preserving limiter implemented in \texttt{WhiskyTHC}.

Our simulations domain covers a cube of 3,024~km in diameter whose
center is at the center of mass of the binary. Our code uses
Berger-Oliger conservative adaptive mesh refinement (AMR;
\citealt{berger:1984zza}) \acused{AMR} with sub-cycling in time and
refluxing \citep{berger:1989a, reisswig:2012nc} as provided by the
\texttt{Carpet} module of the \texttt{Einstein Toolkit}
\citep{schnetter:2003rb}. We setup an \ac{AMR} grid structure with 7
refinement levels. The finest refinement level covers both \acp{NS}
during the inspiral and the remnant after the merger and has a typical
resolution of $h \simeq 185\, {\rm m}$. For selected binaries, we also
perform simulations with finest grid resolution of $123\, {\rm m}$ and
$246 \, {\rm m}$. See Table~\ref{tab:ejecta} for a summary of our
models.

\subsection{Neutrino Leakage Scheme}
\label{sec:leakage}
We treat compositional and energy changes in the material due to weak
reactions using a leakage scheme (\citealt{galeazzi:2013mia,
radice:2016dwd}; see also \citealt{vanriper:1981mko, ruffert:1995fs,
rosswog:2003rv, oconnor:2009iuz, sekiguchi:2010ep, neilsen:2014hha,
perego:2015agy, ardevol-pulpillo:2018btx} for other implementations).
Our scheme tracks reactions involving electron $\nu_e$ and anti-electron
type $\bar{\nu}_e$ neutrinos separately. Heavy-lepton neutrinos are
lumped together in a single effective specie labeled $\nu_x$. We account
for the reactions listed in Table~\ref{tab:leakage}. We use the formulas
from the references listed in the tables to compute the neutrino
production rates $R_{\nu}$, $\nu \in \{ \nu_e, \bar{\nu}_e, \nu_x \}$,
the associated energy release $Q_{\nu}$, and neutrino absorption
$\kappa_{\nu,a}$ and scattering $\kappa_{\nu,s}$ opacities. In doing so,
we assume that the neutrinos follow Fermi-Dirac distributions with
chemical potentials obtained assuming beta-equilibrium with thermalized
neutrinos as in \citet{rosswog:2003rv}. Following
\citet{ruffert:1995fs}, we also distinguish between the number density
weighted opacities $\kappa_{\nu,a}^0$ and $\kappa_{\nu,s}^0$ that
determine the rate at which neutrinos diffuse out of the material, and
the energy density weighted opacities $\kappa_{\nu,a}^1$ and
$\kappa_{\nu,s}^1$ that determine the rate at which energy diffuses out
of the material due to the loss of neutrinos.

\begin{table}
\caption{Weak reaction rates and references for their implementation.
We use the following notation $\nu \in \{\nu_e, \bar{\nu}_e, \nu_{x}\}$
denotes a neutrino, $\nu_{x}$ denote any heavy-lepton neutrino, $N \in
\{n, p\}$ denotes a nucleon, and $A$ denotes a nucleus.}
\label{tab:leakage}
\begin{center}
\begin{tabular}{ll}
\hline\hline
Reaction & Reference \\
\hline
$\nu_e + n \leftrightarrow p + e^-$           & \citet{bruenn:1985en} \\
$\bar{\nu}_{e} + p \leftrightarrow n + e^+$   & \citet{bruenn:1985en} \\
$e^+ + e^- \rightarrow \nu + \bar{\nu}$       & \citet{ruffert:1995fs} \\
$\gamma + \gamma \rightarrow \nu + \bar{\nu}$ & \citet{ruffert:1995fs} \\
$N + N \rightarrow \nu + \bar{\nu} + N  + N$  & \citet{burrows:2004vq} \\
$\nu + N \rightarrow \nu + N$                 & \citet{ruffert:1995fs} \\
$\nu + A \rightarrow \nu + A$                 & \citet{shapiro:1983book} \\
\hline\hline
\end{tabular}
\end{center}
\end{table}

The total neutrino opacities $\kappa_{\nu,a}^\alpha +
\kappa_{\nu,s}^\alpha$, with $\alpha \in \{0,1\}$, are used to compute
an estimate to the optical depth $\tau^\alpha_\nu$ following the scheme
of \citet{neilsen:2014hha}, which is well-suited to the complex
geometries present in \ac{NS} mergers. We use the optical depth to
define effective emission rates as \citet{ruffert:1995fs}:
\begin{equation}\label{eq:R.eff}
  R_\nu^{\rm eff} = \frac{R_\nu}{1 + t_{\rm diff}^0\,
  (t_{\rm loss}^0)^{-1}}\,,
\end{equation}
where we have introduced the effective diffusion time $t_{\rm diff}$
\begin{equation}
  t_{\rm diff}^0 = \mathcal{D} \frac{(\tau^0_\nu)^2}{
  \kappa_{\nu,a}^0 + \kappa_{\nu,s}^0}\,,
\end{equation}
the neutrino emission timescale
\begin{equation}
  t_{\rm loss}^0 = \frac{R_\nu}{n_\nu}\,,
\end{equation}
and $n_\nu$ is the neutrino number density computed assuming
beta-equilibrium with neutrinos. The constant $\mathcal{D}$ is a tuning
parameter that we set to $6$. The effective energy emission rates
$Q_\nu^{\rm eff}$ are computed along the same lines, but using
$\tau^1_\nu$ and $\kappa^{1}_{\nu,a}$, $\kappa^1_{\nu,s}$ instead of
$\tau^0_\nu$, $\kappa^0_{\nu,a}$, and $\kappa^0_{\nu,s}$, respectively.

Neutrinos are split into a trapped component $n_\nu^{\rm trap}$ and a
free-streaming component $n_\nu^{\rm fs}$. Free-streaming neutrinos are
emitted according to the effective rate $R_\nu^{\rm eff}$ of Eq.
(\ref{eq:R.eff}) and with average energy $Q_{\nu}^{\rm eff}/R_{\nu}^{\rm
eff}$ and then evolved according to the M0 scheme we introduced in
\citet{radice:2016dwd} and that is briefly summarized below. In our
implementation the pressure due to the trapped neutrino component is
neglected, since it is found to be unimportant in the conditions
relevant for \ac{NS} mergers \citep{galeazzi:2013mia}.

The M0 scheme evolves the number density of the free-streaming neutrinos
assuming that they move along radial null rays with four vector
$k^\alpha$ normalized so that $k^\alpha u_\alpha = -1$. Under these
assumptions it is possible to show that that the free-streaming
fluid rest frame neutrino number density $n_\nu^{\rm fs}$ satisfies
\citep{radice:2016dwd}
\begin{equation}\label{eq:M0.ndens}
  \nabla_\alpha [n_\nu^{\rm fs} k^\alpha] = R_{\nu}^{\mathrm{eff}} -
  \kappa_{\nu,a}^{\rm eff} n_{\nu}^{\rm fs}\,,
\end{equation}
where $R_\nu^{\rm eff}$ is the effective luminosity from Eq.
(\ref{eq:R.eff}) and the effective absorption rates are defined as
\begin{equation}\label{eq:sigma.eff}
  \kappa_{\nu,a}^{\rm eff} = e^{-\tau_\nu^0}\,
  \left(\frac{E_{\nu}^{\rm fs}}{E_\nu^{\beta}} \right)^2\, \kappa_{\nu,a}^0\,.
\end{equation}
We have also introduced the average energy of free-streaming neutrinos
$E_\nu^{\rm fs}$ and the average neutrino energy in beta-equilibrium
$E_\nu^\beta$, both defined in the fluid rest frame. Note that in the
simulations we reported in \citet{radice:2016dwd} the term in
parenthesis on the right hand side of Eq. (\ref{eq:sigma.eff}) was
neglected. We report a comparison of results obtained with and without
the inclusion of this term in Sec.~\ref{sec:nucleosynthesis.dynamical}.

Our scheme estimates the free-streaming neutrino energy under the
additional assumption, only approximately satisfied in our simulations,
of stationarity of the metric. Accordingly, $\partial_t$ is assumed to
be a Killing vector so that $p^\alpha_\nu (\partial_t)_\alpha$,
$p^\alpha_\nu$ being the neutrinos four-momentum, is a conserved
quantity.  Under this assumption it is possible to show that the average
free-streaming neutrino energy satisfies \citep{radice:2016dwd}
\begin{equation}\label{eq:M0.ene}
  k^t \partial_t (E_\nu^{\rm fs} \chi) +
  k^r \partial_r (E_\nu^{\rm fs} \chi) =
    \frac{\chi}{n_\nu^{\rm fs}} \left(
    Q_\nu^{\rm eff} - E_\nu^{\rm fs} R_\nu^{\rm eff} \right)\,,
\end{equation}
where $\chi = - k^\alpha (\partial_t)_\alpha$.

In the simulations in which it is employed the M0 scheme is switched on
shortly before the two \acp{NS} collide, when neutrino matter
interactions start to become dynamically important. We solve Eqs.
(\ref{eq:M0.ndens}) and (\ref{eq:M0.ene}) on a uniform spherical grid
extending to $512\, G/c^2\, M_\odot \simeq 756\, {\rm km}$ and having
$n_r \times n_\theta \times n_\phi = 3096 \times 32 \times 64$ grid
points.

The coupling between neutrinos and matter is treated using an operator
split approach discussed in detail in \citet{radice:2016dwd}. The source
terms for Eqs. (\ref{eq:number}) and (\ref{eq:euler}) are, respectively,
\begin{equation}\label{eq:number.rhs}
  R_p = ({\kappa}_{{\nu}_e,a}^{\rm eff} n_{\nu_e}^{\rm fs} -
      {\kappa}_{\bar{\nu}_e,a}^{\rm eff} n_{\bar{\nu}_e}^{\rm fs}) -
      (R^\eff_{\nu_e} - R^\eff_{\bar{\nu}_e})\,,
\end{equation}
and
\begin{equation}\label{eq:euler.rhs}
\begin{split}
  Q = ({\kappa}_{\nu_e,a}^{\rm eff} n_{\nu_e}^{\rm fs} E_{\nu_e} &
    + {\kappa}_{\bar{\nu}_e,a}^{\rm eff} n_{\bar{\nu}_e}^{\rm fs} E_{\bar{\nu}_e})
    \\ & - (Q^\eff_{\nu_e} + Q^\eff_{\bar{\nu}_e} + Q^\eff_{\nu_x})\,.
\end{split}
\end{equation}

The M0 scheme is more approximate than the frequency-integrated M1
scheme adopted by, \eg, \citet{sekiguchi:2015dma} and
\citet{foucart:2015vpa}. However it has the advantage of computational
efficiency, it includes gravitational and Doppler effects, albeit in an
approximate way, and does not suffer from unphysical radiation shocks in
the important region above the merger remnant that instead plagues M1
schemes \citep{foucart:2018gis}.

\subsection{Viscosity}
\label{sec:methods.grles}
Our simulations do not include magnetic fields, so we cannot model the
possible emergence of magnetically driven outflows and jets after the
merger \citep{rezzolla:2011da, bucciantini:2011kx, siegel:2014ita,
ruiz:2016rai, metzger:2018uni}. Moreover, we cannot self-consistently
treat the transport of angular momentum in the merger remnant, which is
primarily due to magnetic stresses \citep{duez:2006qe, kiuchi:2014hja,
guilet:2016sqd, kiuchi:2017zzg}. We leave the treatment of jets and
relativistic winds, which necessarily requires high-resolution
\ac{GRMHD} simulations, to future work. On the other hand, we study the
range of possible effects due to the angular momentum transport by means
of an effective viscosity which we include in a subset of our
simulations. The use of this approach has not yet been fully validated
for \ac{NS} merger simulations. However, this approach has been found to
reproduce some of the main features of the MHD dynamics in the context
of post-merger accretion disks \citep{fernandez:2018kax}.

More specifically, we use the general-relativistic large eddy
simulations method (GRLES; \citealt{radice:2017zta}) to explore the
impact of subgrid-scale turbulent angular momentum transport.
Accordingly, we decompose the stress energy tensor of the fluid
$T^{\mu\nu}$ as
\acused{GRLES}
\begin{equation}\label{eq:tmunu.decomp}
  ?[c]T_\mu_\nu? = E n_\mu n_\nu + S_{\mu} n_{\nu} + S_{\nu} n_{\mu}  +
  S_{\mu\nu}\,,
\end{equation}
where
\begin{align}
  &E = T_{\mu\nu} n^\mu n^\nu = \rho h W^2 -p\,,  \\
  &S_\mu = - \gamma_{\mu\alpha} n_\beta T^{\alpha\beta} = \rho h W^2
  v_\mu\,, \\
  &S_{\mu\nu} = \gamma_{\mu\alpha} \gamma_{\mu\beta} T^{\alpha\beta}
  = S_\mu v_\nu + p \gamma_{\mu\nu} + \tau_{\mu\nu}\,,
\end{align}
and $n^\mu$ is the normal to the space-like slice hyper-surface, while
$\gamma_{\mu\nu}$, $v^\mu$, and $W$ are, respectively, the spatial
metric, the three-velocity, and the Lorentz factor. $\tau_{\mu\nu}$ is a
purely spatial tensor representing the effect of subgrid scale
turbulence. As in \citet{radice:2017zta}, we use the following ansatz
for $\tau_{ij}$:
\begin{equation}\label{eq:turb.visc.1}
  \tau_{ij} = - 2 \nu_{_T}  (\rho + p) W^2 \left[
  \frac{1}{2} \big(D_i v_j + D_j v_i\big) -
  \frac{1}{3} D_k v^k \gamma_{ij} \right]\,,
\end{equation}
where $\nu_{_T} = \ell_{\rm mix} c_s$ is the turbulent viscosity, $c_s$
is the sound speed, $D_i$ denotes the covariant derivative compatible
with the spatial metric, and $\ell_{\rm mix}$ is a free parameter we
vary to study the sensitivity of our results to turbulence.  In the
context of accretion disk theory turbulent viscosity is typically
parametrized in terms of a dimensionless constant $\alpha$ linked to
$\ell_{\rm mix}$ through the relation $\ell_{\rm mix} = \alpha\, c_s\,
\Omega^{-1}$, where $\Omega$ is the angular velocity of the fluid
\citep{shakura:1973a}. Recently, \citet{kiuchi:2017zzg} performed very
high resolution \ac{GRMHD} simulations of a \ac{NS} merger with
sufficiently high seed magnetic fields $(10^{15}\, {\rm G})$ to be able
to resolve the \ac{MRI} in the merger remnant and reported averaged
$\alpha$ values for different rest-mass density shells. Combining their
estimate of $\alpha$ with values of $c_s$ and $\Omega$ from our
simulations we find values of $\ell_{\rm mix} = 0{-}30\, {\rm m}$. Here,
we conservatively vary $\ell_{\rm mix}$ between $0$ (default; no subgrid
model) and $50\, {\rm m}$ (very efficient angular momentum transport).

\texttt{WhiskyTHC} consistently includes the contributions of
$\tau_{\mu\nu}$ to the fluid stress energy tensor in the calculation of
the right hand side of the metric and fluid equations. Flux terms
resulting from the inclusion of $\tau_{\mu\nu}$ need to be treated with
care, because a naive application of Godunov-type methods would result
in a scheme suffering from an odd-even decoupling instability
\citep[\eg,][]{lowrie:2001a}. \texttt{WhiskyTHC} uses a proper
combination of left and right biased finite-differencing operators to
discretize terms arising from the derivatives of $\tau_{\mu\nu}$ in a
flux-conservative fashion. The details are given in Appendix
\ref{sec:taudiff}.

\subsection{Models}

\begin{figure}
  \includegraphics[width=0.98\columnwidth]{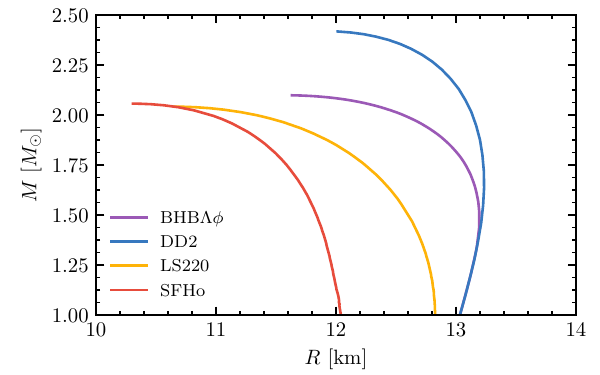}
  \caption{NS masses and radii predicted by the EOSs considered in this
  study. The BHB$\Lambda\phi$ and DD2 EOS predict the same radii for NSs
  less massive than ${\sim}1.5\, M_\odot$. BHB$\Lambda\phi$ predicts
  smaller radii for more massive NSs and a smaller maximum mass. The
  LS220 and SFHo EOSs have similar maximum masses and similar compactness
  close to the maximum mass. However, LS220 predicts radii almost
  $1\, {\rm km}$ larger than SFHo for a canonical $1.4\, M_{\odot}$ NS.}
  \label{fig:tovs}
\end{figure}

For our survey we consider four different nuclear \ac{EOS}: the DD2
\ac{EOS} \citep{typel:2009sy, hempel:2009mc}, the BHB$\Lambda\phi$
\ac{EOS} \citep{banik:2014qja}, the LS220 \ac{EOS}
\citep{lattimer:1991nc}, and the SFHo \ac{EOS} \citep{steiner:2012rk}.
These \acp{EOS} predict \ac{NS} maximum masses and radii within the
range allowed by current astrophysical constraints, including the recent
LIGO/Virgo constraint on tidal deformability
\citep{theligoscientific:2017qsa, abbott:2018wiz, de:2018uhw,
abbott:2018exr}. The LS220 \ac{EOS} is based on a liquid droplet Skyrme
model while the other three \acp{EOS} are based on nuclear statistical
equilibrium with a finite volume correction coupled to a relativistic
mean field theory for treating high-density nuclear matter. DD2 and SFHo
use different parameterizations and values for modeling the mean-field
nuclear interactions. The BHB$\Lambda \phi$ \ac{EOS} uses the same
nucleon interactions as the DD2 \ac{EOS}, but also includes interacting
$\Lambda$ hyperons that can be produced at high density and soften the
\ac{EOS}.

Although these \acp{EOS} differ in many aspects, including their finite
temperature properties and their dependence on the neutron richness of
the system, we can generally characterize them by the TOV solutions they
predict, which we show in Fig.~\ref{fig:tovs}. SFHo, LS220, DD2, and
BHB$\Lambda\phi$ support 2.06, 2.06, 2.42, and 2.11 $M_\odot$ cold,
non-rotating maximum NS masses and have $R_{1.4}$ of 11.9, 12.7, 13.2,
and 13.2 km, respectively. Since \ac{NS} radii correlate with the
pressure at roughly twice saturation density \citep{lattimer:2012nd}, we
refer to \acp{EOS} having smaller $R_{1.4}$ as being ``softer'' and to
\acp{EOS} having larger $R_{1.4}$ as being stiffer.  Although all four
models have saturation density symmetry energies within experimental
bounds, LS220 has a significantly steeper density dependence of its
symmetry energy than the other models. In all models, the finite
temperature behavior of the \ac{EOS} is mainly determined by the nucleon
effective mass, with smaller effective masses leading to higher
temperatures for constant entropy. The LS220 EOS assumes that the
nucleon mass is the bare nucleon mass at all densities, while SFHo has
$m_N^*/m_N = 0.76$ at saturation density, and both DD2 and BHB$\Lambda
\phi$ have $m_N^*/m_N = 0.56$, where $m_N^*$ is the effective nucleon
mass and $m_N$ is the bare nucleon mass.

We considered $35$ distinct binaries with total masses $M \in [2.4,
3.4]\, M_\odot$ and mass ratios in the range $q \in [0.85, 1]$. All
binaries have been simulated using the leakage scheme discussed above,
selected binaries have also been simulated with the M0 scheme, at
different resolutions, and/or with viscosity, for a total of 59
simulations. A summary of all simulations and key results is given in
Table~\ref{tab:ejecta}. Each run is labelled after the employed
\ac{EOS}, the masses of the two \acp{NS} at infinite separation, the
simulated physics, the value of the mixing length $\ell_{\rm mix}$ in
meters, if larger than zero, and, in the case of binaries run at
multiple resolutions, the resolution (\texttt{LR}: low resolution,
\texttt{HR}: high resolution). For example,
\texttt{LS220\_M140120\_M0\_L25} is a binary with \acp{NS} masses $1.4\,
M_\odot$ and $1.2\, M_\odot$ that was simulated with the LS220 \ac{EOS},
employing the M0 scheme, with $\ell_{\rm mix} = 25\, {\rm m}$, and run
at our standard resolution $h = 185\, {\rm m}$. We will make all of our
\ac{GW} waveforms publicly available as part of the \texttt{CoRe}
catalog \citep{dietrich:2018phi}.

We simulate all binaries for at least $20\, {\rm ms}$ after the merger,
or until few milliseconds after \ac{BH} formation if this occurs
earlier. With the exceptions of the runs including viscosity, which we
discuss in more details in a companion paper \citep{radice:2018ghv},
the outflow rate, precisely defined in Sec~\ref{sec:ejecta}, has dropped
to zero at the end of our simulations.  Our models include binaries
spanning all range of possible outcomes predicted for \ac{NS} binary
mergers \citep[\eg,][]{shibata:2016book}. Some of our binaries produce
\acp{BH} promptly at the time of the merger, others produce hypermassive
neutron stars (HMNSs) that collapse on time scales of several
milliseconds \citep{baumgarte:1999cq, shibata:2006nm, baiotti:2008ra,
sekiguchi:2011zd, bernuzzi:2015opx}, or long-lived massive \ac{NS}
remnants, expected to be stable on secular timescales or, in some cases,
indefinitely \citep{giacomazzo:2013uua, foucart:2015gaa,
radice:2018xqa}. Table~\ref{tab:ejecta} reports the time to \ac{BH}
formation in milliseconds from the merger $t_{\rm BH}$, or a lower limit
if the central object does not collapse within the simulation time.
\acused{MNS} \acused{HMNS}

\section{Mass Ejection}
\label{sec:ejecta}
Tidal interactions and shocks exerted on the \acp{NS} close to the time
of merger trigger the ejection of material on a dynamical time scale,
the so called \textit{dynamical ejecta}
\citep[\eg,][]{hotokezaka:2012ze, bauswein:2013yna, radice:2016dwd}.
Dynamical ejecta mass and average properties are reported in
Table~\ref{tab:ejecta}. In simulations that do not include viscosity the
outflow rate drops to zero few milliseconds after the merger. However,
it is expected that more material will become unbound from the remnant
on longer timescales, due to magnetic effects and/or nuclear
recombination \citep[\eg,][]{fernandez:2013tya, perego:2014fma,
siegel:2017nub, fujibayashi:2017puw, fernandez:2018kax}, which we cannot
presently study with our simulations. We refer to this latter component
as the \textit{secular ejecta} to distinguish it from the dynamical
ejecta defined above. The simulations performed with viscosity show the
early development of viscously-driven outflows.  However, due to the high
computational costs, we do not follow the postmerger remnant for
sufficiently long times to study the secular ejecta.

\subsection{Dynamical Ejecta}
\label{sec:ejecta.dynamical}

\begin{figure*}
  \begin{center}
    \includegraphics[width=1.96\columnwidth]{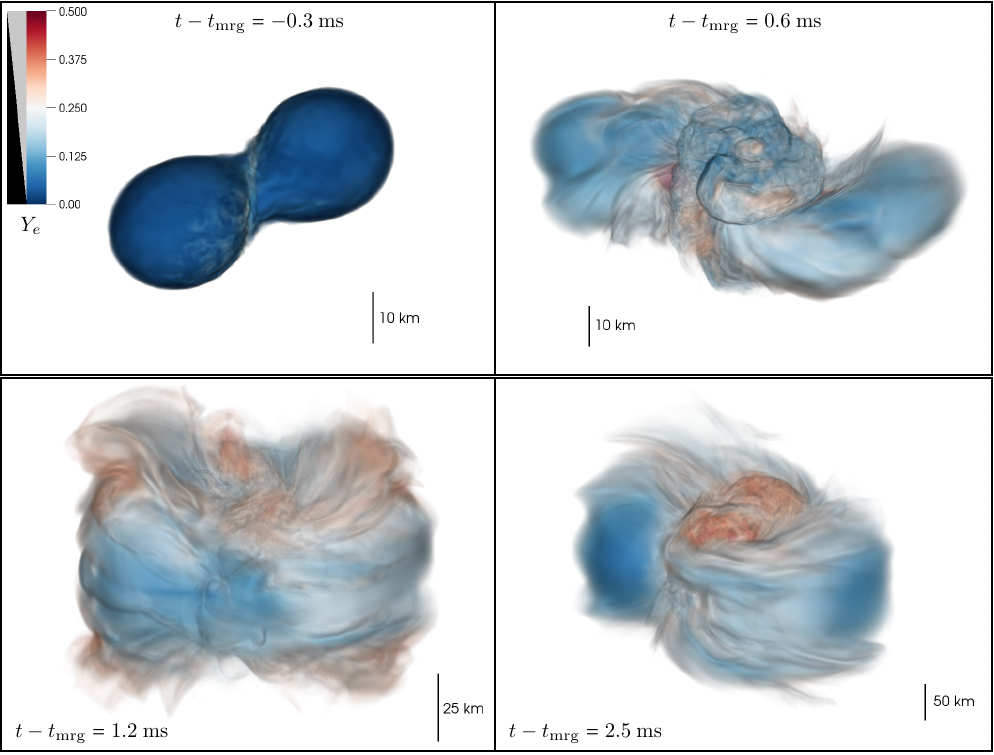}
  \end{center}
  \caption{Volume rendering of the electron fraction of the ejecta for
  the simulation \texttt{SFHo\_M135135\_M0}. The ray-casting opacity is
  linear in the logarithm of the rest-mass density. From the top-left in
  clockwise direction, the transparency minimum -- maximum in the
  opacity scale are $(10^{11} - 10^{14})$ ${\rm g}\ {\rm cm}^{-3}$,
  $(10^8 - 10^{11})$ ${\rm g}\ {\rm cm}^{-3}$, $(10^8 - 10^{11})$ ${\rm
  g}\ {\rm cm}^{-3}$, and $(10^7 - 10^{11})$ ${\rm g}\ {\rm cm}^{-3}$.
  The last panel of this figure should be compared with
  Fig.~\ref{fig:SFHo_M135135_Ye_xz} where we plot a cut of the data in
  the $xz$-plane.}
  \label{fig:SFHo_M135135_Ye_3d}
\end{figure*}

We discuss the qualitative properties of the dynamical ejecta in the case
of the \texttt{SFHo\_M135135\_M0} binary, which we take as fiducial.
Figure~\ref{fig:SFHo_M135135_Ye_3d} summarizes its most salient features.
The bulk of the dynamical outflow is contained within a wide
${\sim}60^\circ$ angle from the orbital plane. In the simulations that do
not account for neutrino absorption the outflow is neutron rich, with
average electron fraction $\langle Y_e \rangle < 0.2$. When neutrino
absorption is included in the simulations, as is the case for the
\texttt{SFHo\_M135135\_M0} binary, the ejecta is reprocessed to higher
values of $Y_e$, but remains neutron rich $\langle Y_e \rangle < 0.25$.

As shown in Fig.~\ref{fig:SFHo_M135135_Ye_3d}, the tidal component of the
dynamical ejecta is emitted first close to the orbital plane, followed
by a more isotropic shock heated ejecta component. However, the shock
heated component generally has higher velocity, so it rapidly overtakes
the tidal component. The two components interact and, as a consequence,
the tidal tail is partially reprocessed by weak reactions to slightly
higher values of $Y_e \simeq 0.1$. Later, we observe the emergence of a
neutrino driven wind component of the outflow, concentrated at high
latitudes. Finally, for this binary, mass ejection shuts off shortly
after the formation of a \ac{BH}.

\begin{figure}
  \includegraphics[width=0.98\columnwidth]{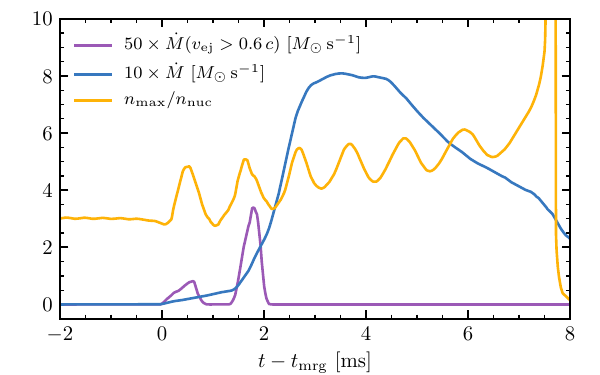}
  \caption{Maximum rest-mass density and outflow rate for the
  \texttt{SFHo\_M135135\_M0} binary. The outflow rate is computed at a
  radius of $R = 443\, {\rm km}$ and shifted in time by $R\, (0.5
  c)^{-1}$, $0.5 c$ being roughly the 99 percentile of the velocity of
  the ejecta at the radius $R$. We show both the total outflow rate as
  well as the outflow rate of only the material with asymptotic velocity
  larger than $0.6\, c$. We find that the bulk of the ejecta is launched
  when the HMNS first bounces back after the merger.}
  \label{fig:ejection.mechanism}
\end{figure}

As the two \acp{NS} merge, their inner cores are violently compressed
against each other. For some of our models this compression is
sufficient to trigger a runaway collapse of the remnant and a \ac{BH}
forms within a single dynamical timescale. However, for most of our
binaries the angular momentum of the remnant is sufficiently large to
prevent the collapse and the remnant's core undergoes a centrifugal
bounce. The bounce starts a cycle of large scale oscillations of the
remnant. As also discussed in detail by \citet{bauswein:2013yna}, for
most binaries the most abundant component of the outflow is triggered
during the first expansion of the remnant. This is demonstrated in
Fig.~\ref{fig:ejection.mechanism}, where we show the maximum density and
the outflow rates measured for our fiducial \texttt{SFHo\_M135135\_M0}
binary. The former is extracted from the flux of unbound material
leaving a coordinate sphere with radius $R = 300\, G/c^2\, M_\odot
\simeq 443\, {\rm km}$ and is retarded according to the velocity of the
tip of the ejecta at the detection sphere. As evidenced in the figure,
and confirmed by a careful inspection of the multidimensional data, the
bulk of the outflow is triggered at the time when the merger remnant
rebounds. However, after the first bounce, the subsequent oscillations
do not produce significant mass ejection. The outflow rate remains
positive for a few milliseconds as slower material, that has also
started expanding during the first ejection episode, reaches the
detector.

\begin{figure*}
  \includegraphics[width=0.98\columnwidth]{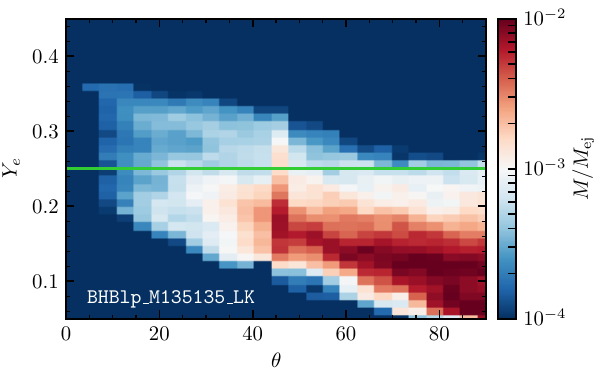}
  \hfill
  \includegraphics[width=0.98\columnwidth]{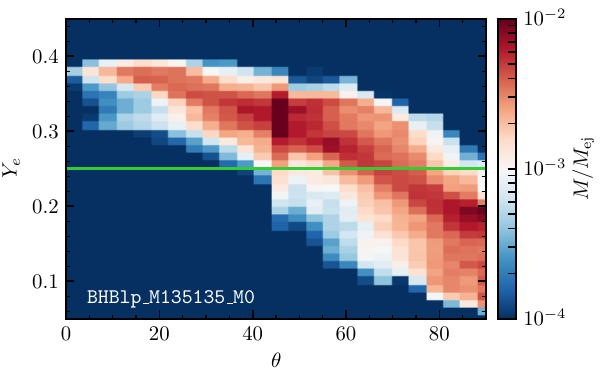}
  \caption{Angular distribution and composition of the ejecta for the
  $1.35\ M_\odot$ vs $1.35\ M_\odot$ binary with BHB$\Lambda\phi$ EOS.
  \textit{Right panel}: neutrino cooling only. \textit{Left panel}:
  simulation with neutrino absorption. Neutrino irradiation is necessary
  to generate high-$Y_e$, polar outflows.}
  \label{fig:ejecta_ye_theta}
\end{figure*}

The inclusion of neutrino re-absorption results in the formation of a new
outflow component. This additional outflow stream is composed of material
that is ablated from the surface of the \ac{HMNS} and its accretion disk.
This material is reprocessed to $Y_e > 0.25$ and channeled along the
polar direction $\theta < 45^\circ$. We remark that this outflow component
is absent if neutrino heating is not included (\citealt{radice:2016dwd,
perego:2017wtu}; see also Fig.~\ref{fig:ejecta_ye_theta} for another
representative case). Furthermore, because in our simulation the tidal
streams interact with the shocked component of the ejecta, the former
also see their $Y_e$ slightly increased with the inclusion of neutrino
absorption, as can be seen from the shift in $Y_e$ of the material close
to the orbital plane (see region $\theta \simeq 90^\circ$ in
Fig.~\ref{fig:ejecta_ye_theta}).

\subsubsection{Outflow Properties}
\label{sec:outflow.properties}

\begin{table*}
\caption{Dynamical ejecta and remnant disks for all simulations. We
report the model name, the grid resolution $h$, the NS masses at
infinite separation $M_a$ and $M_b$, the time of BH formation $t_{\rm
BH}$, the remnant disk masses $M_{\rm disk}$, the total ejected mass
$M_{\mathrm{ej}}$, and the amount of fast moving ejecta $M_{\rm ej}^{v
\geq 0.6c}$. For the dynamical ejecta we report the mass-averaged
electron fraction $\langle Y_{e} \rangle$, specific entropy per baryon
$\langle s \rangle$, and asymptotic velocity $v_{\rm ej}$. We also give
the rms opening angle of the outflow streams from the orbital plane $\theta_{\rm ej} =
\sqrt{\langle \theta^2 \rangle}$, and the total kinetic energy of the
ejecta $T_{\rm ej}$. All ejecta properties are measured on a sphere with
coordinate radius of $300\, G/c^2\, M_\odot \simeq 443\ \mathrm{km}$.}
\vspace{-1em}
\label{tab:ejecta}
\begin{center}
\scalebox{1.00}{
\begin{tabular}{lrd{1.3}d{1.3}rd{2.2}d{1.2}d{1.3}d{1.2}d{2.0}d{1.2}d{2.0}d{1.2}}
\hline\hline
\multirow{2}{*}{Model} &
\myhead{$h$} &
\myhead{$M_a$} &
\myhead{$M_b$} &
\myhead{$t_{\rm BH}$} &
\myhead{$M_{\rm disk}$} &
\myhead{$M_{\rm ej}$} &
\myhead{$M_{\rm ej}^{v\geq 0.6c}$} &
\myhead{$\langle Y_{e}\rangle$} &
\myhead{$\langle s \rangle$} &
\myhead{$v_{\rm ej}$} &
\myhead{$\theta_{\rm ej}$} &
\myhead{$T_{\rm ej}$} \\
&
\myhead{$[{\rm m}]$} &
\myhead{$[M_\odot]$} &
\myhead{$[M_\odot]$} &
\myhead{$[{\rm ms}]$} &
\multicolumn{2}{c}{$[10^{-2}\ M_\odot]$} &
\myhead{$[10^{-5}\ M_\odot]$} &
&
\myhead{$[k_{\rm B}]$} &
\myhead{$[c]$} &
&
\myhead{$[10^{50}\ {\rm erg}]$} \\
\hline
\texttt{BHBlp\_M125125\_LK} & 185 & 1.25 & 1.25 & $>\ $21.9 & \myhead{NA} & 0.13 & 0.000 & 0.13 & 15 & 0.18 & 22 &0.46 \\
\texttt{BHBlp\_M1365125\_LK} & 185 & 1.365 & 1.25 & $>\ $24.0 & 18.73 & 0.06 & 1.367 & 0.14 & 17 & 0.21 & 25 &0.35 \\
\texttt{BHBlp\_M130130\_LK} & 185 & 1.3 & 1.3 & $>\ $26.9 & \myhead{NA} & 0.07 & 0.000 & 0.16 & 22 & 0.12 & 33 &0.11 \\
\texttt{BHBlp\_M135135\_LK} & 185 & 1.35 & 1.35 & $>\ $21.3 & 14.45 & 0.07 & 0.746 & 0.15 & 20 & 0.17 & 28 &0.26 \\
\texttt{BHBlp\_M135135\_LK\_HR} & 123 & 1.35 & 1.35 & $>\ $23.3 & 14.26 & 0.05 & 0.015 & 0.16 & 20 & 0.18 & 24 &0.19 \\
\texttt{BHBlp\_M135135\_M0} & 185 & 1.35 & 1.35 & $>\ $37.0 & 12.75 & 0.14 & 0.283 & 0.26 & 24 & 0.14 & 38 &0.40 \\
\texttt{BHBlp\_M140120\_LK} & 185 & 1.4 & 1.2 & $>\ $23.7 & 20.74 & 0.11 & 0.229 & 0.11 & 13 & 0.16 & 22 &0.36 \\
\texttt{BHBlp\_M140120\_M0} & 185 & 1.4 & 1.2 & $>\ $28.2 & 22.56 & 0.16 & 0.994 & 0.19 & 17 & 0.17 & 30 &0.60 \\
\texttt{BHBlp\_M140140\_LK} & 185 & 1.4 & 1.4 & 12.0 & 7.05 & 0.09 & 0.232 & 0.15 & 18 & 0.17 & 28 &0.34 \\
\texttt{BHBlp\_M140140\_LK\_HR} & 123 & 1.4 & 1.4 & 10.3 & 5.38 & 0.10 & 0.793 & 0.14 & 17 & 0.20 & 26 &0.50 \\
\texttt{BHBlp\_M144139\_LK} & 185 & 1.44 & 1.39 & 10.4 & 8.28 & 0.06 & 0.511 & 0.18 & 22 & 0.20 & 30 &0.30 \\
\texttt{BHBlp\_M150150\_LK} & 185 & 1.5 & 1.5 & \phantom{0}2.3 & 1.93 & 0.05 & 0.727 & 0.17 & 20 & 0.23 & 28 &0.33 \\
\texttt{BHBlp\_M160160\_LK} & 185 & 1.6 & 1.6 & \phantom{0}1.0 & 0.09 & 0.00 & 0.000 & \myhead{$-$} & \myhead{$-$} & \myhead{$-$} & \myhead{$-$} &\myhead{$-$} \\
\texttt{DD2\_M120120\_LK} & 185 & 1.2 & 1.2 & $>\ $24.7 & \myhead{NA} & 0.08 & 0.000 & 0.15 & 21 & 0.14 & 31 &0.16 \\
\texttt{DD2\_M125125\_LK} & 185 & 1.25 & 1.25 & $>\ $32.4 & \myhead{NA} & 0.04 & 0.000 & 0.18 & 27 & 0.15 & 33 &0.10 \\
\texttt{DD2\_M1365125\_LK} & 185 & 1.365 & 1.25 & $>\ $24.2 & 20.83 & 0.04 & 0.443 & 0.15 & 21 & 0.20 & 25 &0.20 \\
\texttt{DD2\_M130130\_LK} & 185 & 1.3 & 1.3 & $>\ $22.9 & \myhead{NA} & 0.12 & 0.005 & 0.13 & 15 & 0.18 & 21 &0.45 \\
\texttt{DD2\_M135135\_LK} & 185 & 1.35 & 1.35 & $>\ $24.4 & 15.69 & 0.03 & 0.024 & 0.18 & 27 & 0.18 & 31 &0.12 \\
\texttt{DD2\_M135135\_LK\_HR} & 123 & 1.35 & 1.35 & $>\ $23.4 & 15.05 & 0.02 & 0.001 & 0.18 & 28 & 0.19 & 30 &0.09 \\
\texttt{DD2\_M135135\_M0} & 185 & 1.35 & 1.35 & $>\ $20.4 & 16.16 & 0.14 & 0.168 & 0.23 & 21 & 0.17 & 31 &0.49 \\
\texttt{DD2\_M140120\_LK} & 185 & 1.4 & 1.2 & $>\ $23.6 & 19.26 & 0.09 & 0.307 & 0.12 & 15 & 0.18 & 23 &0.36 \\
\texttt{DD2\_M140120\_M0} & 185 & 1.4 & 1.2 & $>\ $26.9 & 19.48 & 0.16 & 0.570 & 0.21 & 18 & 0.16 & 34 &0.54 \\
\texttt{DD2\_M140140\_LK} & 185 & 1.4 & 1.4 & $>\ $24.5 & 12.36 & 0.04 & 0.529 & 0.17 & 22 & 0.22 & 28 &0.26 \\
\texttt{DD2\_M140140\_LK\_HR} & 123 & 1.4 & 1.4 & $>\ $24.6 & 16.85 & 0.09 & 0.431 & 0.14 & 17 & 0.18 & 28 &0.39 \\
\texttt{DD2\_M144139\_LK} & 185 & 1.44 & 1.39 & $>\ $23.5 & 14.40 & 0.05 & 0.158 & 0.17 & 22 & 0.20 & 28 &0.26 \\
\texttt{DD2\_M150150\_LK} & 185 & 1.5 & 1.5 & $>\ $23.1 & 16.70 & 0.07 & 0.462 & 0.20 & 23 & 0.17 & 32 &0.26 \\
\texttt{DD2\_M160160\_LK} & 185 & 1.6 & 1.6 & \phantom{0}2.3 & 1.96 & 0.12 & 2.759 & 0.14 & 13 & 0.24 & 20 &0.83 \\
\texttt{LS220\_M120120\_LK} & 185 & 1.2 & 1.2 & $>\ $23.2 & 17.43 & 0.14 & 0.000 & 0.12 & 15 & 0.15 & 25 &0.33 \\
\texttt{LS220\_M1365125\_LK} & 185 & 1.365 & 1.25 & $>\ $26.7 & 16.86 & 0.11 & 0.013 & 0.10 & 13 & 0.16 & 23 &0.32 \\
\texttt{LS220\_M135135\_LK} & 185 & 1.35 & 1.35 & 20.3 & 7.25 & 0.06 & 0.000 & 0.11 & 16 & 0.16 & 23 &0.17 \\
\texttt{LS220\_M135135\_LK\_LR} & 246 & 1.35 & 1.35 & $>\ $27.6 & \myhead{NA} & 0.11 & 0.000 & 0.13 & 16 & 0.16 & 26 &0.32 \\
\texttt{LS220\_M135135\_M0} & 185 & 1.35 & 1.35 & $>\ $22.6 & 9.06 & 0.19 & 0.009 & 0.25 & 19 & 0.15 & 35 &0.50 \\
\texttt{LS220\_M135135\_M0\_L05} & 185 & 1.35 & 1.35 & $>\ $24.5 & 14.07 & 0.27 & 0.061 & 0.27 & 20 & 0.13 & 38 &0.58 \\
\texttt{LS220\_M135135\_M0\_L25} & 185 & 1.35 & 1.35 & 18.1 & 4.65 & 0.20 & 0.395 & 0.26 & 21 & 0.14 & 42 &0.51 \\
\texttt{LS220\_M135135\_M0\_L50} & 185 & 1.35 & 1.35 & $>\ $32.0 & 8.59 & 0.20 & 0.419 & 0.26 & 24 & 0.16 & 44 &0.66 \\
\texttt{LS220\_M135135\_M0\_LTE} & 185 & 1.35 & 1.35 & $>\ $21.1 & 14.24 & 0.11 & 0.000 & 0.20 & 17 & 0.13 & 30 &0.24 \\
\texttt{LS220\_M140120\_LK} & 185 & 1.4 & 1.2 & $>\ $23.5 & 22.82 & 0.19 & 0.000 & 0.09 & 11 & 0.15 & 20 &0.47 \\
\texttt{LS220\_M140120\_M0} & 185 & 1.4 & 1.2 & $>\ $24.8 & 23.38 & 0.24 & 0.001 & 0.18 & 14 & 0.15 & 29 &0.63 \\
\texttt{LS220\_M140120\_M0\_L05} & 185 & 1.4 & 1.2 & $>\ $28.2 & 20.33 & 0.28 & 0.017 & 0.18 & 14 & 0.15 & 29 &0.76 \\
\texttt{LS220\_M140120\_M0\_L25} & 185 & 1.4 & 1.2 & $>\ $28.5 & 12.07 & 0.41 & 0.625 & 0.16 & 13 & 0.20 & 29 &1.97 \\
\texttt{LS220\_M140120\_M0\_L50} & 185 & 1.4 & 1.2 & $>\ $31.2 & 13.97 & 0.70 & 8.289 & 0.18 & 13 & 0.22 & 28 &3.97 \\
\texttt{LS220\_M140140\_LK} & 185 & 1.4 & 1.4 & \phantom{0}9.9 & 4.58 & 0.14 & 0.087 & 0.14 & 16 & 0.17 & 29 &0.48 \\
\texttt{LS220\_M140140\_LK\_HR} & 123 & 1.4 & 1.4 & \phantom{0}9.4 & 2.68 & 0.07 & 0.168 & 0.15 & 17 & 0.17 & 33 &0.26 \\
\texttt{LS220\_M140140\_LK\_LR} & 246 & 1.4 & 1.4 & \phantom{0}8.6 & 3.12 & 0.17 & 0.019 & 0.16 & 15 & 0.19 & 28 &0.68 \\
\texttt{LS220\_M144139\_LK} & 185 & 1.44 & 1.39 & \phantom{0}7.2 & 3.91 & 0.19 & 0.014 & 0.14 & 15 & 0.16 & 30 &0.54 \\
\texttt{LS220\_M145145\_LK} & 185 & 1.45 & 1.45 & \phantom{0}2.3 & 2.05 & 0.16 & 1.230 & 0.14 & 14 & 0.21 & 22 &0.82 \\
\texttt{LS220\_M150150\_LK} & 185 & 1.5 & 1.5 & \phantom{0}0.9 & 0.16 & 0.03 & 0.001 & 0.08 & 11 & 0.19 & 13 &0.13 \\
\texttt{LS220\_M160160\_LK} & 185 & 1.6 & 1.6 & \phantom{0}0.6 & 0.07 & 0.03 & 0.000 & 0.07 &  8 & 0.21 &  8 &0.14 \\
\texttt{LS220\_M171171\_LK} & 185 & 1.71 & 1.71 & \phantom{0}0.5 & 0.06 & 0.03 & 0.000 & 0.08 &  9 & 0.22 &  8 &0.14 \\
\texttt{SFHo\_M1365125\_LK} & 185 & 1.365 & 1.25 & $>\ $26.4 & 8.81 & 0.15 & 1.888 & 0.14 & 14 & 0.23 & 24 &0.92 \\
\texttt{SFHo\_M135135\_LK} & 185 & 1.35 & 1.35 & 12.0 & 6.23 & 0.35 & 1.924 & 0.17 & 14 & 0.24 & 28 &2.24 \\
\texttt{SFHo\_M135135\_LK\_HR} & 123 & 1.35 & 1.35 & \phantom{0}6.9 & 1.78 & 0.23 & 0.982 & 0.17 & 14 & 0.21 & 29 &1.23 \\
\texttt{SFHo\_M135135\_LK\_LR} & 246 & 1.35 & 1.35 & \phantom{0}3.4 & 2.79 & 0.36 & 1.877 & 0.18 & 14 & 0.24 & 27 &2.35 \\
\texttt{SFHo\_M135135\_M0} & 185 & 1.35 & 1.35 & \phantom{0}7.7 & 1.23 & 0.42 & 3.255 & 0.22 & 17 & 0.22 & 33 &2.40 \\
\texttt{SFHo\_M140120\_LK} & 185 & 1.4 & 1.2 & $>\ $24.3 & 11.73 & 0.12 & 1.302 & 0.14 & 14 & 0.20 & 27 &0.59 \\
\texttt{SFHo\_M140120\_M0} & 185 & 1.4 & 1.2 & $>\ $32.7 & 15.65 & 0.30 & 2.368 & 0.22 & 17 & 0.16 & 34 &1.05 \\
\texttt{SFHo\_M140140\_LK} & 185 & 1.4 & 1.4 & \phantom{0}1.1 & 0.01 & 0.04 & 3.853 & 0.19 & 37 & 0.35 & 24 &0.60 \\
\texttt{SFHo\_M144139\_LK} & 185 & 1.44 & 1.39 & \phantom{0}0.9 & 0.09 & 0.04 & 3.056 & 0.18 & 16 & 0.33 & 20 &0.53 \\
\texttt{SFHo\_M146146\_LK} & 185 & 1.46 & 1.46 & \phantom{0}0.7 & 0.02 & 0.00 & 0.000 & \myhead{$-$} & \myhead{$-$} & \myhead{$-$} & \myhead{$-$} &\myhead{$-$} \\
\hline\hline
\end{tabular}
}
\end{center}
\end{table*}

Table~\ref{tab:ejecta} reports ejecta masses, average electron fraction
$\langle Y_e \rangle$, entropy $\langle s \rangle$, asymptotic velocity
$v_{\rm ej}$, and kinetic energy $T_{\rm ej}$ of the ejecta for all
models. We also report the rms opening angle of the outflows about the
equatorial plane $\theta_{\rm rms}$. Note that the definition of
$\theta_{\rm rms}$ implies that at least $75\%$ of the ejecta is
confined within $2 \theta_{\rm rms}$ of the orbital plane.

The dynamical ejecta masses are not fully converged in
our simulations. From the comparison of results obtained at different
resolutions we estimate the relative errors to be of the order of
${\sim} 50\%$. In the following, we will assume the uncertainty in the
ejecta masses to be
\begin{equation}\label{eq:mejerr}
  \Delta M_{\rm ej} = 0.5\, M_{\rm ej} + (5\times 10^{-5})\, M_\odot\,.
\end{equation}
On the other hand, other ejecta properties, such as the average asymptotic
velocities, the entropy, and the composition, appear to be converged
(see Appendix~\ref{sec:resolution} for a more detailed discussion).

\begin{table}
\caption{Dynamical ejecta masses reported in the literature for the
$(1.35 + 1.35)\, M_\odot$ binary simulated with either the DD2, or the
SFHo \ac{EOS}. For each reported value we indicate the reference and
whether the simulation included heating and compositional changes due to
neutrino emission ($\nu$ cool) and/or absorption ($\nu$ abs).
There are differences of factors of a few among published results.}
\label{tab:ejecta.comparison}
\begin{center}
\scalebox{0.95}{
\begin{tabular}{lccd{2.1}l}
  \hline\hline
  EOS  &
  $\nu$ cool. &
  $\nu$ abs. &
  \myhead{$M_{\rm ejecta}$} &
  Ref. \\
  &
  &
  &
  \myhead{$[10^{-3}\, M_\odot]$} & \\
  \hline
  DD2  & $\checkmark$ & $-$          &  $0.3$  & This work \\
  DD2  & $\checkmark$ & $\checkmark$ &  $1.4$  & This work \\
  DD2  & $-$          & $-$          &  $3.1$  & \citet{bauswein:2013yna} \\
  DD2  & $\checkmark$ & $-$          &  $0.6$  & \citet{bovard:2017mvn} \\
  DD2  & $\checkmark$ & $-$          &  $0.4$  & \citet{lehner:2016lxy} \\
  DD2  & $\checkmark$ & $-$          &  $0.9$  & \citet{sekiguchi:2015dma} \\
  DD2  & $\checkmark$ & $\checkmark$ &  $2.1$  & \citet{sekiguchi:2015dma} \\
  \hline
  SFHo & $\checkmark$ & $-$          &  $3.5$  & This work \\
  SFHo & $\checkmark$ & $\checkmark$ &  $4.2$  & This work \\
  SFHo & $-$          & $-$          &  $4.8$  & \citet{bauswein:2013yna} \\
  SFHo & $\checkmark$ & $-$          &  $3.5$  & \citet{bovard:2017mvn} \\
  SFHo & $\checkmark$ & $-$          &  $3.4$  & \citet{lehner:2016lxy} \\
  SFHo & $\checkmark$ & $-$          & $10.0$  & \citet{sekiguchi:2015dma} \\
  SFHo & $\checkmark$ & $\checkmark$ & $11.0$  & \citet{sekiguchi:2015dma} \\
  \hline\hline
\end{tabular}
}
\end{center}
\end{table}

Despite the large inferred numerical uncertainty, we find overall good
qualitative agreement between the ejecta masses estimated from our
simulations and others presented in the literature.  However,
quantitative differences are present. In particular, we compare
dynamical ejecta masses estimated for the $(1.35 + 1.35)\, M_\odot$
binaries with the DD2 and the SFHo \acp{EOS}, which have been considerer
in several previous works and by us. The results are reported in
Table~\ref{tab:ejecta.comparison}. \citet{sekiguchi:2015dma} and
\citet{sekiguchi:2016bjd} considered these binaries with and without the
inclusion of neutrino absorption. In the former case they reported
ejecta masses about ${\sim}3$ times larger than what we find. The
disagreement is somewhat less severe for the simulations with neutrino
absorption included. \citet{lehner:2016lxy} and \citet{bovard:2017mvn}
performed simulations only with neutrino cooling and report ejecta
masses in good agreement with those of our \texttt{DD2\_M135135\_LK} and
\texttt{SFHo\_M135135\_LK} binaries. Note, however, that the latter
study also employed the \texttt{WhiskyTHC} code, albeit with a different
grid setup and lower resolution, so it does not represent a completely
independent confirmation of our results. Finally,
\citet{bauswein:2013yna} performed simulations using a \ac{SPH} code
with an approximate treatment of \ac{GR} and neglected the effect of
neutrinos. They report ejecta mass of $0.48\times 10^{-2}\, M_\odot$ for
the SFHo binary. This is in good agreement with our value of $0.35\times
10^{-2}\, M_\odot$, especially when taking into account that neglecting
neutrino cooling results in an overestimate of the ejecta mass
\citep{radice:2016dwd}. On the other hand, the ejecta mass they report
for the DD2 EOS is a factor ${\sim}10$ larger than what we infer from
our simulations.

\begin{figure*}
  \includegraphics[width=0.98\columnwidth]{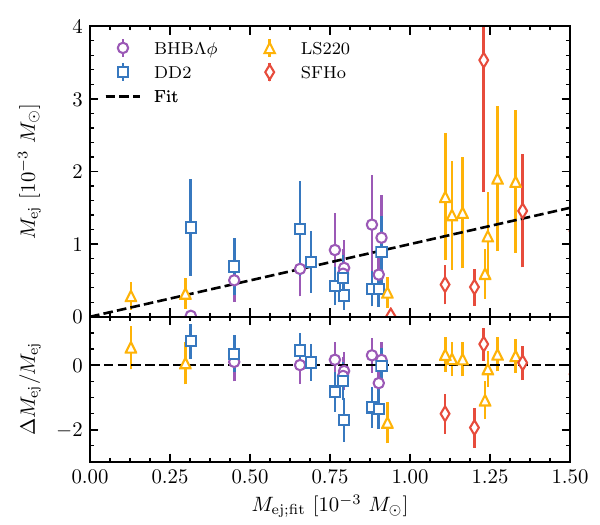}
  \hfill
  \includegraphics[width=0.98\columnwidth]{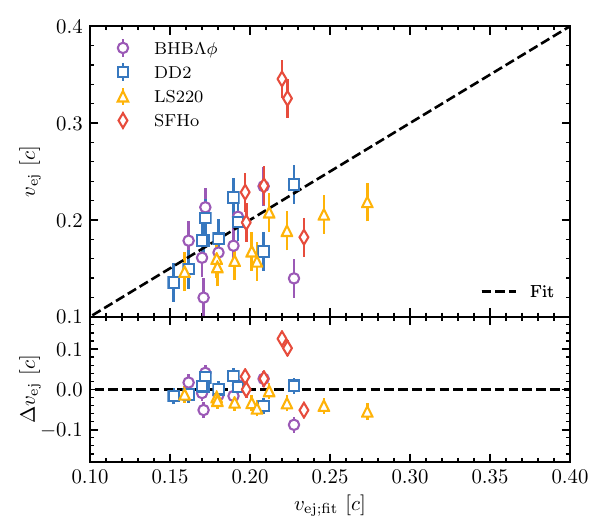}
  \caption{Dynamical ejecta masses (\textit{left panel}) and velocities
  (\textit{right panel}) versus their fit according to
  Eqs.~(\ref{eq:fitmej}) and (\ref{eq:fitvej}). The data points show the
  results from our fiducial subset of simulations, which have also been
  used to re-calibrate the fitting coefficients. For the velocity fit,
  we only include/plot models with total ejecta mass larger than
  $5\times 10^{-5}$ $M_\odot$. There is a correlation between $M_{\rm
  ej}$ and $v_{\rm ej}$ as extracted from our simulations and their
  predictors $M_{\rm ej; fit}$ and $v_{\rm ej; fit}$.}
  \label{fig:ejecta_mass_vel_fit}
\end{figure*}

\citet{dietrich:2016fpt} derived empirical formulas predicting the
ejecta mass and velocity from binary \ac{NS} mergers extending previous
work on the ejecta from BHNS mergers \citep{foucart:2012nc,
kawaguchi:2016ana}\footnote{See \citet{foucart:2018rjc} for updated fits
to BHNS merger data.}. They calibrated their fitting formulas using data
from \citet{hotokezaka:2012ze}, \citet{bauswein:2013yna},
\citet{dietrich:2015iva}, \citet{sekiguchi:2016bjd},
\citet{lehner:2016lxy}, and \citet{dietrich:2016hky}.  Note that most of
these data completely neglected neutrino effects (both emission and
absorption). Moreover, as we discuss above, some of these works predict
rather different ejecta masses for the same binaries. The resulting fits
have been used in \citet{abbott:2017wuw} to estimate the contribution of
the dynamical ejecta to the kilonova that followed GW170817 and, more
recently, by \citet{coughlin:2018miv} to constrain the tidal
deformability of the binary progenitor to GW170817. Note, however that
\citet{coughlin:2018miv} used a modified fit for $\log (M_{\rm
ej}/M_\odot)$.

We recalibrate the model of \citet{dietrich:2016fpt} using our data. We
only consider the simulations performed without neutrino heating and at
our reference resolution $h = 185\, {\rm m}$, so as to have a
homogeneous dataset. We refer to this subset of the runs as being our
fiducial subset of simulations.  The omission of neutrino re-absorption
could result in a systematic underestimate of the ejecta mass by up to
factors of a few (see Table~\ref{tab:ejecta}). However, we expect the
qualitative trends to be robust.

Following \citet{dietrich:2016fpt} we fit the ejecta mass as
\begin{equation}\label{eq:fitmej}
\begin{split}
  & \frac{M_{\rm ej; fit}}{10^{-3} M_\odot}  = \Bigg[
    \alpha \left(\frac{M_b}{M_a}\right)^{1/3} \left( \frac{1 - 2C_a}{C_a} \right) + \\
    & \qquad \beta \left(\frac{M_b}{M_a}\right)^n +
    \gamma \left(1 - \frac{M_a}{M_a^\ast}\right)
  \Bigg] M_a^\ast + (a \leftrightarrow b) + \delta\,,
\end{split}
\end{equation}
where $M_a$, $M_b$ and $M_a^\ast$, $M_b^\ast$ are the \ac{NS}
gravitational and baryonic masses, respectively, and
\begin{equation}
  C_{a,b} = \frac{G M_{a,b}}{c^2 R_{a,b}}
\end{equation}
are the stars' compactnesses. We use Eq.~(\ref{eq:mejerr}) to estimate
the numerical uncertainty in the ejecta mass and we use a standard least
square fit to determine the coefficients $\alpha$, $\beta$, $\gamma$,
$\delta$, and $n$ in Eq.~(\ref{eq:fitmej}). We find
\begin{align}
  & \alpha = -0.657\,, &&
  \beta  = 4.254\,,  &&
  \gamma = -32.61\,, \\
  & \delta = 5.205\,,  &&
  n      = -0.773\,.
\end{align}
We find differences of ${\sim}50{-}100\%$ compared to the values
reported by \citet{dietrich:2016fpt}. In particular, our fit
systematically predicts lower dynamical ejecta masses compared to
\citet{dietrich:2016fpt}.

The right panel of Fig.~\ref{fig:ejecta_mass_vel_fit} shows the ejecta
masses from our fiducial subset of simulations plotted against the predicted
ejecta masses from Eq.~(\ref{eq:fitmej}). Overall, we find that $M_{\rm
ej}$ correlates with $M_{\rm ej;fit}$, suggesting that the model of
\citet{dietrich:2016fpt} captures some of the physics behind the mass
ejection. However, there are differences between the values
measured in simulations and those predicted by the fit as large as
${\sim}200\%$. This is a factor of a few in excess of the
finite-resolution uncertainties we estimate for our simulations. We note
that \citet{dietrich:2016fpt} also reported similarly large fit
residuals. More worrisome is the fact that the model seems to be
systematically missing trends in the data. For example it over predicts
the ejecta for most of the massive DD2 binaries. This suggests that
Eq.~(\ref{eq:fitmej}) does not capture all of the physical effects
relevant for the mass ejection.

We fit the mass-weighted average asymptotic velocity $v_{\rm ej}$ of the
ejecta using an expression similar to the one used by
\citet{dietrich:2016fpt}:
\begin{equation}\label{eq:fitvej}
  v_{\rm ej; fit} = \left[ \alpha \left(\frac{M_a}{M_b}\right) (1 +
  \gamma C_a) \right] + (a \leftrightarrow b) + \beta\,.
\end{equation}
On the basis of the comparison between results obtained at different
resolution, we use a constant value $\Delta v_{\rm ej} = 0.02\, c$ as
fiducial uncertainty for the ejecta velocity and perform a least square
fit of our fiducial subset of simulations using Eq.~(\ref{eq:fitvej}).
When doing so, we exclude simulations with ejecta masses smaller than
$10^{-5}\, M_\odot$, since the ejecta properties cannot be reliably
measured in those cases.  Note that changing the value of $\Delta v_{\rm
ej}$ has no effect on the results of the fitting procedure. Using a
standard least square method we find
\begin{align}
  \alpha = -0.287\,, &&
  \beta  = 0.494\,, &&
  \gamma = -3.000\,.
\end{align}
These coefficients are in good agreement with those reported by
\citet{dietrich:2016fpt} for the ejecta velocity in the orbital plane.
This suggests that the ejecta velocity predicted by the simulations are
robust. We also find good qualitative agreement between the ejecta
velocities from simulations and those predicted by the model, as shown
in the right panel of Fig.~\ref{fig:ejecta_mass_vel_fit}. However, there
is a significant spread with residuals as large as $0.1\, c$.

\begin{figure}
  \includegraphics[width=0.98\columnwidth]{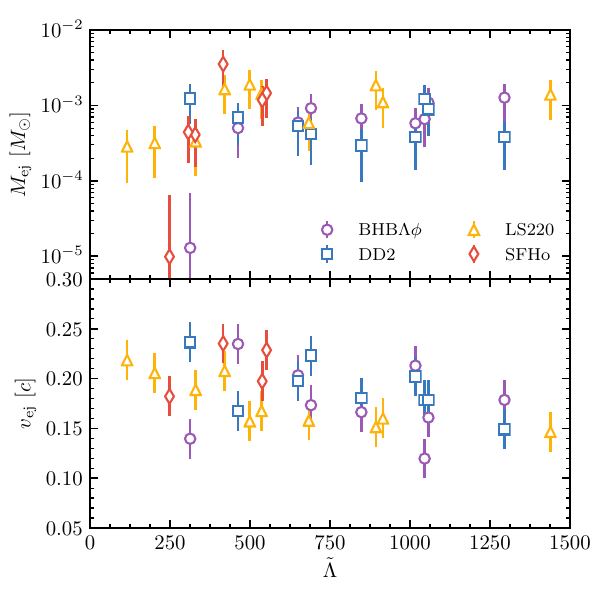}
  \caption{Dynamical ejecta masses $M_{\rm ej}$ and asymptotic
  velocities $v_{\rm ej}$ as a function of the tidal parameter
  $\tilde\Lambda$. The data points show the results from our fiducial
  subset of simulations. No correlation is found between the ejecta mass
  and $\tilde\Lambda$. There is, however, a weak tentative inverse
  correlation between the asymptotic velocity of the ejecta and
  $\tilde\Lambda$.}
  \label{fig:ejecta_vs_Lambda}
\end{figure}

\citet{bauswein:2013yna} compared dynamical ejecta mass obtained for the
$(1.35 + 1.35)\, M_\odot$ binary using different \acp{EOS} and found
that it was inversely proportional to the \ac{NS} radii suggesting that
large ejecta masses could be produced in the case of compact \acp{NS}.
This has been used in \citet{nicholl:2017ahq} to suggest that the large
inferred ejecta mass for GW170817 implies a small \ac{NS} radius
$R_{1.35} \lesssim 12\, {\rm km}$. The strongest support for the
correlation found by \citet{bauswein:2013yna} comes from binaries that
were simulated with an approximate treatment of thermal effects. Our
results (Table~\ref{tab:ejecta}) also indicate that softer \acp{EOS}
might produce more ejecta. However, we do not find evidences for a clear
correlation that would support the conclusions of
\citet{nicholl:2017ahq}.

Figure~\ref{fig:ejecta_vs_Lambda} shows ejecta masses and velocities as a
function of the tidal deformability of the binary $\tilde\Lambda$
\citep[\eg,][]{flanagan:2007ix, favata:2013rwa}.  By comparing results
from simulations with different \acp{EOS} we confirm that there is some
dependency of the ejecta mass on the stiffness of the \ac{EOS}, with
softer \acp{EOS} such as SFHo and LS220 producing more ejecta than
BHB$\Lambda\phi$ and DD2. We also find indications that prompt \ac{BH}
formation results in smaller than average dynamical ejecta masses.
However, we are not able to identify a relation between the ejecta masses
and properties of the \ac{EOS} such as the radius of a reference \ac{NS}
or the tidal deformability. This is not too surprising given that most of
the dynamical ejecta in our simulations is the result of the centrifugal
bounce of the merger remnant. Its characteristics are likely to depend on
thermal effects and details of the \ac{EOS} at high densities that are
not captured by $\tilde\Lambda$. Our data exclude the possibility of
constraining $\tilde\Lambda$ from the measurement of the properties of
the dynamical ejecta alone. On the other hand, it might be possible to
use kilonova observations to constrain other properties of the binary,
for instance using an improved version of Eq.~(\ref{eq:fitmej}), and thus
indirectly improve the bounds on $\tilde\Lambda$ by restricting the
priors used in the \ac{GW} data analysis. However, these applications
would necessarily require more precise theoretical predictions than those
currently available.

\begin{figure}
  \includegraphics[width=0.98\columnwidth]{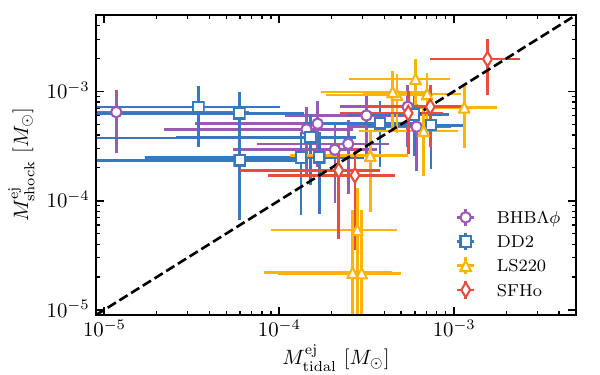}
  \caption{Mass of shock heated ejecta with $s > 10\, k_{\rm B}$ plotted
  against the mass of the cold ejecta. The center-left part of the figure
  is tentatively populated by stiff EOS. Tidally dominated ejecta
  dominates over shock-heated ejecta only for binaries with prompt BH
  formation (lower-middle part of the figure).}
  \label{fig:ejecta_tidal_vs_shock}
\end{figure}

Other properties of the outflow, such as proton fraction and entropy,
also show some dependency on the \ac{EOS}. The reason is that different
\acp{EOS} result in different relative amount of shocked and tidal
ejecta. This is quantified in Fig.~\ref{fig:ejecta_tidal_vs_shock},
where we show the mass of the tidal and shocked ejecta for our fiducial
subset of simulations, \ie, those with only neutrino cooling and $h =
185\, {\rm m}$. For this analysis we tentatively identify as shocked
ejecta all of the unbound material crossing a coordinate sphere with
radius $\simeq 443\, {\rm km}$ and having specific entropy per baryon
larger than $10\, k_{\rm B}$. Material ejected with smaller entropies is
assumed to be originating from tidal interactions. We find that stiff
\acp{EOS}, such as BHB$\Lambda\phi$ and DD2, typically have smaller
$M_{\rm tidal}^{\rm ej}$ than softer \acp{EOS}, such as LS220 and SFHo.
Softer \acp{EOS} also eject more mass overall. The shocked component of
the dynamical ejecta dominates the overall mass ejection for most of the
binaries we have considered. The only exceptions are cases where \ac{BH}
formation occurs shortly after merger ($\lesssim 1\, {\rm ms}$) so that
there is no centrifugal bounce of the remnant.

The balance between shocked and tidal ejecta is also dependent on the
binary mass ratio (see Sec.~\ref{sec:nucleosynthesis.dynamical}). On the
one hand, binaries with larger mass asymmetry produce more massive tidal
outflow streams. On the other hand, asymmetric binaries result in the
partial disruption of the secondary star during merger, less violent
mergers, and smaller amount of shocked ejecta (Table~\ref{tab:ejecta};
\citealt{hotokezaka:2012ze, bauswein:2013yna, lehner:2016lxy,
sekiguchi:2016bjd, dietrich:2016hky, bovard:2017mvn}). We conclude that
the balance between tidal ejecta and shocked ejecta is set by a complex
interplay between microphysical effects and binary properties.

\begin{figure}
  \includegraphics[width=0.98\columnwidth]{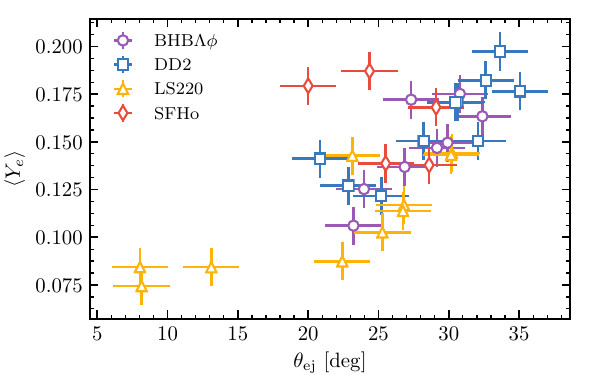}
  \caption{Average electron fraction $\langle Y_e \rangle$ vs.~rms angular
  spread $\theta_{\rm ej}$ of the ejecta. The data points show the
  results from our fiducial subset of simulations. We only include
  models with total ejecta mass larger than $5\times 10^{-5}$ $M_\odot$.
  The shock-heated component of the ejecta is absent in the cases with
  prompt-BH formation.}
  \label{fig:ejecta_theta_vs_ye}
\end{figure}

Tidally-driven outflows have low average $Y_e$ and are preferentially
concentrated close to the orbital plane. Conversely, the shocked ejecta
are spread over a larger angle of the orbital plane and have larger
$\langle Y_e \rangle$, so a correlation between $\theta_{\rm ej}$ and
$\langle Y_e \rangle$ is expected. This is shown in
Fig.~\ref{fig:ejecta_theta_vs_ye}. For clarity, the figure only shows
our fiducial subset of simulations, for which numerical and microphysical
parameters have been set in a consistent way. The binaries in the lower
left corner of the figure are LS220 binaries resulting in prompt \ac{BH}
formation. These are the same binaries appearing in the lower part of
Fig.~\ref{fig:ejecta_tidal_vs_shock}. The outflows from these binaries
are dominated by the tidal ejection of material. The two outliers with
$\theta_{\rm ej} = 20{-}25$ degrees are the \texttt{SFHo\_M140140\_LK}
and \texttt{SFHo\_M144139\_LK}, which also undergo prompt collapse, but
have outflows mostly driven by shocks. With the possible exception of
the binaries resulting in prompt \ac{BH} formation, all others show a
clear correlation between $\langle Y_e \rangle$ and $\theta_{\rm ej}$.
This suggests that a constrain on the opening angle of the dynamical
ejecta, perhaps obtained by combining the observation of multiple
systems with different orientations, could constrain the strength of the
bounce of the massive \ac{NS} after merger and the composition of the
dynamical ejecta.

Overall, we find good qualitative agreement between our results for the
dynamical ejecta and those reported in previous studies that adopted a
more idealized treatment for the \ac{EOS} of \acp{NS} and/or approximate
\ac{GR} \citep{hotokezaka:2013iia, bauswein:2013yna, dietrich:2016hky,
dietrich:2016lyp}. At the same time, there are substantial quantitative
differences in the dynamical ejecta mass and properties from those
studies, as well as with other works that considered a limited number of
binary configurations, but included full-\ac{GR} and neutrino effects
\citep{sekiguchi:2015dma, sekiguchi:2016bjd}. The total ejecta mass is
also not fully converged in our simulations (Table~\ref{tab:ejecta}).
While there appear to be robust trends and correlations between ejecta
properties, binary parameters, and the \ac{EOS} of \acp{NS}, more
comprehensive and better resolved studies would be needed to create
reliable quantitative models of the dynamical ejecta.

\subsubsection{Fast Moving Ejecta}
\label{sec:fast.ejecta}

\begin{figure}
  \includegraphics[width=0.98\columnwidth]{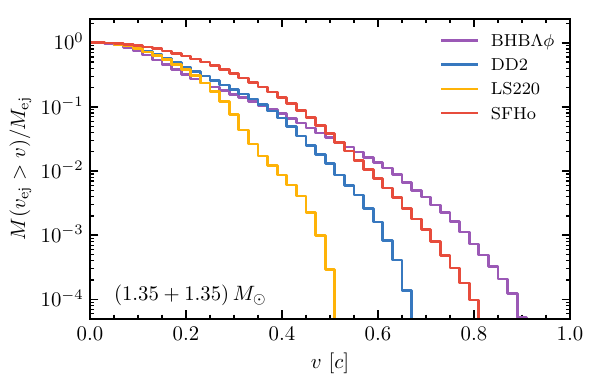}
  \caption{Cumulative distribution of ejecta velocities for the
  $(1.35+1.35)\, M_\odot$ binary with four EOSs and at the reference
  resolution. Neutrino re-absorption has not been included in these
  simulations.}
  \label{fig:ejecta_vel_hist}
\end{figure}

\citet{metzger:2014yda} re-analyzed data from \citet{bauswein:2013yna}
and identified 106 \ac{SPH} particles, corresponding to
${\sim}10^{-4}\, M_\odot$ of material, that were dynamically ejected
with velocities in excess of $0.6\, c$. If indeed present, these fast
moving ejecta would expand sufficiently rapidly to still contain a
significant fraction of free neutrons at freeze out. The decay of the
neutrons in the outermost part of the ejecta would then produce a bright
UV/optical counterpart to the merger on a timescale of several minutes
to an hour \citep{kulkarni:2005jw, metzger:2014yda}. On the other hand,
because of the small number of SPH particles, it cannot be excluded that
this fast component of the ejecta is due to numerical noise, as also
recognized by \citet{metzger:2014yda}.

More recently, a fast moving component of the dynamical ejecta was
proposed as a possible origin for the synchrotron radiation detected
from GW170817 in the first ${\sim}100$~days \citep{mooley:2017enz,
hotokezaka:2018gmo}. This interpretation is currently disfavored on the
light of more recent observations showing an abrupt decline in the
source luminosity in all bands \citep{alexander:2018dcl} and VLBI
observations showing apparent superluminal motion of the radio source
indicative of collimation of the outflow \citep{mooley:2018qfh,
ghirlanda:2018uyx}. Nevertheless, such fast moving component of the
outflows was identified in \citet{hotokezaka:2018gmo} using data from
the simulations of \citet{kiuchi:2017pte}. The latter simulations,
however, employed piecewise polytropic fits to the cold \ac{NS} \ac{EOS}
augmented with an ideal-gas component to describe thermal effects. These
approximations affect the thermodynamical properties of the shocks
responsible for the ejection of this material \citep{bauswein:2013yna},
so they need to be independently verified.

Our previous simulations \citep{radice:2016dwd} did not show evidences
for the presence of a fast moving component of the ejecta. However, in
\citet{radice:2016dwd} we only considered one equal mass configuration
with $M_a = M_b = 1.39\ M_\odot$ simulated using the LS220 \ac{EOS}.
While, \citet{metzger:2014yda} considered a large sample of \acp{EOS}
and binary masses. In our new simulations we find evidence for a fast
moving component of the outflow with asymptotic velocities in excess of
$0.6\, c$ (Table~\ref{tab:ejecta}). The amount of the fast moving ejecta
strongly depends on the \ac{EOS} and other binary parameters. For
instance, for total binary masses up to $2.8\, M_\odot$, with the
exceptions of the simulations performed with viscosity, discussed in a
companion paper \citep{radice:2018ghv}, the LS220 \ac{EOS} does not
seem to predict appreciable amount of fast outflows, in agreement with
\citet{radice:2016dwd}. On the other hand, binaries simulated with the
BHB$\Lambda\phi$, DD2, or SFHo \acp{EOS}, as well as higher-mass LS220
binaries, typically eject ${\sim}10^{-6}{-}10^{-5}\, M_\odot$ of
fast-moving material. This is still one or two orders of magnitude less
than reported in \citet{metzger:2014yda}, but suggest that the ejection
of at least a small amount of fast material does indeed take place
during \ac{NS} mergers. That said, given the small overall mass
involved, we cannot completely exclude its origin as being numerical.
Indeed, we find large variations in the amount of the fast moving ejecta
with resolution and much better resolved simulations would be needed to
fully quantify the mass of these fast outflows. On the other hand, the
smooth distribution of the ejecta in velocity space, shown in
Fig.~\ref{fig:ejecta_vel_hist}, seem to suggests that the properties of
the outflows are reasonably well captured even down to these masses.

We discuss the possible observational consequences of the fast moving
component of the ejecta in Sec.~\ref{sec:em}. Here, we focus on their
origin. \citet{metzger:2014yda} suggested that shocks generated at the
time of the collision between the \acp{NS} could be accelerated by the
steep density gradient at the surface of the remnant and drive outflows
with trans-relativistic velocities. This is a scenario first considered
by \citet{kyutoku:2012fv} who studied its possible high-energy
signatures. The same scenario has also been recently revisited in detail
by \citet{ishii:2018yjg}. They used high-resolution 1D Lagrangian
simulations to resolve the shock acceleration and predict the amount of
ejected material. They found that this mechanism can indeed produce free
neutrons. However, they found the amount of free neutrons that can be
produced in this way to be ${\sim}10^{-7}{-}10^{-6}\, M_\odot$. This
might be insufficient to explain the results of \citet{metzger:2014yda}.
On the other hand, the values found by \citet{ishii:2018yjg} are not
inconsistent with our data.

\begin{figure}
  \begin{center}
    \includegraphics[width=0.98\columnwidth]{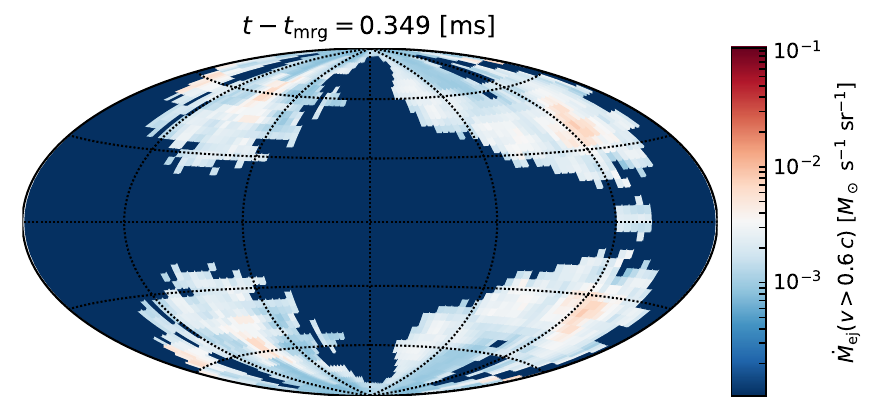}
    \\[1em]
    \includegraphics[width=0.98\columnwidth]{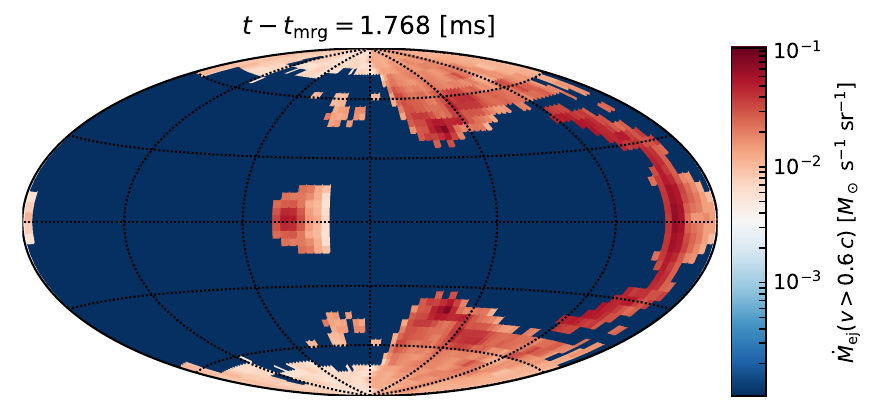}
    \texttt{SFHo\_M135135\_M0}\phantom{ppppppppppp}
  \end{center}
  \caption{Angular distribution of the flux of ejecta with asymptotic
  velocities larger than $0.6\, c$ for the \texttt{SFHo\_M135135\_M0}
  binary. The data is extracted on a coordinate sphere with radius $443\
  \mathrm{km}$. For clarity, the time is retarded in the same way as in
  Fig.~\ref{fig:ejection.mechanism}. We find two distinct episodes
  resulting in the production of fast ejecta: one associated with the
  merger and one associated with the first bounce of the remnant.}
  \label{fig:ejecta_fast_map_sfho_q1}
\end{figure}

Comparable mass binaries simulated with the SFHo \ac{EOS}, the softest
in our set, result in the most violent mergers. As the \acp{NS} collide
material is squeezed out from the collisional interface at large
velocities, as in the simulation discussed by \citet{metzger:2014yda}.
This is evident from Fig.~\ref{fig:ejection.mechanism}, where we show
the outflow rate of the fast-moving component of the ejecta with
asymptotic velocities $v_{\rm ej} > 0.6\, c$. Because of its peculiar
origin, this outflow component is preferentially channeled to high
latitudes and in the directions not obstructed by the \acp{NS}. This is
shown, in the case of the \texttt{SFHo\_M135135\_M0} binary, in the
upper panel of Fig.~\ref{fig:ejecta_fast_map_sfho_q1}. However, this
first clump of material amounts only to about a quarter of all of the
fast ejecta. Most of the fast moving ejecta, instead, appears to
originate when the shock launched from the first bounce of the remnant
breaks out of the forming ejecta cloud, as indicated in
Fig.~\ref{fig:ejection.mechanism}. This second component of the fast
ejecta is also highly anisotropic, but preferentially concentrated close
to the orbital plane, as shown in the lower panel of
Fig.~\ref{fig:ejecta_fast_map_sfho_q1} for the
\texttt{SFHo\_M135135\_M0} binary. This is possibly because of the
oblate shape of the merger debris cloud that favors the acceleration of
the material close to the orbital plane.

\begin{figure*}
  \includegraphics[width=0.98\columnwidth]{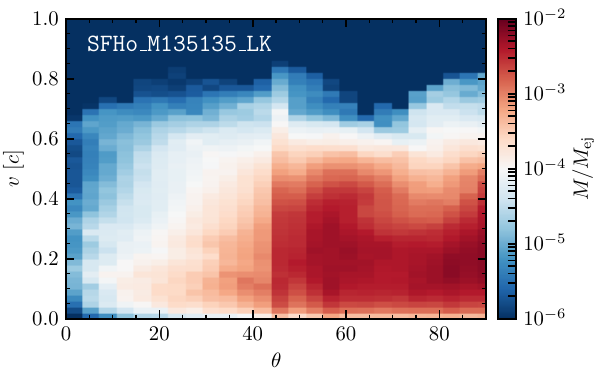}
  \hfill
  \includegraphics[width=0.98\columnwidth]{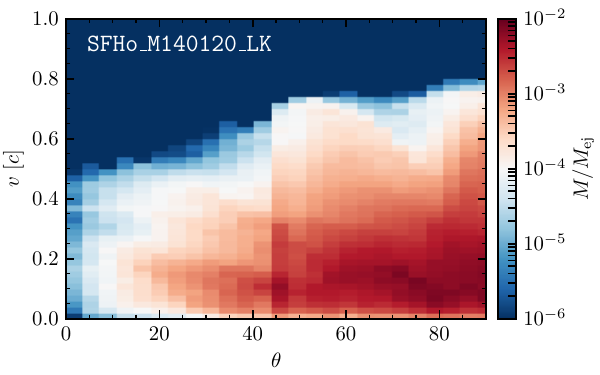}
  \caption{Angular distribution and asymptotic velocity of the ejecta
  for the $(1.35 + 1.35)\, M_\odot$ and $(1.4 + 1.2)\, M_\odot$ binaries
  with the SFHo EOS simulated without neutrino re-absorption. $\theta =
  0$ corresponds to the polar axis, while $\theta = 90^\circ$ is the
  orbital plane. The data is extracted on a coordinate sphere with
  radius $443\ \mathrm{km}$. The fast-moving tail of the ejecta is
  confined to a region close to the orbital plane for the unequal mass
  \texttt{SFHo\_M140120\_LK} binary (\textit{right panel}), but it is
  somewhat more isotropically distributed for the equal mass
  \texttt{SFHo\_M135135\_LK} binary (\textit{left panel}).}
  \label{fig:ejecta_vel_theta}
\end{figure*}

Smaller mass ratio binaries, or binaries simulated with other \acp{EOS},
do not however show evidences for an early fast-ejecta component from
the collisional interface between the \acp{NS}. For these binaries the
fast-moving component of the ejecta is entirely constituted of she
material accelerated after the bounce of the merger remnant.  This
material is preferentially confined to the region close to the orbital
plane, as shown in Fig.~\ref{fig:ejecta_vel_theta}. We stress that this
angular distribution is inconsistent with the ejecta originating at the
collisional interface between the \acp{NS}. The reason for the different
behavior of more unequal mass binaries and/or binaries with stiffer
\acp{EOS} is that the former result in less violent collisions than
those predicted for comparable mass \acp{NS} with the SFHo \ac{EOS},
either because of the larger \ac{NS} radii, or because of the partial
disruption of the secondary star shortly before merger. We note that the
inclusion of neutrino heating only results in a modest quantitative
correction to the distribution of the fast-moving component of the
ejecta, but the overall qualitative picture is unchanged.

\begin{figure}
  \begin{center}
    \includegraphics[width=0.98\columnwidth]{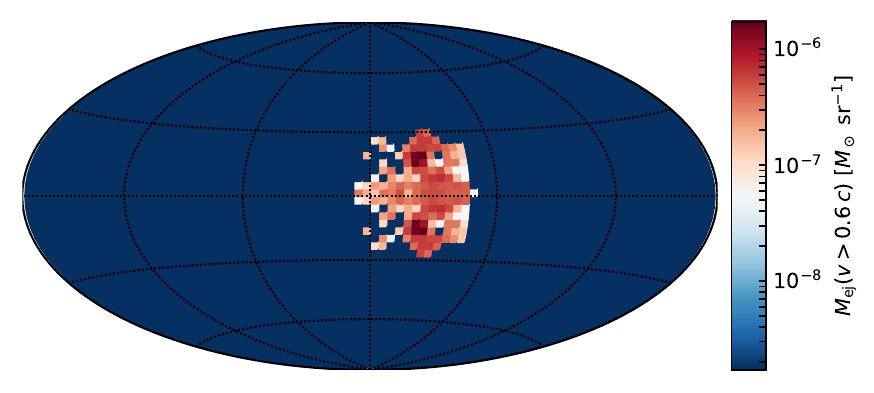}
    \texttt{DD2\_M135135\_LK}\phantom{pppppppppp} \\[1em]
    \includegraphics[width=0.98\columnwidth]{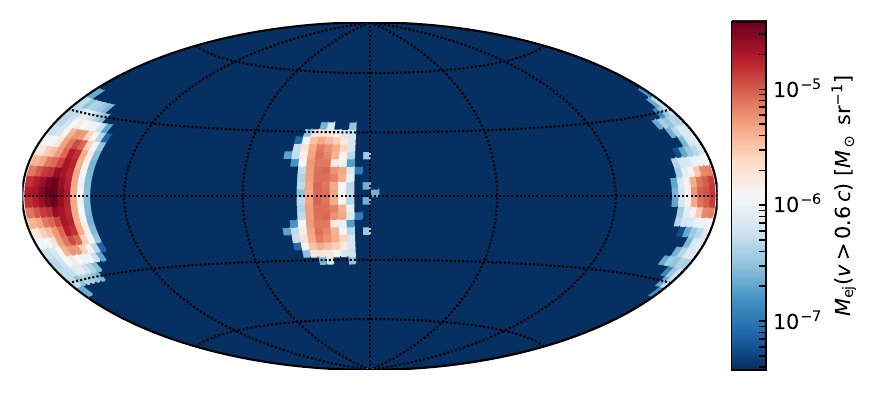}
    \texttt{BHBlp\_M135135\_LK}\phantom{pppppppppppp} \\[1em]
    \includegraphics[width=0.98\columnwidth]{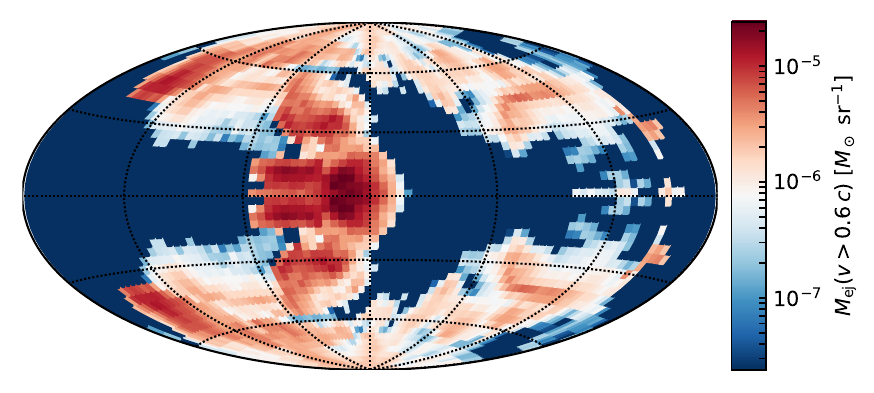}
    \texttt{SFHo\_M135135\_LK}\phantom{ppppppppppp}
  \end{center}
  \caption{Angular distribution of the ejecta with asymptotic velocities
  larger than $0.6\, c$ for the $(1.35 + 1.35)\ M_\odot$ binaries
  simulated without neutrino re-absorption, and at the fiducial
  resolution. The data is extracted on a coordinate sphere with radius
  $443\ \mathrm{km}$. The \texttt{LS220\_M135135\_LK} binary is not
  shown, since it does not produce an appreciable amount of fast moving
  ejecta. The fast-moving component of the outflow is anisotropic and
  covers a relatively small portion of the sky around the binary,
  especially in the case of the BHB$\Lambda\phi$ and DD2 binaries.}
  \label{fig:ejecta_fast_map}
\end{figure}

We find the distribution of the fast-component of the ejecta to be not
only anisotropic in latitude, but also in the azimuthal direction, as
shown in Fig.~\ref{fig:ejecta_fast_map}. The fast-moving material forms
narrow streams that cover only a small fraction of area of a sphere
centered at the location of the binary merger. In contrast, the overall
ejecta distribution is almost uniformly spread in the azimuthal
direction \citep[\eg,][]{bovard:2017mvn}. We remark that non-spinning
equal mass binaries have a discrete rotational symmetry of $180^\circ$
around the orbital angular momentum axis. However, this symmetry is
typically broken at the time of the merger, when turbulence operating in
the shear layer between the two \acp{NS} can exponentially amplify small
initial asymmetries, such as those due to floating point truncation in
the simulations \citep{paschalidis:2015mla, radice:2016gym}. This is
reflected in the asymmetry of the fast-moving tail of the dynamical
ejecta and is another indication of the fact that the bulk of the
fast-moving ejecta is launched with some delay from the merger. Note
that the early fast ejecta for the equal mass binaries with the SFHo
\ac{EOS}, which instead originates at the time of the merger, is
symmetric (see Fig.~\ref{fig:ejecta_fast_map_sfho_q1}).

\begin{figure}
  \includegraphics[width=0.98\columnwidth]{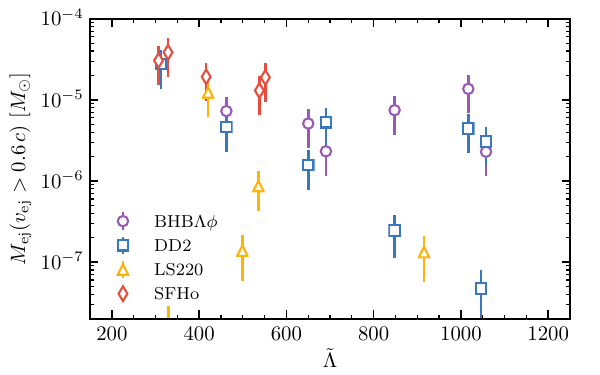}
  \caption{Fast moving component of the ejecta plotted against the
  binary tidal deformability $\tilde\Lambda$. The data points show the
  results from our fiducial subset of simulations. Note that several
  binaries produce no appreaciable amount of fast moving ejecta and are
  not seen in the figure. There is only a tentative correlation between
  the mass of the fast moving ejecta and the tidal deformability of the
  binary.}
  \label{fig:ejecta_Mfast_vs_Lambda}
\end{figure}

We report the total mass of the ejecta with $v_{\rm ej} > 0.6\, c$ in
Fig.~\ref{fig:ejecta_Mfast_vs_Lambda}. The error bars are estimated
following Eq.~(\ref{eq:mejerr}). The motivation for this choice is that,
while we find relatively large numerical uncertainties in the total
ejecta mass, the relative distribution of the ejecta as a function of
velocity appears to be robust (see Sec.~\ref{sec:resolution}).
Consequently, the numerical uncertainty in the total ejecta mass should
be well correlated with that of the fast component. Each simulation is
labelled by the tidal deformability of the binary $\tilde\Lambda$. The
tidal parameter $\tilde\Lambda$ is related to the compactness of the
binary and should correlate with the strength with which different
\acp{NS} binaries collide at the time of merger. We might expect that
$\tilde\Lambda$ should also correlate with the amount of fast ejecta
produced during the collision between the two stars. Indeed, we find
that there is a weak correlation between the two in
Fig.~\ref{fig:ejecta_Mfast_vs_Lambda}. However, there are systematic
effects that are not captured by $\tilde\Lambda$. For instance, the
BHB$\Lambda\phi$ and the DD2 binaries have very close compactnesses,
since the two \acp{EOS} are identical for \acp{NS} with gravitational
mass $M \lesssim 1.5\, M_\odot$. However, the BHB$\Lambda\phi$ binaries
typically produce a significantly larger amount of fast moving ejecta
compared to the DD2 binaries. The reason is likely the more violent
bounce of the merger remnant due to the formation of $\Lambda$ hyperons
in the aftermath of the BHB$\Lambda\phi$ binary mergers
\citep{radice:2016rys}. Perhaps more importantly, there are several
binaries, spanning a wide range of $\tilde\Lambda$, that do not produce
any appreciable amount of fast moving ejecta.

\subsection{Secular Ejecta}
\label{sec:ejecta.secular}

Part of the tidal tails from the \acp{NS} remains bound to form a
rotationally supported disk around a central remnant more precisely
defined below. In the cases in which the former survives for more than
about one millisecond, we observe the formation of hot ${\sim}10{-}20\,
{\rm MeV}$ streams of material expelled from what is originally the
interface region between the \acp{NS}. This material assembles into an
excretion disk. Consequently, there is a correlation between the life
time of the remnant and the remnant disk mass \citep{radice:2017lry},
see also Table~\ref{tab:ejecta}.

\begin{figure}
  \includegraphics[width=\columnwidth]{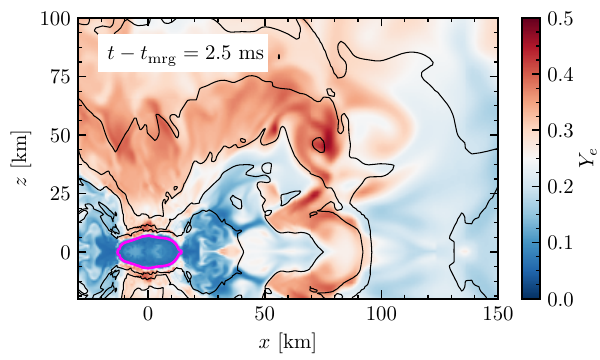}
  \caption{Electron fraction and density in the xz-plane for the
  \texttt{SFHo\_M135135\_M0} simulation. The thin-black contours enclose
  regions with density larger than $10^6$, $10^7$, $10^8$, $10^9$,
  $10^{10}$, and $10^{11}$ ${\rm g}\ {\rm cm}^{-3}$. The thick magenta
  light encloses the region with density larger than $10^{13}$ ${\rm g}\
  {\rm cm}^{-3}$. This figure should be compared with the last panel of
  Fig.~\ref{fig:SFHo_M135135_Ye_3d} where a volume rendering of the same
  data is shown.}
  \label{fig:SFHo_M135135_Ye_xz}
\end{figure}

The disks are geometrically thick and moderately neutron rich
(see Fig.~\ref{fig:SFHo_M135135_Ye_xz}; \citealt{siegel:2017jug}). We
observe the propagation of $m=2$ spiral density waves originated as the
streams from the massive \ac{NS} remnant impact the disk. After the
first $10{-}20$~ms from the merger, these streams subside, and $m=1$
spiral density waves, induced by the one-armed spiral instability of the
remnant \ac{NS} \citep{paschalidis:2015mla, east:2015vix,
radice:2016gym}, become dominant.

If the merger does not result in the formation of a \ac{BH} within few
dynamical timescales, the remnant is composed of a relatively slowly
rotating inner core surrounded by a rotationally supported envelope
\citep{shibata:2005ss, kastaun:2016yaf, hanauske:2016gia,
ciolfi:2017uak}. In particular, regions with rest-mass densities below
$\simeq 10^{13}\, {\rm g}\, {\rm cm}^{-3}$ are mostly rotationally
supported \citep{hanauske:2016gia}. Note that the rotational structure of
the remnant might be strongly affected by the effective shear viscosity
arising due to the turbulent fluid motion in the remnant
\citep{radice:2017zta, shibata:2017jyf, kiuchi:2017zzg,
fujibayashi:2017puw, radice:2018xqa}. In our analysis we define as the
central part of the remnant the region with $\rho_0 \geq 10^{13}\,{\rm
g}\ {\rm cm}^{-3}$, and we estimate the remnant disk mass $M_{\rm disk}$
from the integral of the rest-mass density of the region with $\rho_0 <
10^{13}\ {\rm g}\ {\rm cm}^{-3}$.  We remark that the same criterion has
also been adopted by \citet{shibata:2017xdx}. Furthermore, as shown in
Fig.~\ref{fig:SFHo_M135135_Ye_xz}, the density threshold $10^{13}\, {\rm
g}\, {\rm cm}^{-3}$ does indeed approximately corresponds to the boundary
of the centrally condensed remnant. Our results are given in
Table~\ref{tab:ejecta}. Unfortunately, the 3D output necessary to estimate
the disk mass in post processing has been accidentally deleted for six of
our simulations. For those cases the disk masses are not given.

Figure~\ref{fig:SFHo_M135135_Ye_xz} shows the electron fraction and the
density contours shortly after merger, during the formation of the
accretion disk. At this time the disk has not yet reached its maximum
extent and is expanding behind the cloud of the dynamical ejecta. The
former can be recognized for their higher electron fraction and are
located at radii $\gtrsim 75\, {\rm km}$. The neutron rich outflow
visible at radii $\gtrsim 100$~km is part of the tidal tail, while the
higher $Y_e$ material between the forming accretion disk and the tidal
tail is part of the outflow generated during the first bounce of the
merger remnant.

\begin{figure}
  \includegraphics[width=0.98\columnwidth]{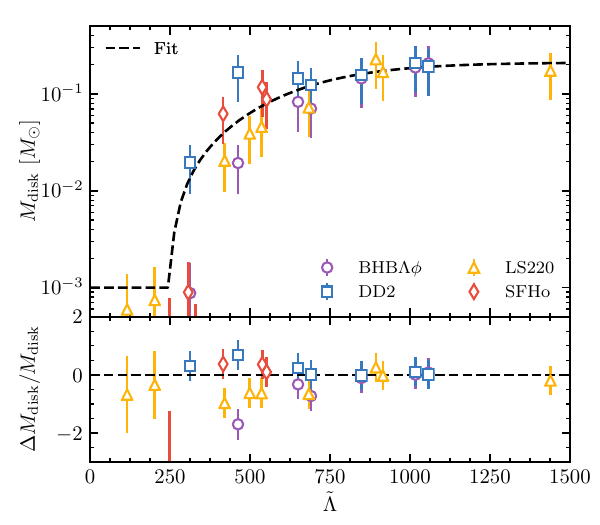}
  \caption{Remnant disk masses as a function of the tidal parameter
  $\tilde\Lambda$. The data points show the results from our fiducial
  subset of simulations. Disk formation is suppressed in the case of
  prompt BH formation, corresponding to small $\tilde\Lambda$'s. The
  final disk masses saturate for large values of $\tilde\Lambda$.}
  \label{fig:disk_mass_fit}

\end{figure}

We find that the remnant disk masses tightly correlated with the tidal
deformability of the binary $\tilde\Lambda$. This is shown in
Fig.~\ref{fig:disk_mass_fit} where we plot the disk masses for our
fiducial models. The error bars are estimated from the comparison of
simulations performed at different resolutions (see
Table~\ref{tab:ejecta}). In particular, we estimate the uncertainty on
the disk mass due to numerical errors to be ${\sim}30\%$.  To be
conservative, we estimate the uncertainties on the disk masses as
\begin{equation}
  \Delta M_{\rm disk} = 0.5\, M_{\rm disk} + (5 \times 10^{-4})\,
  M_\odot
\end{equation}
We remark that the error bars only account for finite-resolution
uncertainties and we cannot exclude that missing physics, or more
extreme binaries with larger mass asymmetries and/or extreme spins,
could deviate from the trend shown in Fig.~\ref{fig:disk_mass_fit}.
Moreover, because a significant fraction (${\sim}30{-}90\%$) of the disk
is accreted promptly after \ac{BH} formation, the transition between low
and high mass disks visible in Fig.~\ref{fig:disk_mass_fit} would likely
become sharper if we could evolve all of the hypermassive remnants to
collapse.

Notwithstanding these caveats, we find that our data is reasonably well
fitted with the simple expression
\begin{equation}\label{eq:mdisk_fit}
  \frac{M_{\rm disk}}{M_\odot} = \max \left\{ 10^{-3},
    \alpha + \beta \tanh\left( \frac{\tilde\Lambda - \gamma}{\delta} \right)
    \right\}
\end{equation}
with fitting coefficients $\alpha = 0.084$, $\beta = 0.127$, $\gamma =
567.1$, and $\delta = 405.14$. In the case of GW170817, the amount of
ejecta estimated from the modeling of the kilonova signal ${\sim}0.05\
M_\odot$ is about an order of magnitude too large to be explained by the
dynamical ejecta. This suggests that most of the material powering the
kilonova should have been produced during the viscous evolution of the
accretion disk.

This finding is in agreement with the results from long-term GRMHD
simulations of postmerger disks that show that up to ${\sim}40\%$ of the
accretion disk can be ejected over the viscous timescale
\citep{siegel:2017nub, fernandez:2018kax}. According to this scenario,
$M_{\rm disk}$ should have been larger than at least ${\sim}0.1\
M_\odot$. This, in turn, implies that $\tilde\Lambda$ for GW170817
should have been larger than about $400$. Another possibility, that,
however, we cannot presently test with our data, is that the progenitor
binary to GW170817 had a large mass asymmetry. Under these conditions it
has been suggested that the merger could still produce a massive disk
even for compact configurations \citep{shibata:2003ga, shibata:2006nm,
rezzolla:2010fd}.  On the other hand, large mass asymmetries are
disfavored on the basis of mass measurements for galactic double pulsars
\citep{ozel:2016oaf}.  Moreover, \citet{shibata:2006nm} found that the
accretion disk mass only increases by about an order of magnitude for
extremely asymmetric binaries.

\begin{figure}
  \includegraphics[width=0.98\columnwidth]{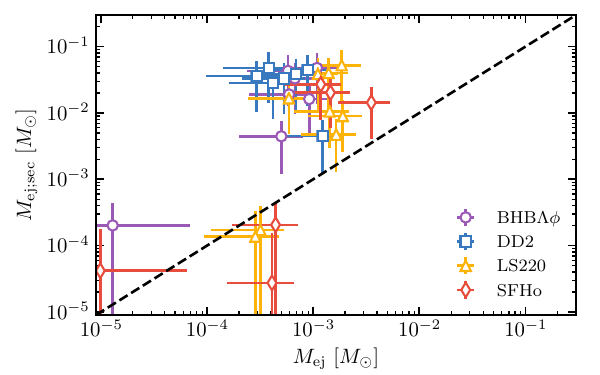}
  \caption{Dynamical ejecta $M_{\rm ej; dyn}$ versus secular ejecta
  masses $M_{\rm ej; sec}$. With the exception of the prompt BH
  formation cases that are able to expel at least a few $10^{-4}\,
  M_\odot$ in dynamical ejecta, the secular ejecta dominate over the
  dynamical ejecta.}
  \label{fig:dyn_ejecta_vs_wind_ejecta}
\end{figure}

We speculate on the total amount of mass expelled in a BNS merger.  We
consider three different ejection channels.  In addition to the
dynamical ejecta, directly extracted from the simulations, we include
the possible presence of neutrino- and magnetically-driven winds from
the disk, which develop on a time scale of a few tens of ms.  We
parametrize their contribution to the ejecta as a fraction of the disk
mass. The neutrino-driven wind is estimated as $M_{\rm ej;wind} =
\xi_{\rm wind} M_{\rm disk}$, with $\xi_{\rm wind} = 0.03 \pm 0.015$.
The uncertainty includes variations due to the possible presence of a
longer-lived remnant \citep[\eg,][]{martin:2015hxa, just:2014fka}. The
viscous ejecta is taken to be $M_{\rm ej; vis} = \xi_{\rm vis} M_{\rm
disk}$ with $\xi_{\rm vis} = 0.2 \pm 0.1$, a range including the results
from most postmerger simulations \citep[\eg,][]{metzger:2014ila,
just:2014fka, siegel:2017jug, fujibayashi:2017puw, fernandez:2018kax}.

In Fig.~\ref{fig:dyn_ejecta_vs_wind_ejecta} we compare dynamical and
secular ejecta masses, the latter estimated as $M_{\rm ej; sec} = M_{\rm
ej;wind}+M_{\rm ej;vis}$. With the exception of the prompt-BH formation
cases that produce at least $10^{-4}\, M_\odot$ of dynamical ejecta, the
total ejecta is largely dominated by the disk ejecta. As a consequence
of the tight correlation between the disk mass and $\tilde{\Lambda}$, we
expect a correlation between the total ejecta and $\tilde{\Lambda}$.
Indeed, our estimates for $M_{\rm ej}+M_{\rm ej;wind}+M_{\rm ej;vis}$
are reasonably well fitted by the same simple formula used for $M_{\rm
disk}$, Eq.~(\ref{eq:mdisk_fit}), but assuming a floor value of $5
\times 10^{-4}M_{\odot}$ and with fitting coefficients $\alpha = 0.0202
$, $\beta = 0.0341$, $\gamma = 538.8$, and $\delta = 439.4$. In most of
the cases, the relative error between the total ejecta mass and the fit
is below 50\%. When prompt BH formation occurs, we do not expect the
total ejecta to be larger than $\sim 10^{-3} M_{\odot}$. On the other
hand, for long-lived remnants the mass of the unbound material can span
a broad range of masses: from a few times $10^{-3} M_{\odot}$ to $0.1
M_{\odot}$, increasing with $\tilde{\Lambda}$.

In comparison to previous studies (\citealt{oechslin:2006uk,
hotokezaka:2012ze}, as reported in \citealt{wu:2016pnw}) we find that
disk winds and viscous outflows should contribute a significantly larger
fraction of the overall ejecta. The difference can be explained in part
by the fact that previous simulations did not include the effects of
neutrino cooling, or used approximate treatments for the gravity, and in
part by the fact that we assume that a larger fraction of the accretion
disk can be unbound secularly compared to \citet{wu:2016pnw}. The latter
is motivated by recent GRMHD simulations that showed that up to
${\sim}40\%$ of the disk can be unbound secularly \citep{siegel:2017nub,
fernandez:2018kax}. Our study differs from some of the previous studies
also in the numerical setup and analysis methodology.

\section{Nucleosynthesis}
\label{sec:nucleosynthesis}
The discovery of a kilonova counterpart to GW170817 \citep{gbm:2017lvd,
arcavi:2017a, chornock:2017sdf, cowperthwaite:2017dyu, coulter:2017wya,
drout:2017ijr, evans:2017mmy, kasliwal:2017ngb, nicholl:2017ahq,
smartt:2017fuw, soares-santos:2017lru, tanvir:2017pws, tanaka:2017qxj}
provided compelling evidence that \ac{NS} mergers are one of the main
sources of $r$-process elements in the Universe \citep{kasen:2017sxr,
rosswog:2017sdn, hotokezaka:2018aui}. However, the question of whether
\ac{NS} mergers are the only source of $r$-process elements or whether
is a contribution from other sources is still not completely settled.
Part of the uncertainty is due to the lack of a full theoretical
understanding of the nucleosynthetic yields from mergers. Here, we study
in detail the dependency of the $r$-process nucleosynthesis on the
properties of the binary, mostly focusing on the dynamical ejecta.

\subsection{Dynamical Ejecta}

\label{sec:nucleosynthesis.dynamical}
For simplicity, we perform most of our nucleosynthesis calculations
using the approach we developed in \citet{radice:2016dwd} and that we
briefly recall. We extract electron fraction $Y_{e,R}$, specific entropy
$s_R$, velocity $v_R$, and rest-mass density $\rho_{0,R}$ of the unbound
material crossing a coordinate sphere surface with radius $R = 300\,
G/c^2\, M_\odot \simeq 443\, {\rm km}$. A fluid element is considered to be
unbound if its kinetic energy is sufficient to overcome the
gravitational potential well, that is, $u_t \leq -1$, where we have
assumed a nearly stationary metric. See \citet{kastaun:2014fna} and
\citet{bovard:2017mvn} for possible alternative criteria. For the
nucleosynthesis calculations we further assume that the outflow is
undergoing an homologous expansion
\begin{equation}\label{eq:homologous.flow}
  \rho_0(t) = \rho_{0,R} \left(\frac{v_R}{c R} t\right)^{-3}\,.
\end{equation}
This density history is matched with the expansion histories used in the
parametrized $r$-process calculations of \citet{lippuner:2015gwa}:
\begin{equation}
  \rho_0(t) = \rho_0\big(s, Y_e, T = 6\, {\rm GK}\big)
    \left( \frac{3\tau}{e t} \right)^3\,,
\end{equation}
where $e \simeq 2.718$ is Euler's number. From the matching we extract
the expansion timescale $\tau$. To compute the nucleosynthesis yields we
bin the ejecta according to their density, entropy, and expansion time
scale. The nucleosynthesis yields in each bin can be pre-computed,
because, under the assumption of homologous expansion, the $r$-process
outcome depends only on $\rho_{0,R}$, $s_R$ $Y_{e,R}$, and $\tau_R$.
Then we compute the full nucleosynthetic yields by summing up the
contributions from all bins.  The uncertainty due to this procedure and
the comparison with nucleosynthetic calculations performed using
Lagrangian tracer particles are discussed in
Sec.~\ref{sec:nucleosynthesis.tracers}.

\begin{figure*}
  \begin{minipage}{0.49\textwidth}
  \includegraphics[width=0.98\columnwidth]{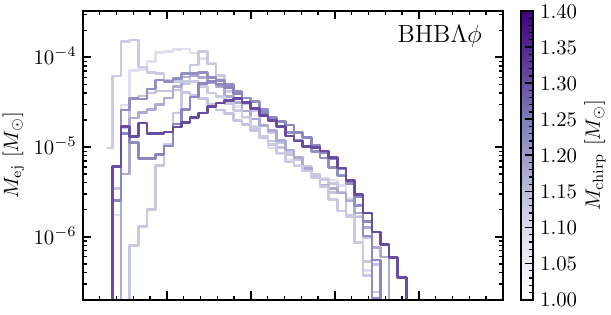}
  \includegraphics[width=0.98\columnwidth]{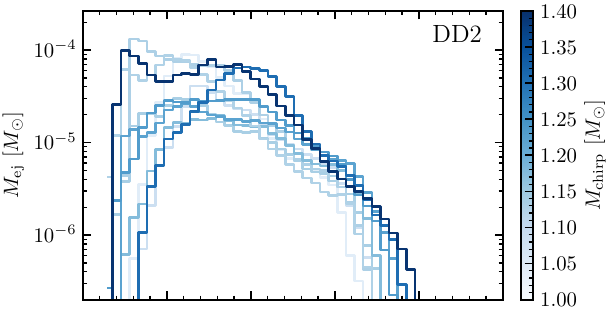}
  \includegraphics[width=0.98\columnwidth]{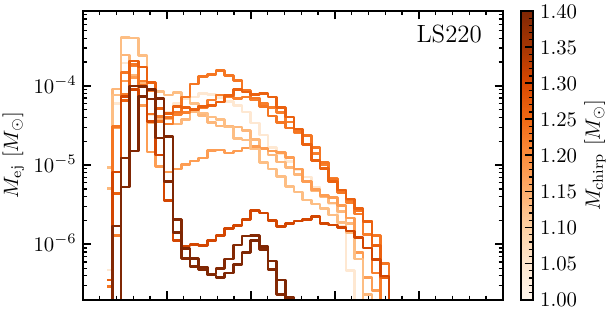}
  \includegraphics[width=0.98\columnwidth]{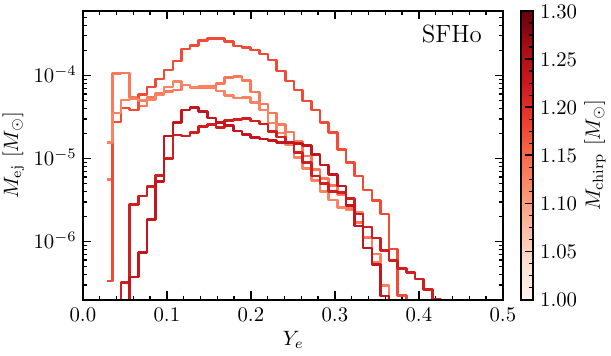}
  \end{minipage}
  \hfill
  \begin{minipage}{0.49\textwidth}
  \includegraphics[width=0.98\columnwidth]{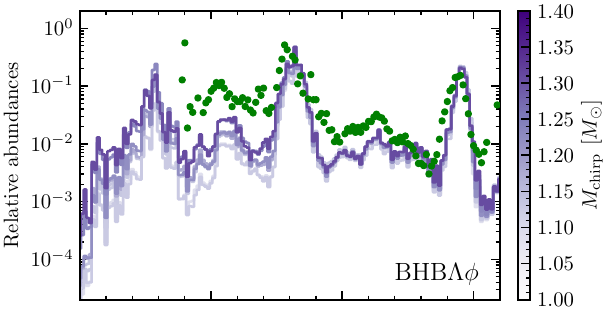}
  \includegraphics[width=0.98\columnwidth]{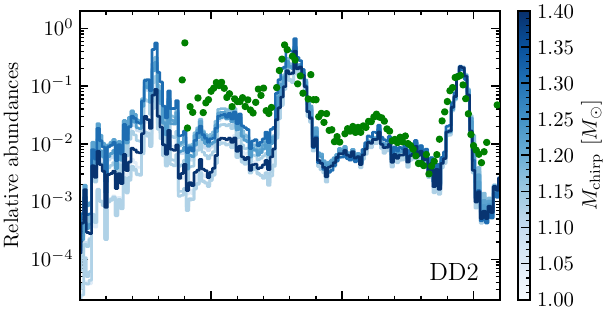}
  \includegraphics[width=0.98\columnwidth]{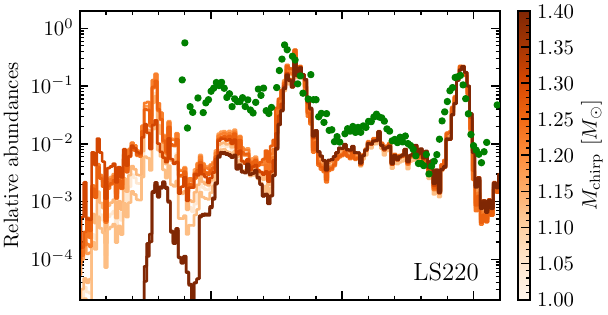}
  \includegraphics[width=0.98\columnwidth]{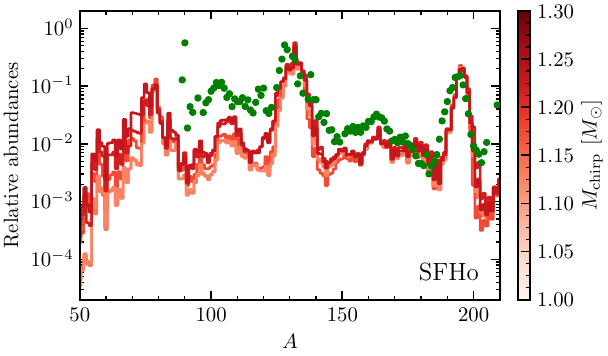}
  \end{minipage}
  \caption{Electron fraction of the ejecta before the activation of the
  $r$-process (\textit{left panel}) and final nucleosynthetic abundances
  (\textit{right panel}). The histograms show the results from our
  fiducial subset of simulations. We only include models with total
  ejecta mass larger than $5\times 10^{-5}$ $M_\odot$. The green dots in
  the right panel show the solar abundances from
  \citet{arlandini:1999an}. All abundance curves are normalized by
  fixing the overall fraction of elements with $180 \leq A \leq 200$.}
  \label{fig:ejecta_ye_yield}
\end{figure*}

We discuss first the systematics of the nucleosynthetic yields for our
fiducial subset of runs, for which neutrino re-absorption has been
neglected. The effect of neutrino re-absorption is discussed below.
Fig.~\ref{fig:ejecta_ye_yield} reports histograms of the electron
fraction $Y_e$ of the ejecta and the relative abundances of different
isotopes synthesized by the $r$-process 32 years after the merger. We
show the electron fraction since it is the most important variable
determining the outcome of the $r$-process in the conditions relevant
for \ac{NS} mergers \citep[\eg,][]{lippuner:2015gwa, radice:2016dwd}.
In the figure we compare isotopic abundances of $r$-process elements
estimated from our simulation with the solar abundances from
\citet{arlandini:1999an}.  Both abundances are normalized by fixing the
overall fraction of elements with $180 \leq A \leq 200$.

In absence of neutrino heating, the distribution of the ejecta as a
function of $Y_e$ drops rapidly for $Y_e > 0.15{-}0.2$. We find
systematic changes in the composition of the outflows as the total mass
of the binary increases. However, these variations are inconsistent
between the different \acp{EOS}. The BHB$\Lambda\phi$ and SFHo binaries
produce higher $Y_e$ ejecta as the binary mass is increased. Presumably
because the larger compactness of the more massive binaries inhibits the
production of tidally-driven ejecta. As a consequence, these binaries
produce more of the light $r$-process elements, \ie, the isotopes with
$90 \lesssim A \lesssim 125$, when approaching the prompt
\ac{BH}-formation threshold. Conversely, the relative amount of shocked
ejecta from the LS220 binaries progressively decreases as the binary
mass increases and the ejecta for the most massive binaries appears to
be dominated by the tidal tail. Consequently, the relative abundance of
light $r$-process elements decreases with the total binary mass for the
LS220 binaries. Finally, the behavior of the DD2 binaries is non
monotonic with mass: for chirp masses up to ${\sim}1.25\, M_\odot$ the
average electron fraction $\langle Y_e \rangle$ increases with the mass.
However, the highest mass binary \texttt{DD2\_M160160\_LK}, has the
lowest $\langle Y_e \rangle$ and the lowest relative abundance of
light $r$-process elements.

Overall, we find that, irrespective of binary parameters and \ac{EOS},
the dynamical ejecta robustly produce second- and third-peak elements,
\ie, elements with $125 \lesssim A \lesssim 145$ and $185 \lesssim A
\lesssim 210$, respectively. The reason for this is that the bulk of the
outflows are always very neutron rich, with $\langle Y_e \rangle < 0.2$.
On the other hand, we observe variance in the production of the light
$r$-process elements.

The normalization condition we impose effectively scales the second and
third peaks of the solar abundances to the second and third peaks of the
calculated abundances (see Figure \ref{fig:ejecta_ye_yield}). With this
normalization, it is clear that our calculations underproduce the
material in the mass ${\sim} 140$ region beyond the second $r$-process
peak.  We note that although different models produce a wide range of
$Y_e$ distributions, this underproduction is robust.  Therefore, it is
more likely due to the choice of nuclear input data used in our network
calculations. In particular, the choice of fission fragment
distributions and neutron-induced fission rates can have a significant impact on
abundances in this region \citep{eichler:2014kma}, and the symmetric
fission fragment distributions used in our calculations likely
underproduce material in this region.

\begin{figure*}
  \includegraphics[width=0.98\columnwidth]{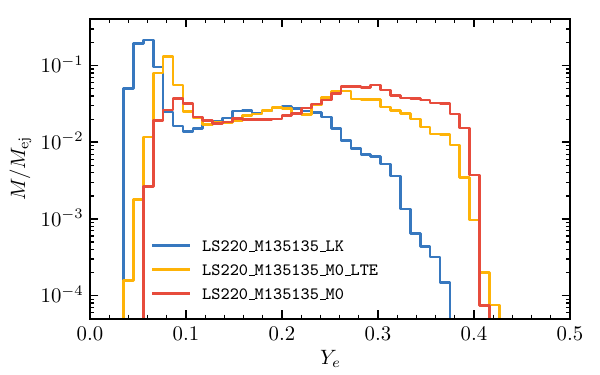}
  \hfill
  \includegraphics[width=0.98\columnwidth]{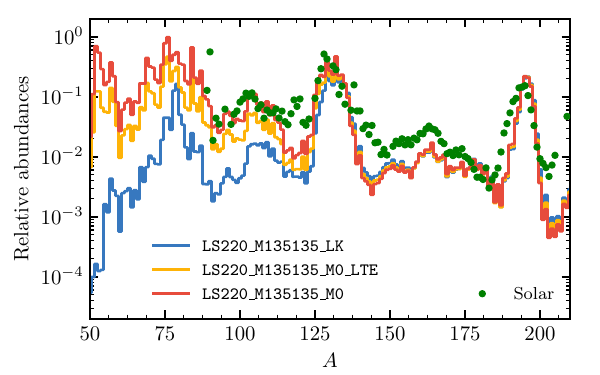}
  \caption{Dynamical ejecta sensitivity to the neutrino treatment.
  \textit{Left panel}: electron fraction. \textit{Right panel}:
  nucleosynthetic yields. All abundance curves are normalized by fixing
  the overall fraction of elements with $180 \leq A \leq 200$. Neutrino
  absorption can have a strong impact on the ejecta composition and on
  the nucleosynthetic yields around the light $r$-process elements.}
  \label{fig:ejecta_dep_nu}
\end{figure*}

Neutrino-matter interaction rates roughly scale with the square of the
incoming neutrino energy. Consequently, the composition, nucleosynthetic
yield, and \ac{EM} opacity of the ejecta depend on the details of the
neutrino radiation spectra \citep{foucart:2016rxm}. The determination of
the latter is not possible using an energy integrated scheme as is the
one adopted for this work. To quantify the relative uncertainty, we
simulate the $(1.35 + 1.35)\, M_\odot$ binary using the LS220 \ac{EOS}
and two different schemes for the calculation of the absorption
opacities. In addition to the standard treatment in which we
approximatively account for the incoming neutrino energy using
Eq.~(\ref{eq:sigma.eff}), we also perform a simulation in which the
absorption cross-section is calculated assuming thermal equilibration
between matter and neutrinos. The latter simulation is labelled
\texttt{LS220\_M135135\_M0\_LTE}.

The results of this comparison are shown in
Fig.~\ref{fig:ejecta_dep_nu}. As anticipated, the inclusion of
compositional changes and heating due to the absorption of neutrinos
results in a significant shift in the ejecta $Y_e$ distribution. The
tidal tail is reprocessed to slightly higher values of $Y_e$ and a new
ejecta component, with $Y_e$ up to $0.4$ is found. The approximate
inclusion of non-LTE effects results in a slight increase in the
absorption rates. This is expected because the incoming neutrinos
originally decoupled from the central regions of the remnant and the
accretion disk at higher temperatures than those found in the ejecta.
Overall, we find that the dynamical ejecta, especially in the regions
close to the orbital plane, robustly produces heavy $r$-process elements
with relative abundances close to solar. On the other hand, the neutrino
re-absorption significantly affects the production of elements in the
region close to the first peak of the $r$-process relative abundance
pattern, \ie, around $A = 80$. The treatment of the absorption opacities
can result in changes in the relative abundance of elements in this
region of up to factors of a few.  This points to the need for more
sophisticated simulations with spectral neutrino transport.

We do not consider the effect of neutrino oscillations in this work.
However, we point out that recent studies have shown that \ac{NS} merger
remnants might host ideal conditions for resonant oscillations and fast
flavor conversion \citep{zhu:2016mwa, frensel:2016fge, deaton:2018ser}.
This might have an impact on the nucleosynthesis of light $r$-process
elements \citep{wu:2017drk}. The investigation of these effects in the
context of dynamical simulations is urgent.

\begin{figure*}
  \includegraphics[width=0.98\columnwidth]{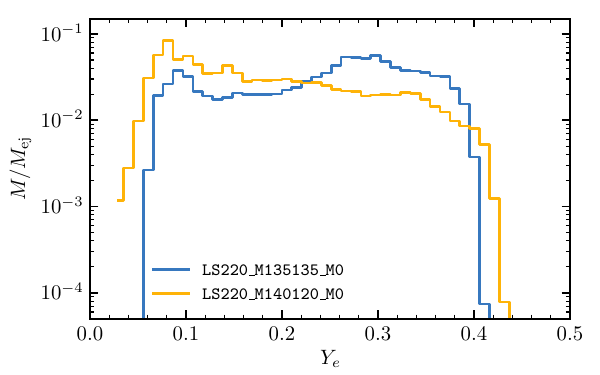}
  \hfill
  \includegraphics[width=0.98\columnwidth]{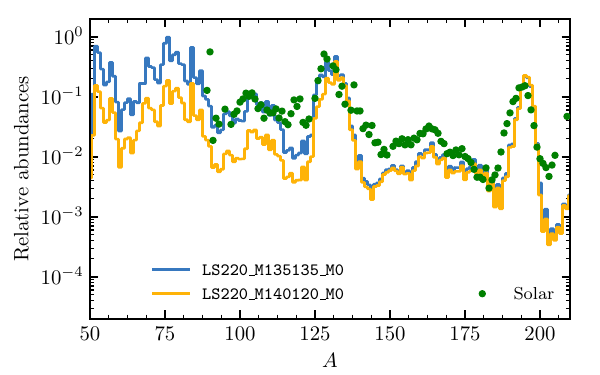}
  \caption{Dynamical ejecta sensitivity to the binary mass ratio.
  \textit{Left panel}: electron fraction. \textit{Right panel}:
  nucleosynthetic yields. All abundance curves are normalized by fixing
  the overall fraction of elements with $180 \leq A \leq 200$.  The
  nucleosynthetic yield from the dynamical ejecta is sensitive to the
  mass ratio in the region of the light $r$-process elements $A \lesssim
  120$.}
  \label{fig:ejecta_dep_q}
\end{figure*}

The nucleosynthetic yield of the dynamical ejecta is also sensitive to
the mass ratio of the binary, especially in the region of light
$r$-process elements. This is shown in Fig.~\ref{fig:ejecta_dep_q} where
we compare composition and nucleosynthesis yeids between the
\texttt{LS220\_M135135\_M0} and \texttt{LS220\_M140120\_M0} binaries
which are representative of the general trend with mass ratio. The
relative abundances are normalized as in Fig.~\ref{fig:ejecta_ye_yield}
over the mass range $180 \leq A \leq 200$. Because of the more abundant
tidal tail the relative composition of the ejecta is shifted to lower
$Y_e$. This affects the relative abundances of first-peak $r$-process
elements which are reduced by factors of several. A similar trend is
also present in the simulations that do not include neutrino
re-absorption, see Table~\ref{tab:ejecta}.

\begin{figure*}
  \includegraphics[width=0.98\columnwidth]{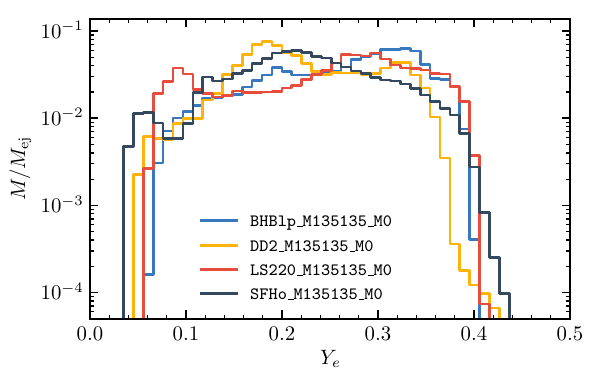}
  \hfill
  \includegraphics[width=0.98\columnwidth]{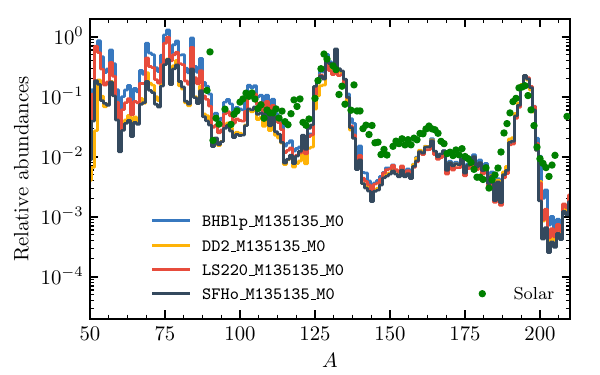}
  \caption{Dynamical ejecta sensitivity to the EOS. \textit{Left panel}:
  electron fraction. \textit{Right panel}: nucleosynthetic yields. All
  abundance curves are normalized by fixing the overall fraction of
  elements with $180 \leq A \leq 200$. The total ejecta mass shows
  depends sensitively on the EOS, however electron fraction and
  nucleosynthetic yields appear to be insensitive to the EOS.}
  \label{fig:ejecta_dep_eos}
\end{figure*}

Even though there are differences in the quality and quantity of the
dynamical ejecta, the $r$-process nucleosynthesis appears to be only
weakly sensitive to the \ac{EOS}. In Fig.~\ref{fig:ejecta_dep_eos} we
show electron fraction distributions and final isotopic abundances for
the dynamical ejecta from the $(1.35 + 1.35)\, M_\odot$ binary simulated
with different \acp{EOS}. Neutrino re-absorption has been included in
these simulations using the M0 scheme. Irrespective of the \ac{EOS}, the
dynamical ejecta is neutron rich $\langle Y_e \rangle \lesssim 0.25$,
but has a broad distribution in $Y_e$, with a significant amount of
ejecta with $Y_e$ as large as $0.4$. Consequently, both heavy and light
$r$-process elements are produced. For comparable mass binaries, as are
those shown in Fig.~\ref{fig:ejecta_dep_eos}, we find that the dynamical
ejecta robustly synthesizes $r$-process elements with isotopic
abundances close to solar. Conversely, the dynamical ejecta from
binaries with large mass asymmetry underproduces light $r$-process
elements. These conclusions are not altered with the inclusion of
viscosity. Indeed, while viscosity impacts the overall ejecta mass
(Table~\ref{tab:ejecta} and \citealt{radice:2018ghv}), it does not
affect the electron fraction and the isotopic abundances of the
$r$-process in the dynamical ejecta. Overall, our simulations suggest
that the \ac{EOS} of dense matter controls the amount of the dynamical
ejecta, but that the relative isotopic abundances are most sensitive to
weak reaction rates, mass ratio, and total binary mass.

\subsection{Secular Ejecta}
\label{sec:nucleosynthesis.secular}

The nucleosynthetic yields of the secular ejecta have been the subject
of several recent studies. In the case of neutrino-driven winds,
neutrino irradiation of the expanding ejecta drives the electron
fraction towards higher values than those of the original cold weak
equilibrium configuration. The electron fraction of the material in the
wind depends on the relative timescales for weak re-equilibration and
expansion. If the expansion is not too rapid, then the material achieves
electron fractions given by the weak equilibrium in optically thin
conditions with neutrinos \citep{qian:1996xt}:
\begin{equation}
 \left( Y_e \right)_{\rm eq} = \left[ 1 + \frac{L_{\bar{\nu}_e} \left(
 \epsilon_{\bar{\nu}_e } - 2 \Delta + 1.2 \Delta^2/
 \epsilon_{\bar{\nu}_e }  \right)} {L_{{\nu}_e}     \left(
 \epsilon_{{\nu}_e }     + 2 \Delta + 1.2 \Delta^2/ \epsilon_{{\nu}_e }
 \right) } \right]^{-1}\,,
\end{equation}
where $L_{\nu_e}$ and $L_{\bar{\nu}_e}$ are the luminosities for
electron neutrinos and antineutrinos, $ \epsilon_{\nu} \equiv \langle
E^2_\nu \rangle / \langle E_\nu \rangle $, $E_{\nu}$ is the neutrino
energy, $\langle x \rangle$ the average of $x$ over the neutrino
distribution function, and $\Delta$ is the mass difference between
neutrons and protons in vacuum. During the early post merger phase,
$L_{\bar{\nu}_e} \gtrsim L_{\nu_e}$, while electron antineutrinos are
significantly hotter than neutrinos ($\langle E_{\bar{\nu}_e} \rangle
\approx 15~{\rm MeV} > \langle E_{\nu_e} \rangle \approx 10~{\rm MeV}$,
e.g.  \citealt{dessart:2008zd,perego:2014fma,foucart:2016rxm}). Thus,
$\left( Y_e \right)_{\rm eq} \lesssim 0.45$ and $r$-process
nucleosynthesis can occurs. However, due to the small neutron-to-seed
ratio, only nuclei with $A \lesssim 130$ can be synthetized and this
ejecta is expected to contribute to the first $r$-process peak. If the
dynamical ejecta is dominated by the tidal component and is poor of
first peak $r$-process elements, the neutrino-driven wind can complement
the nucleosynthesis and lead to the production of all $r$-process
elements \citep{martin:2015hxa}.

If a massive disk forms after the merger, viscously-driven ejection,
occurring over the disk lifetime, can become the dominant source of
ejecta from a BNS merger (see Sec.~\ref{sec:ejecta.secular}). Viscous
hydrodynamics simulations of the long term disk evolution, mainly
performed in axisymmetry assuming a BH-torus system, showed that, if the
disk becomes transparent to neutrinos, the rapid decrease in temperature
makes neutrino cooling inefficient and the disk becomes convectively
unstable \citep[e.g.][]{fernandez:2013tya,just:2014fka}. The resulting
large scale mixing, combined with the long time scale over which
neutrino-matter interactions can occur, produces a rather uniform, broad
distribution of $Y_e$ in the ejecta. In particular, it was found that
the resulting distribution has $0.1 \lesssim Y_e \lesssim 0.45$ and all
$r$-process elements from the first to the third peak, as well as
Uranium and Thorium, can be synthesized in proportions close to solar
\citep[e.g.][]{wu:2016pnw}.

Recent 3D GRMHD simulations of a BH disk torus \citep{siegel:2017nub,
fernandez:2018kax} confirmed the presence of a self-regulating mechanism
based on electron degeneracy in the disk mid-plane that ensures the
presence of a reservoir of neutron rich material ($Y_e \sim 0.1$). This
results in the production of neutron rich outflows ($\langle Y_e \rangle
\sim 0.2$). The resulting nucleosynthesis yields all $r$-process
elements between the second and the third peak. If the kinetic energy
dissipation in the innermost part of the disk results in a significant
neutrino luminosity, neutrino absorption increases the electron fraction
of the ejecta, producing also elements down to the first peak. Neutrino
influence on the properties of the viscous ejecta can be even more
relevant in the presence of a long-lived massive neutron star. This is
expected to emit a large amount of neutrinos over the diffusion time
scale \citep[a few seconds; \eg,][]{dessart:2008zd,perego:2017fho}.  The
high neutrino flux is expected to unbind matter in a neutrino-driven
wind during the first tens of ms after the merger
\citep{dessart:2008zd,perego:2014fma,fujibayashi:2017xsz} and can
further increase the electron fraction of the viscous ejecta. For a very
long lived massive neutron star, the properties of the neutrino- and
viscously-driven ejecta could become similar and the resulting
$r$-process nucleosynthesis in the viscous ejecta could be limited to
the first and second $r$-process peaks \citep{lippuner:2017bfm}.

\subsection{Effect of the Thermodynamic History of the Ejecta}
\label{sec:nucleosynthesis.tracers}

In principle, nucleosynthesis in the ejecta should be calculated by
following the non-equilibrium evolution of the materials composition as
it is advected along with the fluid flow and potentially undergoes
mixing. Such an approach would require tracking the large number of
isotopic abundances needed to follow the $r$-process flow along with
adding stiff, coupled source terms to the composition equations. Such an
approach is too computationally expensive, but it would include the
possible feedback on the ejecta dynamics of nuclear heating and
composition based changes to the EOS \citep{metzger:2010aa}. The next
level of approximation to the nucleosynthesis in the outflows would be
following nucleosynthesis in Lagrangian tracers of the flow, while
ignoring the backreaction of nucleosynthesis on the flow dynamics. This
is likely to be a very reasonable approximation, since the nuclear
energy release is only likely to be important to the dynamics of a small
amount of marginally bound material.  Nevertheless, nuclear burning will
produce entropy in these fluid elements which may change the nuclear
flow. Therefore, most calculations of nucleosynthesis in binary NS
merger ejecta involve taking density histories, $\rho(t)$, of Lagrangian
tracers and evolving the composition and entropy of the material in time
starting at $t_0$, with entropy $s_0$ and electron fraction $Y_{e,0}$
extracted from the simulation output using a self-heating nuclear
reaction network \citep{freiburghaus:1999a}.

\begin{figure}
 \includegraphics[width=0.98\columnwidth]{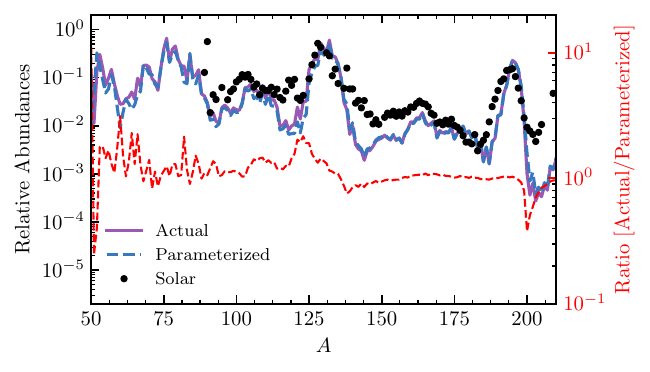}
 \caption{Integrated nucleosnythesis from a simulation using the actual
 density history from the simulation extracted from tracer particles
 (dashed line) and using the integrated mass flux approximation
 described in the text (solid line). The thin dashed line shows the
 ratio of the two calculations.}
 \label{fig:abundance_calculation_comparison}
\end{figure}

For the nucleosynthesis results discussed above in this section, an even
more approximate method has been used in which we assume $\rho(t)
\approx \rho_0(t; \tau_d, s_R, Y_{e,R})$, where $Y_{e,R}$ and $s_{R}$
are the electron fraction and entropy at of an ejected fluid element
when it crosses a radius $R$ and $\tau_d = e R/(3 v_R)
(\rho_R/\rho_0)^{1/3}$ as described above. This allows us to rapidly
calculate nucleosynthesis without including Lagrangian tracer particles
in the simulations, but this approximation needs to be tested.
Therefore, we have run one simulation with a large number of tracer
particles and calculated nucleosynthesis in these tracers using their
actual density histories. The results of this calculation compared to
our approximate method of calculating nucleosynthesis are shown in
Fig.~\ref{fig:abundance_calculation_comparison}. At all mass numbers,
the calculations agree to within a factor of two and in most regions to
better than 20\%. This agreement is good enough to give us confidence in
our approximation.  Nevertheless, we would like to understand where the
variations come from.

\begin{figure}
 \includegraphics[width=0.98\columnwidth]{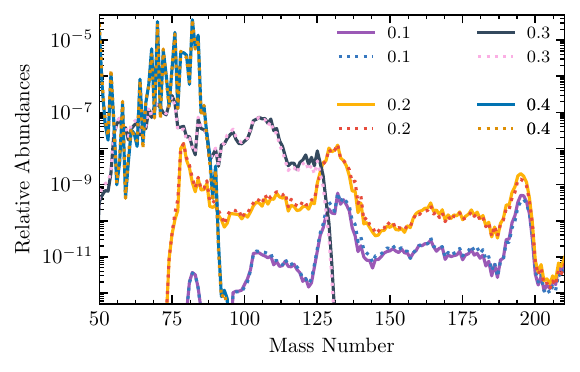}
 \caption{Nucleosynthesis for four selected tracer particles calculated
 using the actual density history from the simulation (solid lines) and
 using a parameterized density history extracted as described in the
 text (dashed lines). The legend labels the $Y_e$ of each separate
 trajectory. We see that there is generally good agreement between the
 nucleosynthesis predictions with the different density histories across
 a broad range of $Y_e$. The abundance curves are scaled by arbitrary
 factors for clarity.}
 \label{fig:density_history_comparison}
\end{figure}

The first possible source of error in our approximation is that going
from $\rho(t) \rightarrow \rho_0(t; \tau_d, s_{R}, Y_{e,R})$ introduces
errors into the calculation. We test this for individual tracer
particles by running nucleosynthesis calculations using both the actual
density history $\rho(t)$ and using $\rho_0(t)$. The nucleosynthesis
results for these particles are shown in
Fig.~\ref{fig:density_history_comparison}. Our approximate functional
form for the density captures most of the behavior of the actual
particles and we recover very similar nucleosynthesis in both cases. All
abundances above $10^{-5}$ agree within a factor of two between the two
calculations for all $Y_e$.

\begin{figure}
 \includegraphics[width=0.98\columnwidth]{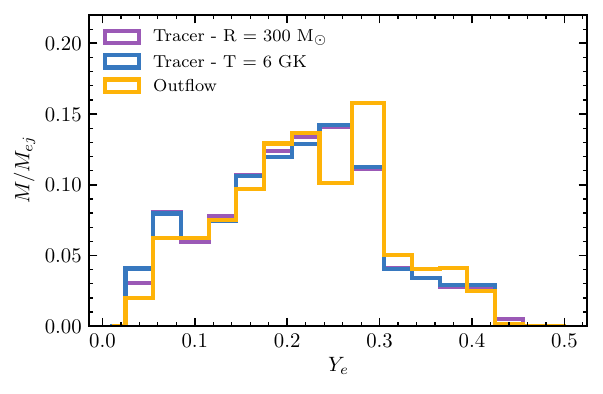}
 \caption{Distribution of the electron fraction in the ejecta as
 inferred from the tracer particles when they reach a radius of $300 \,
 G/c^2\, M_\odot$, when they reach a temperature of $6 \, \textrm{GK}$,
 and from the integrated mass outflow at $300\, G/c^2\, M_\odot$
 calculated on the Eulerian grid as described above.}
 \label{fig:ye_dist_comparison}
\end{figure}

The second possible source of error comes from sampling error in the
Lagrangian tracer particles. To test how well they represent the
underlying Eulerian flow, we compare the distribution of $Y_{e,R}$ and
$s_R$ inferred from integrating the flux using the Eulerian outflow and
sampling the conditions in the Lagrangian particles at $R$. These are
shown in Fig. \ref{fig:ye_dist_comparison}.  The $Y_e$ and entropy of
the particle may change between $R=300 \, G/c^2\, M_\odot$ and the time
nucleosynthesis begins at ${\sim} 6 \, \textrm{GK}$. The average
electron for the Eulerian outflow is $\langle Y_e \rangle = 0.22$, while
it is $\langle Y_e \rangle = 0.20$ at both measurement points in the
tracer particles. By comparing the tracer $Y_e$ distributions at these
two points, we can see that this approximation introduces at most a
$25\%$ error in the $Y_e$ distribution (at $Y_e = 0.04$) and
substantially less error at most $Y_e$. The difference between either
tracer $Y_e$ distribution and the integrated Eulerian outflow $Y_e$
distribution is substantially larger, almost $40\%$ at
$Y_e = 0.25$. We have checked that this error is not due to undersampling
by the Lagrangian tracer particles.

\section{Electromagnetic Signatures}
\label{sec:em}
The radioactive decay of freshly synthesized $r$-process nuclei in
the expanding ejecta powers a quasi-thermal emission known as kilonova
\citep{li:1998bw,kulkarni:2005jw,metzger:2010sy}. The properties of the
emission crucially depend on the amount of mass, on the expansion
velocity, and on the detailed composition of the ejecta.  The latter, in
particular, determines the photon opacity, which can substantially
differ from the opacity of iron group elements in the presence of a
significant fraction of lanthanides \citep[\eg,][]{kasen:2013xka,
tanaka:2013ana}.

Eleven hours after the detection of GW170817, a kilonova transient
consistent with a binary NS merger was detected and intensively followed
up for a few weeks \citep[AT2017gfo;][]{gbm:2017lvd,arcavi:2017a,
coulter:2017wya, drout:2017ijr, evans:2017mmy, kasliwal:2017ngb,
nicholl:2017ahq, smartt:2017fuw, soares-santos:2017lru, tanvir:2017pws}.
The emission peaked in the UV and visible bands within the first day.
Subsequently, the kilonova reddened and peaked in the near infrared
(NIR) on a time scale of a few days.

A long-lasting synchrotron remnant is suggested as another promising
electromagnetic counterparts to neutron star mergers
\citep{nakar:2011cw, hotokezaka:2016clu}. Previous studies showed that
such signals can be produced by several different ejecta components,
\eg, short GRB jet, cocoon, and dynamical ejecta \citep{vaneerten:2011,
piran:2013, hotokezaka:2015eja}. A synchrotron remnant has been detected
in X-ray, optical, and radio bands for GW170817 \citep{haggard:2017qne,
hallinan:2017woc, margutti:2017cjl, mooley:2017enz,
ruan:2017bha,troja:2017nqp, alexander:2018dcl, dobie:2018,
lyman:2018qjg, margutti:2018xqd}. Recently, \citet{mooley:2018qfh} and
\citet{ghirlanda:2018uyx} observed the superluminal motion of a compact
radio emission region in GW170817 and provided us with direct evidence
of the existence of a narrowly collimated off-axis jet.

\subsection{Kilonovae}

We compute synthetic kilonova light curves for the non-viscous, standard
resolution models presented in Table~\ref{tab:ejecta} for which disk
masses are available. We use the semi-analytical kilonova model
presented in \citet{perego:2017wtu}. The model accounts for different
ejecta components and their possible anisotropies. It assumes azimuthal
symmetry around the rotational axis of the binary and reflection
symmetry with respect to the orbital plane. The polar angle $\theta$ is
discretized in 30 angular bins equally spaced in $\cos{\theta}$. Within
each polar ray, the luminosity, radius, and temperature at the outer
photosphere are related through Stefan-Boltzmann's law. Finally, we
compute AB magnitudes in different UV/visible/IR bands assuming the
distance of the source to be 40~Mpc.

Following \cite{villar:2017wcc}, we include a floor temperature $T_{\rm
c}$ for the photosphere. We assume all material with temperature less
than $T_{\rm c}$ to be transparent so that the effective photosphere
temperature is always larger or equal to $T_{\rm c}$ \citep[see,
\eg,][]{barnes:2013wka}. Guided by the values of $T_{\rm c}$ obtained by
\citet{villar:2017wcc} for AT2017gfo for different ejecta components, we
assume a composition dependent $T_{\rm c}$. For material having $Y_e >
0.25$ we set $T_{\rm c} = T_{\rm c,blue} = 3000\, {\rm K}$, otherwise we
set $T_{\rm c} = T_{\rm c,red}=1000\, {\rm K}$.

Angular profile, expansion velocity, and electron fraction of the
dynamical ejecta are directly extracted from the simulations. Where
$Y_e$ is larger than $0.25$, the kilonova model assumes a
Lanthanide-free gray photon opacity $\kappa = \kappa_{\rm blue} =
1.0~{\rm cm^2~g^{-1}}$. Otherwise we set $\kappa = \kappa_{\rm red} =
30~{\rm cm^2~g^{-1}}$.

For the secular ejecta we include both neutrino-driven and
viscously-driven winds. The mass of these two components are assumed to
be fixed fractions of the remnant accretion disk, as outlined in
Sec.~\ref{sec:ejecta.secular}.

The neutrino-driven wind part of the secular ejecta is assumed to be
uniformly distributed from the pole ($\theta = 0$) to $\theta =
60^\circ$. This ejecta component is assumed to expand radially with
$v_{\rm rms} = 0.08~c$. We assume low opacity $\kappa = \kappa_{\rm
blue}$ for the part of the neutrino driven wind at $\theta \leq
45^\circ$, which is found to be less neutron rich in postmerger
simulations \citep[\eg,][]{perego:2014fma, martin:2015hxa}, while we set
the opacity of the rest of the neutrino-driven wind to be $\kappa =
\kappa_{\rm purple} = 5.0\, {\rm cm^2\, g^{-1}}$, since this part of the
wind is expected to have a broader distribution of $Y_e$'s.

The viscously-driven wind is assumed to have a $\sin^2{\theta}$
distribution in mass as a function of the polar angle and to expand
homologously with $v_{\rm rms} = 0.06~c$. Due to the expected broad and
homogeneous $Y_e$ distribution of this outflow component
\citep[\eg,][]{metzger:2014ila, fujibayashi:2017puw}, we assume $\kappa
= \kappa_{\rm purple}$ at all angles.

The model parameters discussed above, as well as other parameters not
explicitly mentioned, have been calibrated using AT2017gfo in
\citet{perego:2017wtu}. We refer to that study for a more detailed
account of the procedure used to generate synthetic kilonova
lightcurves.

\begin{figure*}
  \includegraphics[width=\linewidth]{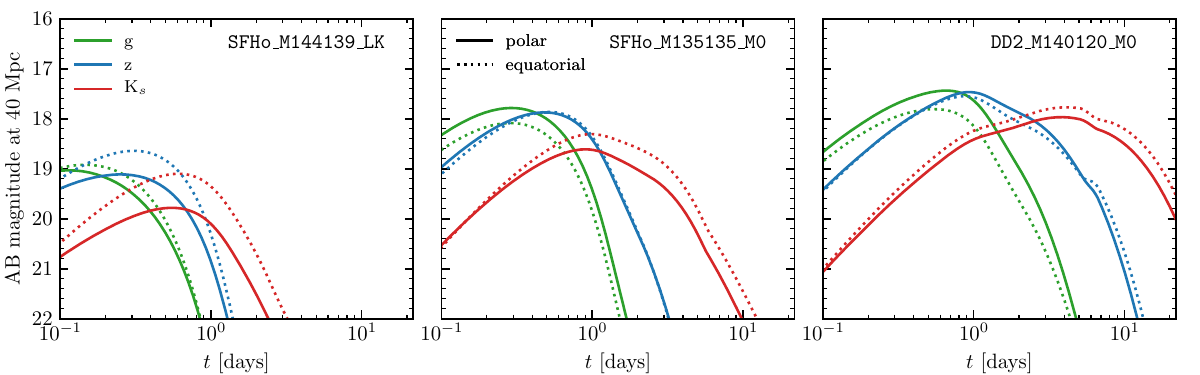}
  \caption{Kilonova synthetic light curves in three different bands
  ($g$, $z$, and $K_s$) for three representative models. The left panel
  shows a binary with prompt BH formation. The middle panel shows a
  binary forming an hypermassive massive NS ($t_{\rm BH} = 7.7~{\rm
  ms}$). Finally, the right panel shows a binary forming a long-lived
  supramassive NS ($t_{\rm BH} > 34.2~{\rm ms}$). Solid and dashed lines
  correspond to polar and equatorial viewing angles, respectively.
  Prompt BH formation binaries do not form massive accretion disk and
  result in faint and fast transients. Longer-lived NS remnants are
  associated with the formation of more massive disks that are source of
  more abundant secular outflows. They result in brighter kilonovae that
  evolve on longer timescales.}
  \label{fig:kilonova.representative}
\end{figure*}

In Fig.~\ref{fig:kilonova.representative} we present synthetic light
curves for three representative binaries in three different bands. The
left panel shows a prompt BH formation case, \texttt{SFHo\_M144139\_LK},
the middle panel shows a short-lived HMNS case,
\texttt{SFHo\_M135135\_M0}, and the right panel shows a case where a
black hole has not formed by the end of the simulation,
\texttt{DD2\_M140120\_M0}. In the latter case we assume that the massive
NS will not collapse within the accretion disk lifetime. In a previous
work, we have already speculated on the possible implications of a
long-lived or even stable massive NS on the kilonova light curves
\citep{radice:2018xqa}. We compute light curves in a band in the visible
part of the \ac{EM} spectrum ($g$) and in two NIR bands ($z$ and $K_s$).
These are chosen because their effective wavelength midpoints satisfy
the proportion 1:2:4, and because they span a significant fraction of
expected detection range.

The \texttt{SFHo\_M144139\_LK} binary undergoes prompt collapse and
forms a \ac{BH} surrounded by a light accretion disk with $M_{\rm disk}
= 9 \times 10^{-4}~M_{\odot}$. In this case the ejecta is predominantly
of dynamical origin ($M_{\rm ej}=4\times 10^{-4}M_{\odot}$). The outflow
is neutron rich ($\langle Y_e \rangle = 0.18$) and expands rapidly
($v_{\rm ej} = 0.33\, c$). Despite the large effective photon opacity of
the ejecta, because of the small mass and fast expansion of the
outflows, this binary produces a rapid, but faint transient peaking
within the first day of the merger in all bands. Polar observers
preferentially receive radiation from lower opacity, but also lower
density material. For these observers the kilonova fades even more
rapidly and is bluer.

The \texttt{SFHo\_M135135\_M0} binary produces a short-lived HMNS. After
its collapse a \ac{BH}-torus system with a disk of $0.0123~M_{\odot}$ is
formed. We estimate that this binary will eject $\lesssim 3 \times
10^{-3}~M_{\odot}$ of material in the form of secular ejecta, an amount
comparable to that of the dynamical ejecta, ${\sim} 4.2 \times 10^{-3}
M_{\odot}$. The resulting kilonova transient peaks within a day in both
$g$ and $z$ bands and subsequently reddens. Because of the higher $Y_e$
of the dynamical ejecta in the polar region, and because of the presence
of the neutrino-driven wind from the disk, polar observers will see a
marginally brighter kilonova in the UV/visible bands compared to
equatorial observers. Conversely, due to the larger amount of dynamical
and viscous ejecta close the orbital plane, equatorial observers will
receive a larger NIR flux.

The results are qualitatively different for the third binary,
\texttt{DD2\_M140120\_M0}, which forms a long-lived remnant. While the
mass and the characteristics of the dynamical ejecta from this binary
are in line with that of the \texttt{SFHo\_M135135\_M0} binary, the
former produces a much more massive disk ($M_{\rm
disk}\approx0.19~M_{\odot}$). Consequently, the secular ejecta dominates
over the dynamical ejecta in this case and the outflow is overall more
massive and has a lower expansion velocity. The ejecta are optically
thick for days and this results in more slowly evolving light curves in
all bands. The light curves in both the $z$ and the $K_s$ bands present
a double peak structure with a first peak around 1 day and a second one
at around 5-7 days depending on the inclination.  The $z$ band shows a
kink instead of a second peak, and both peak and kink are shifted to
earlier times compared to the two peaks redder filters. These features
are the result of the presence of different ejecta components: the
dynamical and neutrino-driven wind ejecta, which mostly determine the
early light curve, and the viscously-driven wind, which dominates at
late times.

\begin{figure}
  \includegraphics[width=0.98\columnwidth]{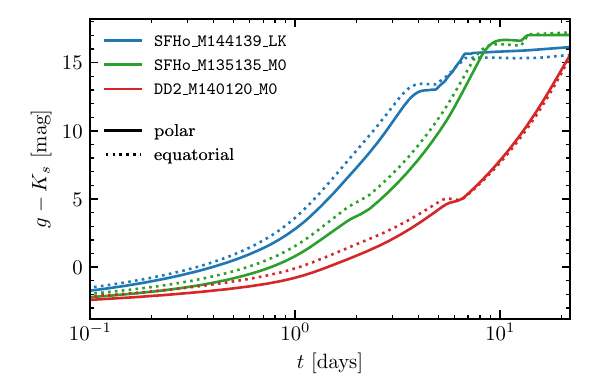}
  \caption{Color evolution of the kilonova light curves computed as the
  difference in the AB magnitude between $g$ and $K_s$ bands for three
  representative models. Binaries forming HMNSs or dynamically stable
  remnants power significantly bluer transients.}
  \label{fig:kilonova.representative.color}
\end{figure}

The color evolutions for these models, exemplified by the difference in
magnitude between the $g$ and $K_s$ bands, are shown in
Fig.~\ref{fig:kilonova.representative.color}. All of the synthetic
kilonova signals show a fast evolution towards the infrared after about
a day from the merger. However, there are are significant difference,
between them. This suggests that the color of the kilonova signal could
be used to infer the outcome of the merger. We remark that a similar
point was also made by \citet{metzger:2014ila} and
\citet{lippuner:2017bfm}. However, in our case the difference in the
color is purely due to the differences in dynamical ejecta and disk
masses, since we do not assume that long-lived remnant produce disk
winds with different compositions, as instead argued by those authors.

\begin{figure*}
  \includegraphics[width=\linewidth]{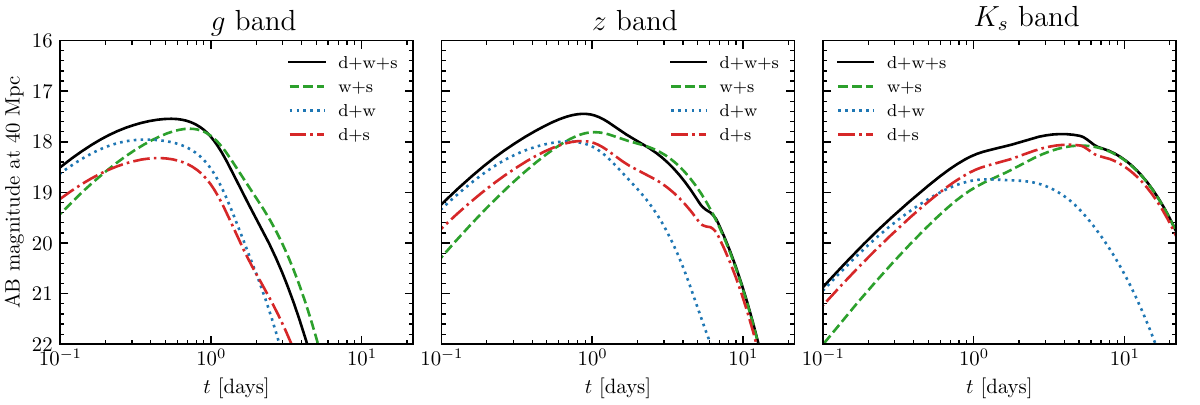}
  \caption{Synthetic kilonova light curves for an observer located at
  $45^\circ$ in the $g$ ({\it left}), $z$ ({\it middle}) and $K_s$ ({\it
  right}) bands for the \texttt{DD2\_M140120\_M0} binary computed using
  different combinations of ejecta components: dynamical, ``d'',
  neutrino-driven wind, ``w'', and viscously-driven wind, ``s''. The
  comparison between the light curve obtained including all ejecta
  componetns (d+w+s) and light curves obtained by neglecting in turn one
  of the components reveals the different roles of the components in
  shaping the light curves, at different times and in different bands.}
  \label{fig:test.2.components}
\end{figure*}

We study the relative impact that each of the ejecta components has on
the kilonova light curve in the case of the \texttt{DD2\_M140120\_M0}
binary. In addition to the ``full'' light curve, which includes all
ejecta components, we generate light curves in which, in turn, one of
the ejecta components has been removed. In doing so we assume the
viewing angle to be $45^\circ$. The results are shown in
Fig.~\ref{fig:test.2.components}. We remind that our model does not
simply add contributions from the different components, but also
accounts for irradiation and reprocessing effects. Starting from a few
days from the merger the light curves are dominated by the viscous
ejecta, especially in the NIR bands. This is not surprising, since this
component dominates the overall mass ejection. However, we find that
both the neutrino-driven wind and the dynamical ejecta play an important
role in determining the visible and NIR light curves in the first days.
The dynamical ejecta is most important in the very first day, when the
viscous ejecta is still too opaque to contribute significantly. All
ejecta components are necessary to reproduce the light curve properties
in the UV and visible bands.

\begin{figure}
  \includegraphics[width=0.98\columnwidth]{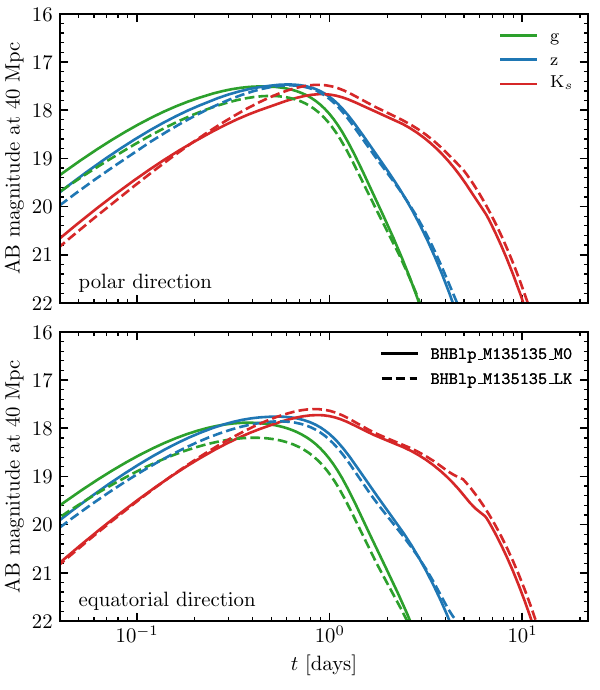}
  \caption{Synthetic kilonova light curves in the $g$, $z$
  and $K_s$ bands for the \texttt{BHBlp\_M135135\_M0} and
  \texttt{BHBlp\_M135135\_LK} binaries, both in the polar
  (\textit{top panel}) and in the equatorial (\textit{bottom panel})
  directions. Because neutrino-driven winds from the disk dominate the
  early light curve, the presence of a neutrino-irradiate component of
  the dynamical ejecta only yields a modest $\lesssim 0.5$~mag
  change of the color light curve in the first few hours after merger.}
  \label{fig:kilonova.nu.impact}
\end{figure}

The presence of a polar component of the dynamical ejecta irradiated by
neutrinos could, in principle, impact the properties of the kilonova
emission. We study this in Fig.~\ref{fig:kilonova.nu.impact} where we
compare light curves obtained from two binaries having the same NS
masses and EOS, but different neutrino treatments:
\texttt{BHBlp\_M135135\_LK} and \texttt{BHBlp\_M135135\_M0}. The angular
distribution and composition of dynamical ejecta of these binaries are
shown in Fig.~\ref{fig:ejecta_ye_theta}. The inclusion of neutrino
re-absorption results in an increase of the magnitude in the UV and
visible bands within the first few hours of the merger. However, these
differences amount only to less than 0.5 mag, even for an observer
located along the polar direction. The reason is that the overall
UV/optical signal is actually dominated by the secular neutrino-driven
wind, while the dynamical ejecta is less important.

In general, we find that the high-latitude neutrino-irradiated part of
the dynamical ejecta has a very modest impact on the kilonova light
curve. Instead the UV/optical signal is typically dominated by the secular
neutrino-driven winds. There are two reasons for this. First, the
secular neutrino-driven wind dominates the overall outflow of high-$Y_e$
material at high latitudes. Indeed, only a small fraction of the
dynamical ejecta is directed far from the orbital plane and experiences
significant neutrino irradiation. Second, because of its fast expansion,
the lanthanides-free part of the dynamical ejecta becomes transparent
already a few hours after merger and does not contribute significantly to
the kilonova light curve afterwards. The situation is slightly different
for prompt collapse binaries for which the accretion disk masses are
modest. However, in these cases the effect of neutrino irradiation of
the dynamical ejecta is expected to be negligible.

\begin{figure*}
  \includegraphics[width=\linewidth]{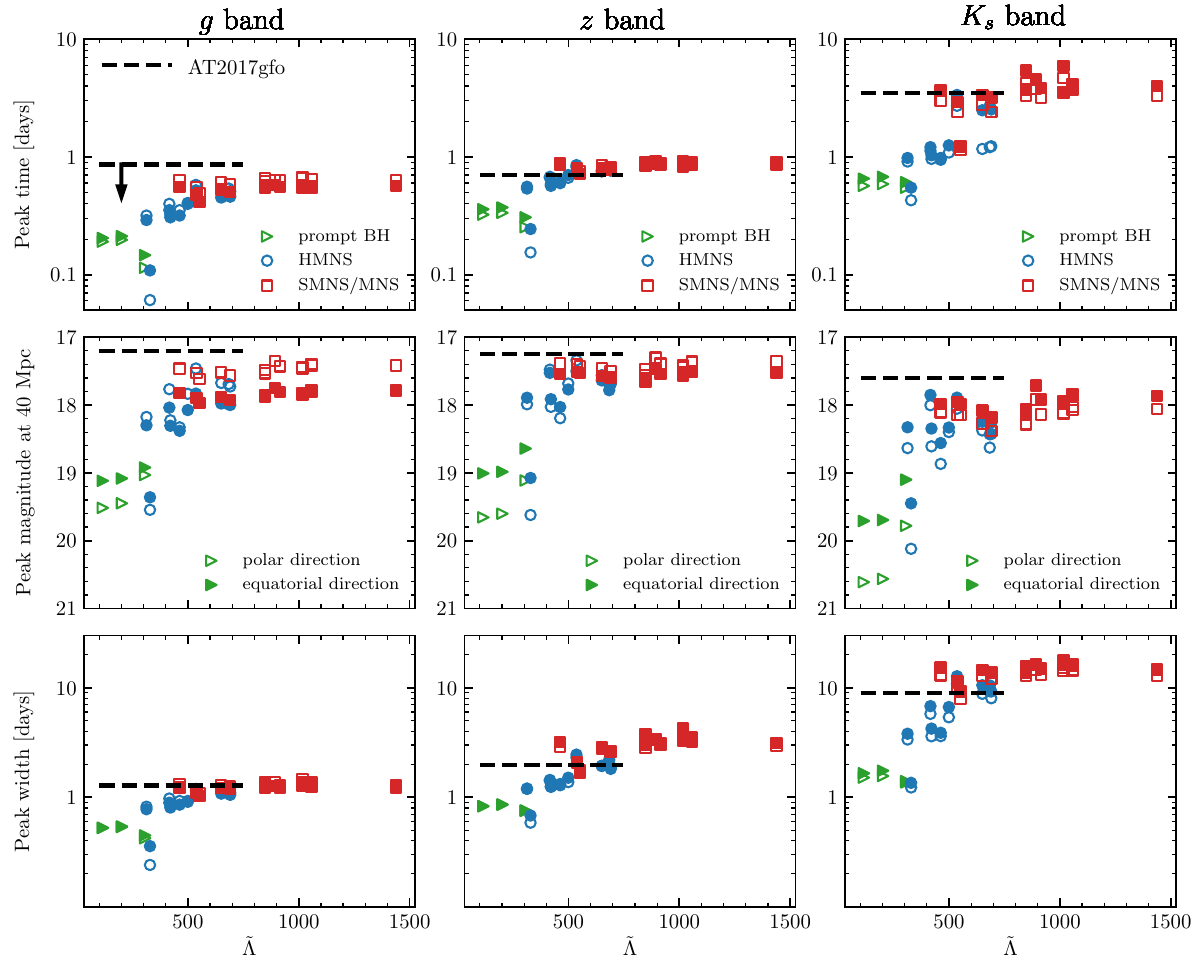}
  \caption{Peak times ({\it top}), AB magnitudes ({\it middle}), and
  widths ({\it bottom}) in three phomometric bands ($g$, $z$, and $K_s$,
  moving from left to right) of the kilonova synthetic light curves for
  our fiducial subset of simulations assuming a distance of 40 Mpc for
  the sources.  The peak width is defined as the time interval over
  which the AB magnitude increases by one across the peak. The dashed
  black lines correspond to the values (or limits) obtained for
  AT2017gfo, the EM counterpart to GW170817. The horizontal extension of
  the lines correspond to the 90\% highest posterior density interval
  $300^{+420}_{-230}$ for $\tilde\Lambda$ obtained assuming low-spin
  priors by \cite{abbott:2018wiz}. The peak times, magnitudes, and
  widths of AT2017gfo are obtained from \cite{villar:2017wcc},
  \citet{coulter:2017wya}, \citet{smartt:2017fuw}, and
  \citet{tanvir:2017pws}. BH formation (small $\tilde{\Lambda}$) results
  in faint and rapid transients. Longer-lived remnants (large
  $\tilde{\Lambda}$) are associated with brighter luminosity in the
  UV/visible bands and longer lasting, more slowly evolving NIR signals.}
  \label{fig:kilonova.peak.properties}
\end{figure*}

A summary of the most important features of the kilonova light curves in
different bands is presented in Fig.~\ref{fig:kilonova.peak.properties}.
For each binary in our fiducial subset of simulations we present the
time, the magnitude, and the width of the kilonova peak as a function of
$\tilde{\Lambda}$ in three different bands: $g$, $z$, and $K_s$.  The
peak width is defined as the time interval around the peak over which
$\Delta M = + 1$. We further distinguish between polar and equatorial
observers. Different symbols denote different merger outcomes.  There is
a clear dependence of the kilonova properties on the nature of the
merger remnant and on $\tilde{\Lambda}$. These trends, already
anticipated by the three representative cases of
Fig.~\ref{fig:kilonova.representative}, are a consequence of the
dependence of the disk mass on $\tilde{\Lambda}$ (see
Fig.~\ref{fig:disk_mass_fit}) and of the dominant role of the secular
ejecta in determining the properties of the kilonova. Faint and rapidly
decreasing kilonovae indicate the formation of BH via prompt collapse,
while bright and long lasting light curves in the NIR bands are
signatures of longer-lived remnants.

Figure~\ref{fig:kilonova.peak.properties} also reports the values or
upper limits inferred for AT2017gfo. The properties of the kilonova are
compatible with the formation of a massive \ac{NS} that survived for at
least several milliseconds after merger, as also argued by
\citet{shibata:2017xdx}. However, it is important to emphasize that our
theoretical models are based on a number of assumptions that still need
to be verified with first-principle simulations, most notably the
assumption that a fixed fraction of the accretion disk is unbound by
winds.

The differences in the kilonova brightnesses have important consequences
for their detectability. We consider the limiting magnitudes reported in
\citep{Rosswog:2016dhy} for the Large Synoptic Survey Telescope (LSST)
in the $g$ and $z$ bands, and for the Visual and Infrared Survey
Telescope for Astronomy (VISTA) in the $K_{\rm s}$ band and assume
exposure times of 60 seconds. Considering the detection horizon for
Advanced LIGO to be of 120-170 Mpc \citep{Aasi:2013wya}, we estimate
that kilonovae counterparts would be detectable for a large fraction of
the NS merger GW events. If a long-lived massive NS is formed, then the
kilonova would be dectable for all \ac{GW} events in both the $g$ and
$z$ bands up to the horizon distance of Advanced LIGO, and up to a
luminosity distance of ${\sim}90$ Mpc in the $K_{\rm s}$ band. If, on
the other hand, a \ac{BH} is rapidly formed after merger, then the
kilonova would only be detectable in the $K_{\rm s}$ band to distances
of 30-50 Mpc, depending on the orientation of the binary. Kilonovae
associated with prompt BH formation are detactable in the $z$ and $g$
bands in the entire Advanced LIGO and Virgo volume, however their
observation will be challenging given their fast decline.

\begin{figure}
  \includegraphics[width=0.98\columnwidth]{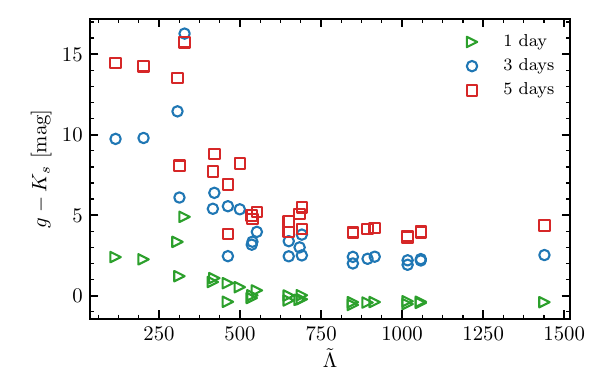}
  \caption{Color of the kilonova light curves for all the fiducial
  subset of simulations at three different times, as a function of the
  binary tidal parameter $\tilde{\Lambda}$.  Light curves are computed
  for an observer located at a polar angle of $45^\circ$. The color of
  the transient at different times shows a significant correlation with
  $\tilde{\Lambda}$.}
  \label{fig:kilonova.full.sample.color}
\end{figure}

In Fig.~\ref{fig:kilonova.full.sample.color} we show the difference in
magnitude between the $g$ and $K_s$ bands for kilonovae at three
different epocs and as a function of $\tilde{\Lambda}$. Our results show
that the color evolution and properties previously described for the
three representative models are generic. Moreover, there is a clear
correlation between $g - K_s$ and $\tilde{\Lambda}$ which suggests that
it might be possible to constrain the EOS of NS matter with \ac{EM}
observations in different bands.

We want to stress that, while the correlations between kilonova light
curve properties and $\tilde\Lambda$ document in
Figs.~\ref{fig:kilonova.peak.properties} and
\ref{fig:kilonova.full.sample.color} appear robust, their quantitative
determination must wait until simulations self-consistently including
dynamical and secular ejecta from \ac{NS} mergers become available. The
impact of viscosity on the kilonova lightcurves is discussed in a
companion paper \citep{radice:2018ghv}.

\subsection{Synchrotron remnants}

We calculate the light curve of the synchrotron radiation arising from
the dynamical ejecta with the semi-analytic method of
\citet{hotokezaka:2015eja}. According to this model electrons
accelerated in the shock between the ejecta and the ISM emit synchrotron
radiation in the amplified magnetic filed. The total flux density is
calculated by integrating the radiation flux from each solid angle over
the equal-arrival time surface. The ISM number density $n$ is a
parameter of the model. The conversion efficiencies of the internal
energy of the shock to the energy of the accelerated electrons and
amplified magnetic field are assumed to be $\epsilon_e$ and
$\epsilon_B$, respectively. The initial velocity profile of the ejecta
is given by mapping the three dimensional structure of the ejecta to a
one dimensional structure, then the ejecta are evolved adiabatically.

Before discussing the numerical results, here we give briefly the
scalings of the peak time and peak flux with the relevant physical
quantities. The peak time of the light curve can be estimated from the
deceleration time of the ejecta \citep{nakar:2011cw}:
\begin{eqnarray}
  t_{\rm dec} &\sim &3 \cdot 10^3\,{\rm day} \, \left(\frac{T_{\rm
  ej}}{10^{50}{\rm erg}} \right)^{1/3} \\ & & ~~~\times
  \left(\frac{n}{10^{-3}}{\rm cm^{-3}} \right)^{-1/3} \left(\frac{v_{\rm
  ej}}{0.3c} \right)^{-5/3}, \nonumber
\end{eqnarray}
where $E$ and $v$ are the ejecta kinetic energy and
initial velocity\footnote{This deceleration time is measured in the
merger rest frame. The difference between this time and observer time is
about a factor of a few.}. The peak flux can be estimated as
\begin{eqnarray}
  & F_{\nu} & \sim  10\,{\mu}{\rm Jy}\,\left(\frac{T_{\rm ej}}{10^{50}{\rm
  erg}}\right)\left(\frac{n}{10^{-3} {\rm cm^{-3}}}
  \right)^{\frac{p+2}{4}} \left(\frac{\epsilon_B}{10^{-2}}
  \right)^{\frac{p+2}{4}}\\ \label{eq:flux} & & \times
  \left(\frac{\epsilon_e}{10^{-1}} \right)^{p-1} \left(\frac{v_{\rm ej}}{0.3c}
  \right)^{\frac{5p-7}{2}} \left(\frac{D}{40\,{\rm Mpc}} \right)^{-2}
  \left(\frac{\nu}{3\,{\rm GHz}} \right)^{-\frac{p-1}{2}},\nonumber
\end{eqnarray}
where $p$ is the spectral index of the non-thermal electrons.

\begin{figure}
  \includegraphics[width=0.98\columnwidth]{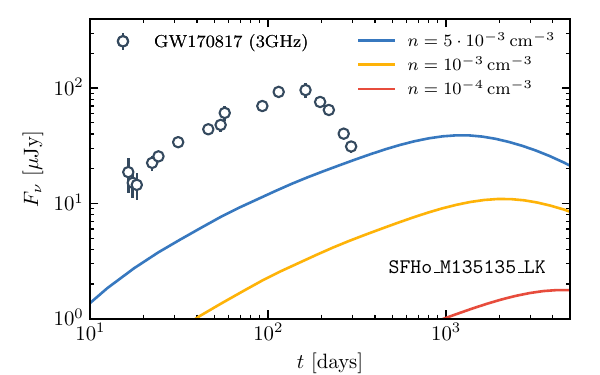}
  \caption{Radio light curves of the dynamical ejecta of
  \texttt{SFHo\_M135135\_LK} at $3$ GHz.  Here we assume the
  microphysics parameters to be $\epsilon_e=0.1$, $\epsilon_B=0.01$, and
  $p=2.16$. Also shown as open circles are the observed flux densities
  at $3$\,GHz of the afterglow in GW170817
  \citep{hallinan:2017woc,mooley:2017enz,mooley:2018qfh}.}
  \label{fig:radio.sfho}
\end{figure}

As in the previous work by \citet{hotokezaka:2018gmo}, who used the
results of high resolution merger simulations performed by
\cite{kiuchi:2017pte}, we also find that it is the fast component of the
dynamical ejecta with velocities of ${\sim} 0.3{-}0.8\, c$ to
predominantly produce a synchrotron signal. Figure \ref{fig:radio.sfho}
shows the expected radio signals of the dynamical ejecta of
\texttt{SFHo\_M135135\_LK} compared with GW170817. Here we calculate the light
curve with a similar method to \cite{hotokezaka:2018gmo} and choose the
number density of the ISM to be $10^{-4}$--$5\cdot 10^{-3}\,{\rm
cm^{-3}}$ as suggested by \cite{mooley:2018qfh}. While the dynamical
ejecta component is fainter than the observed flux densities until
${\sim} 300$ days, this component can be detectable in radio and X-rays
on time scales of a year to ten years in the optimistic case. The peak
flux density depends also on the EOS and it is fainter for the cases of
DD2, BHB$\Lambda \phi$, and LS200 because the kinetic energy in the high
velocity component is lower than that of SFHo (see Eq. \ref{eq:flux} for
the scaling).

\begin{figure}
  \includegraphics[width=0.98\columnwidth]{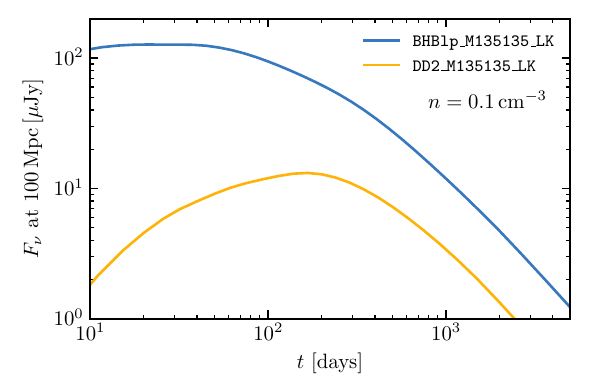}
  \caption{Radio light curves of \texttt{DD2\_M135135\_LK} and
  \texttt{BHBlp\_M135135\_LK} at $3$ GHz.  The distance to the source
  and ISM density are assumed to be $100$ Mpc and $0.1\,{\rm cm^{-3}}$,
  respectively. Here the  microphysics parameters are set to be
  $\epsilon_e=\epsilon_B=0.1$ and $p=2.5$.}
  \label{fig:radio.dd2.vs.bhblp}
\end{figure}

The synchrotron remnant arising from the dynamical ejecta may be
detectable in future GW events if the ISM density is large enough. For
instance, radio counterparts with a flux of $\gtrsim 100\,{\rm \mu Jy}$
can be detectable by blind survey with VLA, ASKAP, and MeerKAT when the
GW localization area is better than ${\sim} 30\,{\rm deg^2}$
\citep{hotokezaka:2016clu}. Of course, if the host galaxy is identified
by finding other EM counterparts, the detection limit is reduced to a
few tens of ${\rm \mu Jy}$. Figure \ref{fig:radio.dd2.vs.bhblp} shows
the expected light curves for \texttt{DD2\_M135135\_LK} and
\texttt{BHBlp\_M135135\_LK} at $3$ GHz, assuming  a distance of $100$
Mpc and density of $0.1\,{\rm cm^{-3}}$.  Note that the radio flux of
\texttt{SFHo\_M135135\_LK} is brighter than both models under the same
condition (density, distance and microphysics parameters). Therefore,
the radio afterglow arising from the dynamical ejecta can be detectable
for \texttt{SFHo\_M135135\_LK} and \texttt{BHBlp\_M135135\_LK} if a
merger occurs within $100$ Mpc and the surrounding ISM density of
$\gtrsim 0.1\,{\rm cm^{-3}}$.

It is worth noting that differences in the high density part of the EOSs
lead to large differences in the expected light curves. For example, as
shown in Fig.~\ref{fig:radio.dd2.vs.bhblp}, \texttt{BHBlp\_M135135\_LK}
results in the radio flux that is much brighter than that of
\texttt{DD2\_M135135\_LK}, even though these EOSs share the same form up
to around the nuclear saturation density. The reason is that
\texttt{BHBlp\_M135135\_LK} produces a larger amount of fast-moving
ejecta because of the more violent merger dynamics due to the appearance
of $\Lambda$ hyperons at high densities (see
Fig.~\ref{fig:ejecta_vel_hist}). Therefore, we may be able to constrain
the neutron star EOS using the observed light curves of the synchrotron
remnants of future GW events.

The impact of viscosity on the synchrotron lightcurves is discussed in a
companion paper \citep{radice:2018ghv}.










\section{Conclusions}
\label{sec:conclusions}
We have performed 59 full-GR \ac{NS} merger simulations employing a
microphysical treatment of the \ac{NS} matter and including compositional
and energy changes due to the emission of neutrinos. A subset of our
simulations also included a treatment of neutrino re-absorption and/or of
the effective viscosity due to MHD turbulence in the stars. This is
the largest set of \ac{NS} simulations with realistic microphysics to
date. Our studies focused on the ejection of material during and after
the mergers, and on the associated nucleosynthetic and \ac{EM}
signatures.

\subsection{Mass Ejection}
We find that material is ejected on a dynamical timescale during the
mergers due to the combined effects of tidal torques, shocks, and
neutrino heating. Tidally-driven ejecta flow close to the orbital plane
of the binary, have low entropies (${\sim} 10\, k_{\rm B}$ per baryon),
and are very neutron rich, with electron fraction ${\sim} 0.1$.
Shock-driven ejecta have broad distributions in entropy and electron
fractions, and are distributed over a broad ${\sim}60^\circ$ angle from
the orbital plane. Neutrino driven winds are emitted preferentially
close to the polar axis and have electron fractions in excess of $0.25$.
In accordance with previous studies using approximate GR and/or
approximate microphysics \citep[\eg,][]{hotokezaka:2012ze,
bauswein:2013yna}, we find that the most conspicuous episode of mass
ejection is triggered by the centrifugal bounce of the merger remnant
shortly after merger. Overall, we find dynamical ejecta masses of few
times $10^{-3}\, M_\odot$ with average electron fractions $\lesssim
0.25$, and average entropies in the range $10{-}30\, k_{\rm B}$ per
baryon.

The mass of the dynamical ejecta has a large numerical uncertainty. From
the comparison of low- and high-resolution data, we estimate the relative
error in its determination to be as large as 50\%. Even larger
discrepancies of a factor of a few are found when comparing with and
among results published by different groups \citep{bauswein:2013yna,
sekiguchi:2015dma, lehner:2016lxy, bovard:2017mvn}. On the other hand,
the intensive properties of the ejecta, \eg, asymptotic velocity,
composition, etc., appear to be robust with resolution.

We have studied the dependency of the outflow properties on the \ac{NS}
masses and \ac{EOS}. Following \citet{dietrich:2016fpt} we have
constructed fits of the ejecta mass and velocity in terms of the \ac{NS}
masses and compactnesses. These capture some of the qualitative trend of
our data reasonably well, especially for the ejecta velocity.  However,
the fitting coefficients we infer from our data are significantly
different from those of \citet{dietrich:2016fpt}. This is not too
surprising given that most of the simulations used by
\citet{dietrich:2016fpt} employed idealized microphysics and neglected
weak reactions, which likely resulted in the overestimation of the
ejecta mass \citep{radice:2016dwd}.  We also find that there are
systematic effects not captured by these fits. One of the main reasons
is that the strength of the bounce of the merger remnant, an important
parameter determining the ejecta mass, depends on details of the
\ac{EOS} at higher densities and temperatures than those determining the
fitting parameters. Accurate models of the dynamical ejecta mass could
have important applications to multimessenger astronomy and, indeed, the
fits of \citet{dietrich:2016fpt} have already been applied, with small
variations, to GW170817 \citep{abbott:2017wuw, coughlin:2018miv}.
However, our results indicate that more work and better simulations will
be required to build truly quantitative ejecta models.

Our results indicate that soft \ac{NS} \acp{EOS} predict larger ejecta
masses, at least in the range of mass ratios we have probed, in
accordance with previous results \citep{hotokezaka:2012ze,
bauswein:2013yna, lehner:2016lxy, sekiguchi:2016bjd, dietrich:2016hky,
bovard:2017mvn}. \ac{EOS} effects are also imprinted in the ratio
between the masses of the tidal and shock-heated components of the
ejecta.  However, we do not find any clear correlation between ejecta
mass and velocity and the tidal deformability of the binary
$\tilde\Lambda$.  Consequently, it appears to be impossible to
directly constrain the \ac{NS} \ac{EOS} from measurements of the
dynamical ejecta. On the other hand, dynamical ejecta measurements might
help to break degeneracies in the other binary parameters and, in this
way, they might improve the constraints on the tidal deformability of
\acp{NS} indirectly.

We have analyzed the geometry of the outflows and found that it depends
on the relative amount of tidal- and shock-driven ejecta, as does the
average electron fraction. Consequently, we find that there is a
correlation between the rms opening angle of the ejecta and their average
electron fraction. This suggests that it might be possible to indirectly
constrain the composition, and hence the nucleosynthetic yields, of the
dynamical ejecta by combining observations of similar binary systems with
different orientations.

We have computed the velocity distribution of the dynamical ejecta and
found that shocks during and shortly after merger can accelerate a small
fraction of the ejecta ${\sim}10^{-6}\, M_\odot$ to asymptotic
velocities in excess to $0.6\, c$. This fast moving material is
preferentially located close to the orbital plane, presumably because of
the oblate shape of the merger debris cloud into which the accelerating
shocks propagates. It is also distributed in a very anisotropic fashion.
The amount of material reaching these high velocities depends on the
binary parameters and on the \ac{EOS}. Binaries with compact \acp{NS}
typically produce significantly more of this fast-moving component of
the outflow.  When comparing binaries simulated with the
BHB$\Lambda\phi$ and DD2 \acp{EOS}, we find that the former
systematically predicts a larger mass of the fast moving ejecta by a
factor of a few. This is because, even though BHB$\Lambda\phi$ and DD2
predict identical \ac{NS} structures for most of the binaries we have
considered, the former softens due to the appearance of
$\Lambda$-hyperons after merger and predicts more violent bounces
\citep{radice:2016rys}.

We find that binaries with remnants that are stable for at least several
rotation periods after merger result in the formation of massive
${\sim}0.1{-}0.2\, M_\odot$ accretion disks. Conversely, \acp{BH} formed
promptly after merger, \ie, within ${\sim}1\, {\rm ms}$, are endowed
with rather light accretion disks ${\sim}10^{-3}\, M_\odot$. More in
general, the accretion disk masses are found to depend on the lifetime of
the remnant and to correlate with the binary's tidal deformability
\citep{radice:2017lry}. It is expected that neutrino- and viscously-driven
outflows will carry away a significant fraction ${\sim}10{-}50\%$ of
these accretion disks on a timescale of a few seconds after the end of our
calculations \citep{metzger:2014ila, just:2014fka, siegel:2017jug,
fujibayashi:2017puw, fernandez:2018kax}. Accordingly, these secular
outflows are expected to be the dominant component of the ejecta. For
this reason, kilonova observations can be used to constrain the
accretion disk masses \citep{perego:2017wtu} and binary tidal
deformabilities \citep{radice:2017lry}.

\subsection{Nucleosynthesis}
We have computed the nucleosynthetic yields of the dynamical ejecta. We
find that second and third $r$-process peak isotopes, \ie, with $A
\gtrsim 125$, are robustly produced with relative abundance close to
solar. However, the relative abundance of light $r$-process elements,
\ie, with $90 \lesssim A \lesssim 125$, and second and third peak
isotopes are sensitive to the binary masses, weak reactions, and, to a
lesser extend, to the \ac{NS} \ac{EOS}.

We find that, for binaries close to the prompt \ac{BH} formation
threshold, different \ac{EOS} predict different ratios of tidal- and
shock-driven dynamical ejecta and, consequently, different relative
abundances of light $r$-process elements and second and third peak
$r$-process elements. In particular, the LS220 and DD2 \acp{EOS}
predicts that high mass binaries should produce a smaller relative
fraction of light elements compared to low mass binaries, while the
BHB$\Lambda\phi$ and SFHo \acp{EOS} have the opposite trend.

The absorption of neutrinos has a large impact on the composition and
yields of the outflow. When neutrino absorption is included, the ejecta
distribution in electron fraction become broad and $Y_e$'s of up to
$0.4$ are reached. The dynamical ejecta from comparable mass binaries in
simulations that included neutrino re-absorption produce $r$-process
isotopic abundances close to solar. If neutrino re-absorption is not
included, instead, light $r$-process elements are underproduced in our
simulations. The isotopic abundances in the region of the first
$r$-process peak are also sensitive to details of the neutrino
treatment, such as the assumptions made for the incoming neutrino
energy. This points to the need for simulations including energy
dependent neutrino-radiation treatments. This will be the object of
future work.

The binary mass ratio also impacts the production of first-peak
elements, with unequal mass systems underproducing elements with $90
\lesssim A \lesssim 125$. We find the relative yields to depend on the
mass ratio almost as sensitively as on the neutrino treatment.

The secular ejecta is expected to be an important, if not dominant,
component of the overall outflow, especially in the case of long-lived
remnants or when massive disks are formed. Consequently, the secular
ejecta should be taken into account when estimating the nucleosynthetic
yields from \ac{NS} mergers. Presently, however, we can only speculate
on the nucleosynthesis from this component. Long term merger and
postmerger simulations will be required to construct self-consistent
models. We leave this to future work.

We have estimated the $r$-process nucleosynthesis yields using the
simulation data extracted at a fixed radius of $300\, G/c^2\, M_\odot
\simeq 450\, {\rm km}$ and used precomputed yields from parametrized
trajectories. This is the same strategy we employed in
\citet{radice:2016dwd}. As part of our analysis, we have now validated
this procedure by comparing with a more computationally expensive
analysis done with Lagrangian tracer particles. We find the
nucleosynthesis to be rather insensitive to the detailed thermodynamical
history of the tracer particles. The electron fraction of the material
does not significantly evolve between the time we record it at $300\,
G/c^2\, M_\odot$ and when the temperature drops below $6\, {\rm GK}$ and
the $r$-process begins. On the other hand, we find that, due to advection
errors in the position of the tracer particles, the $Y_e$ distribution
inferred from the tracers can differ quantitatively from that inferred
from the grid data at the same radius. Overall, however, the differences
between the results obtained with the two methods are well below those
associated with other sources of uncertainty, most notably the treatment
of neutrino radiation.

\subsection{Electromagnetic Counterparts}
We have computed synthetic kilonova light curves for all our models
using the model of \citet{perego:2017wtu}. Our light curve model is
informed using the detailed angular structure of the outflows, density,
velocity, and composition, as extracted from the simulations. These have
been augmented with the inclusion of secular ejecta composed of
neutrino-driven and viscously-driven winds which we assume to entrain 3\%
and 20\% of the disk, respectively. These values are motivated by recent
long-term postmerger simulations \citep{perego:2014fma, martin:2015hxa,
just:2014fka, metzger:2014ila, just:2014fka, siegel:2017jug,
fujibayashi:2017xsz, fujibayashi:2017puw, fernandez:2018kax}.

Binaries resulting in prompt \ac{BH} formation have kilonovae dominated
by the dynamical ejecta. These peak at about a day after merger and then
rapidly decline. They are also very red, with $g - K_s \simeq 10$
mag three days after the merger. As the merger remnant lifetime
increases, so does the disk mass, and hence the associated kilonovae
become increasingly dominated by radiation coming from the secular
ejecta. These kilonovae peak on longer timescales of few days to a week
and are bluer, with $g - K_s \simeq 3{-}7$ mag three days after the
merger. This suggests that kilonovae observations could be used to probe
the lifetime of the merger remnant. A similar suggestion was also put
forward by \citet{metzger:2014ila} and, more recently, by
\citet{lippuner:2017bfm}. They argued that longer lived remnants should
result in bluer kilonovae because of the neutrino irradiation of
the ejecta from the central remnant. This is expected to lower the
electron fraction of the outflow below the threshold needed for the
production of lanthanides resulting in a drop of the photon opacity of
the material. The phenomenon we find goes in the same direction, but is
physically distinct since it arises from the correlation between disk
masses and remnant lifetimes and not from changes in the ejecta
photon opacity, which we have kept constant instead.

Since for most binaries the kilonova signal is dominated by the secular
ejecta, the effect of the inclusion of neutrino re-absorption in the
modeling of the dynamical ejecta has only a modest impact on the
kilonova signal. Moreover, since remnants disk masses strongly correlate
with binary tidal deformabilities, the peak properties of the kilonovae,
peak time, magnitude, width, are found to depend sensitively on
$\tilde\Lambda$. This suggests that kilonovae observations could be used
directly to probe the \ac{EOS} of \acp{NS}. However, it is important to
stress than this correlation is in part due to our assumption that a
fixed fraction of the accretion disk is ejected after merger. Long-term
simulations of the evolution of postmerger remnants are necessary to
verify if this assumption is valid, or whether light and massive
postmerger disks evolve in quantitatively different ways.

We have computed the synchrotron radio signal expected from the
interaction of the ejecta with the ISM using the model of
\citet{hotokezaka:2015eja}. The radio fluence and the timescales over
which the radio remnant evolves depend sensitively on the ISM density,
and on the kinetic energy and the detailed velocity distribution of the
ejecta.

We find that some of our models predict a rebrightening of the
synchrotron signal from GW170817 on a timescale of months to years from
the merger, when the emission from the ejecta will start to dominate
over the emission coming from the \ac{SGRB} jet.

Depending on the ISM density and the orientation of the binary, it could
be possible to detect the radio signal due to the interaction of the
ejecta with the ISM on timescales of weeks to months after merger at
distances of up to ${\sim}100$ Mpc. These observations would allow us to
probe the velocity distribution of the ejecta and, hence, the violence
with which the merged objects bounces back after the \ac{NS} collision.
The latter, in turn, depends on the \ac{EOS} of matter at several times
nuclear density. For instance, due to the appearance of $\Lambda$
hyperons after the merger, the BHB$\Lambda\phi$ \ac{EOS} predicts more
violent mergers and brighter radio flares by up to two orders of
magnitude compared with the DD2 EOS from which it only differs at high
densities due to the inclusion of hyperons. This suggests that radio
flares could be used to probe the \ac{EOS} of \acp{NS} in a regime that
is not accessible with \ac{GW} observations of the inspiral
alone\footnote{The high density \ac{EOS} could also be constrained with
\ac{GW} observations of the postmerger signal
\citep{bernuzzi:2015rla,radice:2016rys, most:2018eaw,
bauswein:2018bma}.}.


\begin{acknowledgments}
It is a pleasure to acknowledge A.~Burrows for the many stimulating
discussions.
DR gratefully acknowledges support from a Frank and Peggy Taplin
Membership at the Institute for Advanced Study and the
Max-Planck/Princeton Center (MPPC) for Plasma Physics (NSF PHY-1523261).
AP acknowledges support from the INFN initiative "High Performance data
Network" funded by CIPE.  DR and AP acknowledge support from the
Institute for Nuclear Theory (17-2b program). AP thanks the Institute
for Advanced Study for its hospitality and support. SAF acknowledges
support from the United States Department of Energy through the
Computational Science Graduate Fellowship, grant number DE-SC0019323.
SB acknowledge support by the EU H2020 under ERC Starting Grant,
no.~BinGraSp-714626.  LFR acknowledges support from U.S. Department of
Energy through the award number DE-SC0017955.
The simulations were performed on BlueWaters, Bridges, Comet, and
Stampede, and were enabled by the NSF PRAC program (ACI-1440083 and
AWD-1811236) and the NSF XSEDE program (TG-PHY160025). The analysis
employed computational resources provided by both the TIGRESS high
performance computer center at Princeton University, which is jointly
supported by the Princeton Institute for Computational Science and
Engineering (PICSciE) and the Princeton University Office of Information
Technology, and the Institute for Cyber-Enabled Research, which is
supported by Michigan State University.
\end{acknowledgments}

\paragraph{Software}
\texttt{WhiskyTHC} \citep{thccode}, \texttt{Lorene} \citep{lorenecode},
Matplotlib \citep{hunter:2007a}, NumPy \citep{Oliphant:2015a}, SciPy
\citep{jones:2001a}, Pandas \citep{mckinney:2010a}, iPython
\citep{perez:2007a}, VisIt \citep{visitcode}.

\bibliography{references,references_KH}

\appendix

\section{Finite-Resolution Effects}
\label{sec:resolution}
Here we discuss uncertainties in the ejecta mass, velocity, and
composition due to finite-resolution effects. The mass of the dynamical
ejecta has not yet reached convergence at the resolution adopted in our
simulations. As can be seen in Table~\ref{tab:ejecta}, the dynamical
ejecta mass shows moderate variations and changes in a non monotonic way
as we vary the grid resolution. Consequently, our estimate for the
ejecta mass should only be considered as semi-quantitative. We estimate
the relative uncertainty in the ejecta mass due to finite-resolution
effects alone to be as large as 50\%. We remark that, as discussed in
Sec.~\ref{sec:ejecta.dynamical}, the differences between ours and other
published results, as well as between results published by different
groups, are even larger (see also Table~\ref{tab:ejecta.comparison}).

\begin{figure*}
  \includegraphics[width=3.5in]{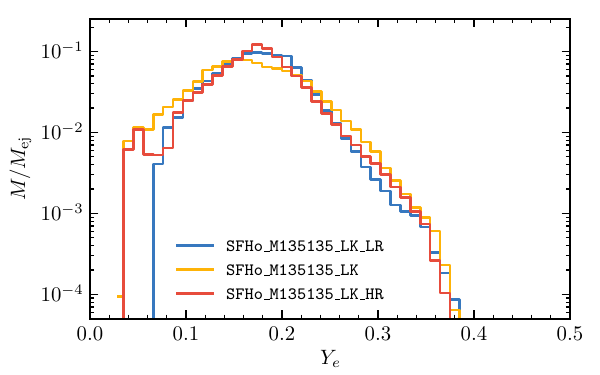}
  \hfill
  \includegraphics[width=3.5in]{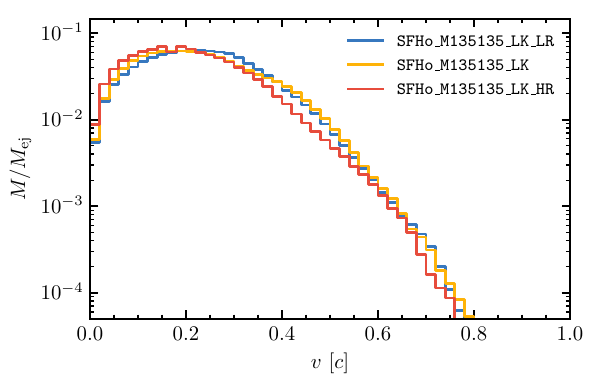}
  \caption{Dynamical ejecta sensitivity to resolution. \textit{Left
  panel}: electron fraction. \textit{Right panel}: asymptotic velocity.
  The total ejecta mass shows large fractional variation with
  resolution, however electron fraction, velocity distribution, and
  nucleosynthetic yields appear to be robust.}
  \label{fig:ejecta_dep_res}
\end{figure*}

Despite the significant uncertainty in the total ejecta mass, we find
the other properties of the outflow, velocity, composition, entropy, to
be robust with resolution. As an example, we show in
Fig.~\ref{fig:ejecta_dep_res} histograms of the ejecta as a function of
$Y_e$ and of the asymptotic velocity for the \texttt{SFHo\_M135135\_LK}
binary. The robustness of the ejecta properties suggests that, while the
overall morphology of the outflow streams are well captured by the
simulations, there are stochastic variations, for example resulting in
the fallback of some of the material in these streams due to interaction
with the merger debris, that affect the overall amount of matter
entrained by these outflows (see also the discussion in
\citealt{bauswein:2013yna}).

\section{Treatment of the Viscous Fluxes}
\label{sec:taudiff}
The \ac{GRLES} approach requires the inclusion of the term $\partial_j
(\sqrt{-g} \tau^{ij})$ in the fluid momentum equations. This numerical
derivative needs to be carefully treated to avoid the odd-even
decoupling instability. To this aim we introduce the left and right
biased finite differencing operators
\begin{equation}
  \partial_i^\pm u = \pm\frac{u(x \pm h \mathbf{e}_i) -
  u(x)}{h}\,,
\end{equation}
where $?[c]e_i^j? = ?[c]\delta_i^j?$ are the coordinate basis vectors
and $h$ is the grid spacing. For the covariant derivatives we use the
finite-differencing operators
\begin{equation}
  D_i^\pm u^j = \partial_i^\pm u^j + ?[c]\Gamma^j_{ik}? u^k\,,
\end{equation}
where $?[c]\Gamma^j_{ik}?$ are the Christoffel symbols of the Levi-Civita
connection associated with the spatial metric $\gamma_{ij}$. These are
computed using a standard centered 2nd order finite-differencing
operator.

Using the biased finite-differencing operators we define
\begin{equation}
  \tau_{ij}^\pm = - 2\, \nu_{_T}\,  (\rho + p)\, W^2 \left[
  \frac{1}{2} \big(D_i^\pm v_j + D_j^\pm v_i\big) -
  \frac{1}{3} D_k^\pm v^k \gamma_{ij} \right]\,.
\end{equation}
Then we discretize the divergence of $\tau^{ij}$ as
\begin{equation}
  \partial_j (\sqrt{-g} \tau^{ij}) \simeq
    \frac{1}{2}  \left\{
      \partial_j^- \big[\sqrt{-g}\, (\tau^+)^{ij}\big] +
      \partial_j^+ \big[\sqrt{-g}\, (\tau^-)^{ij}\big]
    \right\}\,.
\end{equation}
It is easy to verify that this yields a flux-conservative scheme. Exact
conservation is also ensured at refinement level boundaries using the
flux correction algorithm of \citet{berger:1989a} as implemented in
\citet{reisswig:2012nc}.

\acrodef{ADM}{Arnowitt-Deser-Misner}
\acrodef{AMR}{adaptive mesh-refinement}
\acrodef{BH}{black hole}
\acrodef{BBH}{binary black-hole}
\acrodef{BHNS}{black-hole neutron-star}
\acrodef{BNS}{binary neutron star}
\acrodef{CCSN}{core-collapse supernova}
\acrodefplural{CCSN}[CCSNe]{core-collapse supernovae}
\acrodef{CMA}{consistent multi-fluid advection}
\acrodef{CFL}{Courant-Friedrichs-Lewy}
\acrodef{DG}{discontinuous Galerkin}
\acrodef{HMNS}{hypermassive neutron star}
\acrodef{EM}{electromagnetic}
\acrodef{ET}{Einstein Telescope}
\acrodef{EOB}{effective-one-body}
\acrodef{EOS}{equation of state}
\acrodef{FF}{fitting factor}
\acrodef{GR}{general-relativistic}
\acrodef{GRLES}{general-relativistic large-eddy simulation}
\acrodef{GRHD}{general-relativistic hydrodynamics}
\acrodef{GRMHD}{general-relativistic magnetohydrodynamics}
\acrodef{GW}{gravitational wave}
\acrodef{ILES}{implicit large-eddy simulations}
\acrodef{LIA}{linear interaction analysis}
\acrodef{LES}{large-eddy simulation}
\acrodefplural{LES}[LES]{large-eddy simulations}
\acrodef{MRI}{magnetorotational instability}
\acrodef{NR}{numerical relativity}
\acrodef{NS}{neutron star}
\acrodef{PNS}{protoneutron star}
\acrodef{SASI}{standing accretion shock instability}
\acrodef{SGRB}{short $\gamma$-ray burst}
\acrodef{SPH}{smoothed particle hydrodynamics}
\acrodef{SN}{supernova}
\acrodefplural{SN}[SNe]{supernovae}
\acrodef{SNR}{signal-to-noise ratio}

\end{document}